\documentclass{pasj01}
\onecolumn
\draft
\Received{$\langle$2019 April 8$\rangle$}
\Accepted{$\langle$2019 October 29$\rangle$}
\Published{$\langle$publication date$\rangle$}
\usepackage{natbib}
\usepackage{rotating}

\newcommand{\fermi}{\textit{Fermi}}
\newcommand{\gr}{$\gamma$-ray}

\begin{document}

\title{Property Studies of ``Loner" Flares of Gamma-Ray Blazars}

\author{Gege Wang\altaffilmark{1,2}, Zhongxiang Wang\altaffilmark{1}, 
Liang Chen\altaffilmark{1},
Jianeng Zhou\altaffilmark{1}, \& Yi Xing\altaffilmark{1}}

\altaffiltext{1}{\footnotesize 
Shanghai Astronomical Observatory, Chinese Academy of Sciences, Shanghai 200030, China}

\altaffiltext{2}{\footnotesize 
Graduate University of the Chinese Academy of Sciences, No. 19A, Yuquan Road, Beijing 100049, China}


\maketitle

\begin{abstract}
We search through $\gamma$-ray data obtained with the Large Area Telescope
(LAT) onboard the {\it Fermi Gamma-Ray Space Telescope} and find 
24 blazars (or candidates) that have a single clear flare event in
their 9.5 year long-term light curves. 
We define these events as loner flares since each flare stands out
significantly above the relatively stable, low-flux light curve.
We analyze the LAT data in detail for these 24 sources. The flares
in ten of them are primarily due to a single sharp peak, for which we 
study by fitting
with two different analytic functions. The time durations thus determined 
for the sharp 
peaks are in a range of 4--25 days. The $\gamma$-ray spectra of the 24
blazar sources
can be described with a power-law or a log-parabola function. We obtain their
spectral properties in the flaring and quiescent states, and find that
in the flares 16 of the sources have harder emission and three have softer 
emission while the other five keep the same emission.
We discuss a possible correlation between the differences in photon index 
in the quiescent and flaring states and photon indices in quiescence. In 
addition, the sharp peak flares seem to have a tendency of having long
time durations and hard emission, possibly related to their physical origin
in a blazar jet. Studies of more similar flares will help establish
these possible features.

\end{abstract}

\section{Introduction}
Since its launch in June 2008, the Large Area Telescope (LAT) on board the
{\it Fermi Gamma-ray Space Telescope (Fermi)} has conducted observations 
at \gr\ energies of 0.1--300 GeV over 10 years by scanning the whole sky 
every three
hours. More than 3000 sources have been reported in the \fermi\ LAT third
source catalog (3FGL; \citealt{ace+15}), and nearly 5000 sources have
been very recently
listed in the LAT  fourth source catalog (4FGL; \citealt{4fgl19}).
From the observations, large amount of data are collected for 
the detected sources, now allowing detailed analysis for the purpose of 
studying different properties of the sources.

From \fermi\ LAT, it has been well established that the dominant \gr\ sources 
in the sky are blazars \citep{3fagn15,4fagn19}. These subclass sources
of the Active Galactic Nuclei (AGN)
have a jet pointing close to the line of sight. 
Due to the relativistic beaming effect, emission from the jet dominates over
that from the host galaxy over nearly the entire electromagnetic spectrum,
and thus blazars are characterized by rapid and 
large-amplitude flux variations (e.g., \citealt{umu97}).
The all-sky
monitoring capability of the \fermi\ LAT has provided us with light curve
and spectral data for more than 1800 identified and candidate 
blazars (\citealt{3fagn15};  and $\sim$2800 in the very recently updated 
fourth 
catalog of AGN detected with LAT; \citealt{4fagn19}). Many studies have been 
carried out,
while focusing on bright \gr\ blazars and their flares, and related fast 
variability and spectral changes (some combined with multi-band observations; 
e.g., \citealt{nal13,hcd14,kss14,hay+15,fsub16,msb19}; also for recent reviews
of AGN at \gr\ energies, see \citealt{dg16,ms16}). The studies have provided
constraints on
the physical properties of the jet emission zones and emission 
processes.

 Generally blazars appear to vary randomly, with their power spectral 
densities (PSDs) often described with
a power law (e.g., \citealt{cha+08,abd+10a,cha+12}).
The noise-like fluctuations can be explained as due to the presence of
many turbulent cells, driven by instability or magnetic reconnection.
Sometimes a significantly bright flare may be seen above minor fluctuations
for a blazar,
and it is intriguing to check the properties of such flares, how they arise
and decay and whether their spectra have significant changes. The properties
help understand the underlying physical processes.
In our systematic analysis of \fermi\ LAT data for obtaining long-term
light curve 
information for \gr\ blazars,  which was based on 3FGL,  
we have noted one type of ``loner" flares. They appeared as one single 
event lasting comparatively short (most with
a time duration of three months) while in most
time the blazar source has been in a  relatively quiescent state 
without other strong 
flaring activity. More specifically, in our 90-day binned light curve data, 
one to three data points stand out significantly as a flare 
above the otherwise low-flux flat light curve. 
We have selected these loner flare sources as our targets, for the purpose
of studying the properties of the flares and compare them with those
of the quiescent states. For these sources, 
the flaring and quiescent states can be clearly defined. 
Detailed flux variations during a flare and related spectral changes have
been obtained.
In this paper, we report 
the results from the studies. In Section 1.1, we describe our target 
selection. We present
the data analysis and results in Section 2 and 3 respectively, and discuss
the results in Section 4.
\begin{figure}
\centering
\includegraphics[scale=0.5]{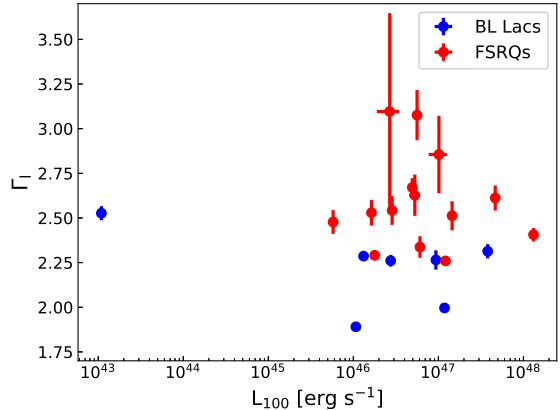}
\caption{Photon indices ($\Gamma$ in a power law or $\alpha$ in a log-parabola)
and 0.1--100 GeV luminosities 
of 21 blazar targets (with redshifts) in the quiescent state.}
\label{fig:gl}
\end{figure}

\subsection{Target Selection}
Using the \fermi\ LAT data, we obtained 90-day binned light curves (9.5 years
long) for more 
than 1800 identified and candidate blazars \citep{ace+15} for our systematic
study of blazar variability.  The choice of 90-day per bin was more for
the consideration of computational time required but without losing 
sensitivity to large flares. For the data selection and analysis, see below
Section 2. We went over all the light curves and found  nearly 30 loner
flare sources. Among them, several events were not blazar flares, 
but instead were caused by solar flares or \gr\ bursts occurring in the field
(with references to \gr\ solar flare observations\footnote{\footnotesize http://hesperia.gsfc.nasa.gov/fermi\_solar/} and {\it Fermi}-LAT GRB List of detections\footnote{\footnotesize www.asdc.asi.it/grblat/}).
We excluded those sources, and in the end found 24 blazars with a loner flare.
 The blazars are listed in Table~\ref{tab:src}.
Most of the flares of the blazars are 10$\sigma$--19$\sigma$ above 
the quiescent light 
curves, while in the cases of J0055.1$-$1219, J0303.4$-$2407, J1040.5+0617, and J2134.2$-$0154, the significances compared to the quiescent light curves
are 6.0$\sigma$, 7.7$\sigma$, 7.1$\sigma$, and 9.3$\sigma$, respectively.

\begin{figure*}
\centering
\includegraphics[width=0.3\textwidth]{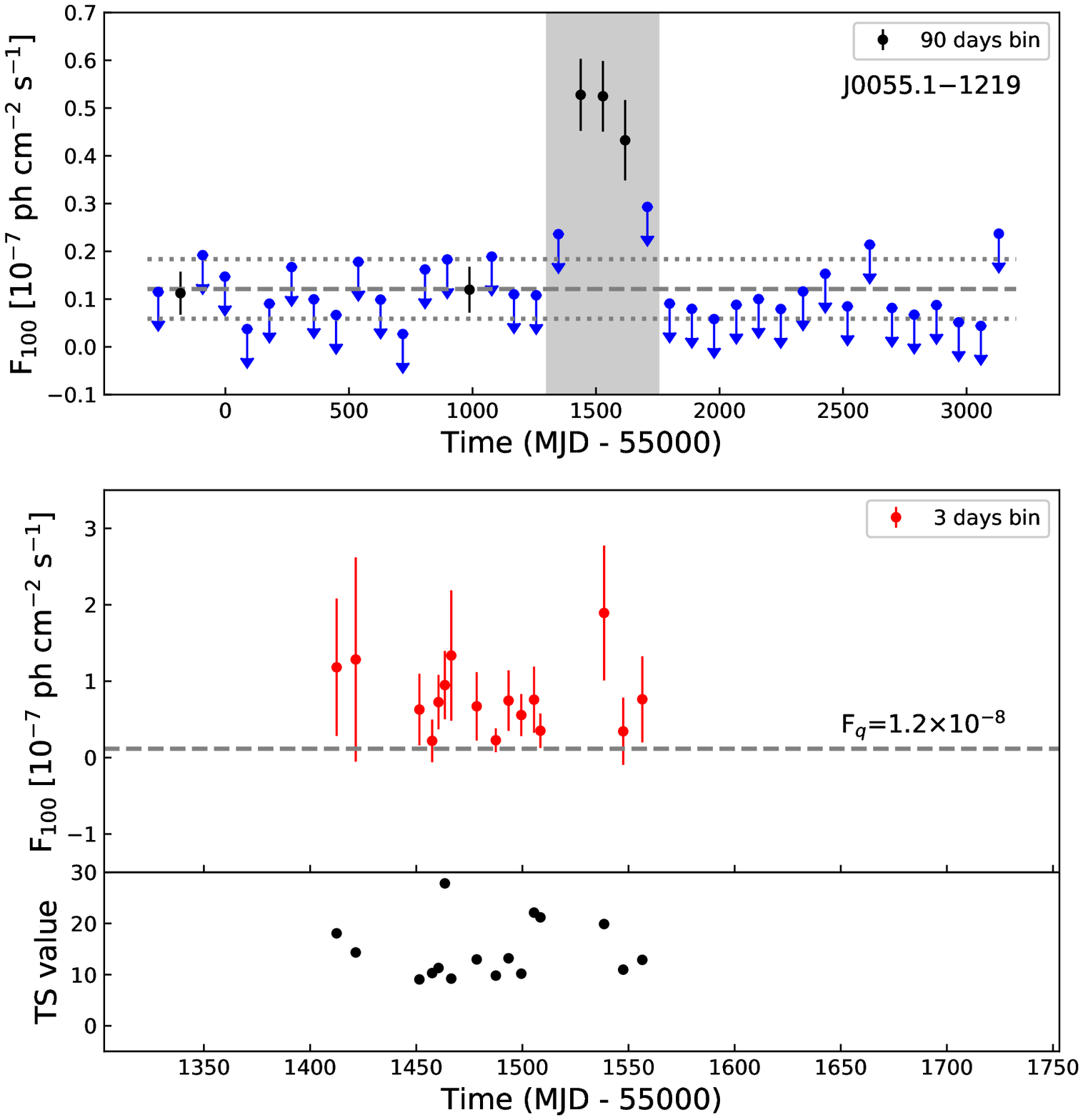}
\includegraphics[width=0.3\textwidth]{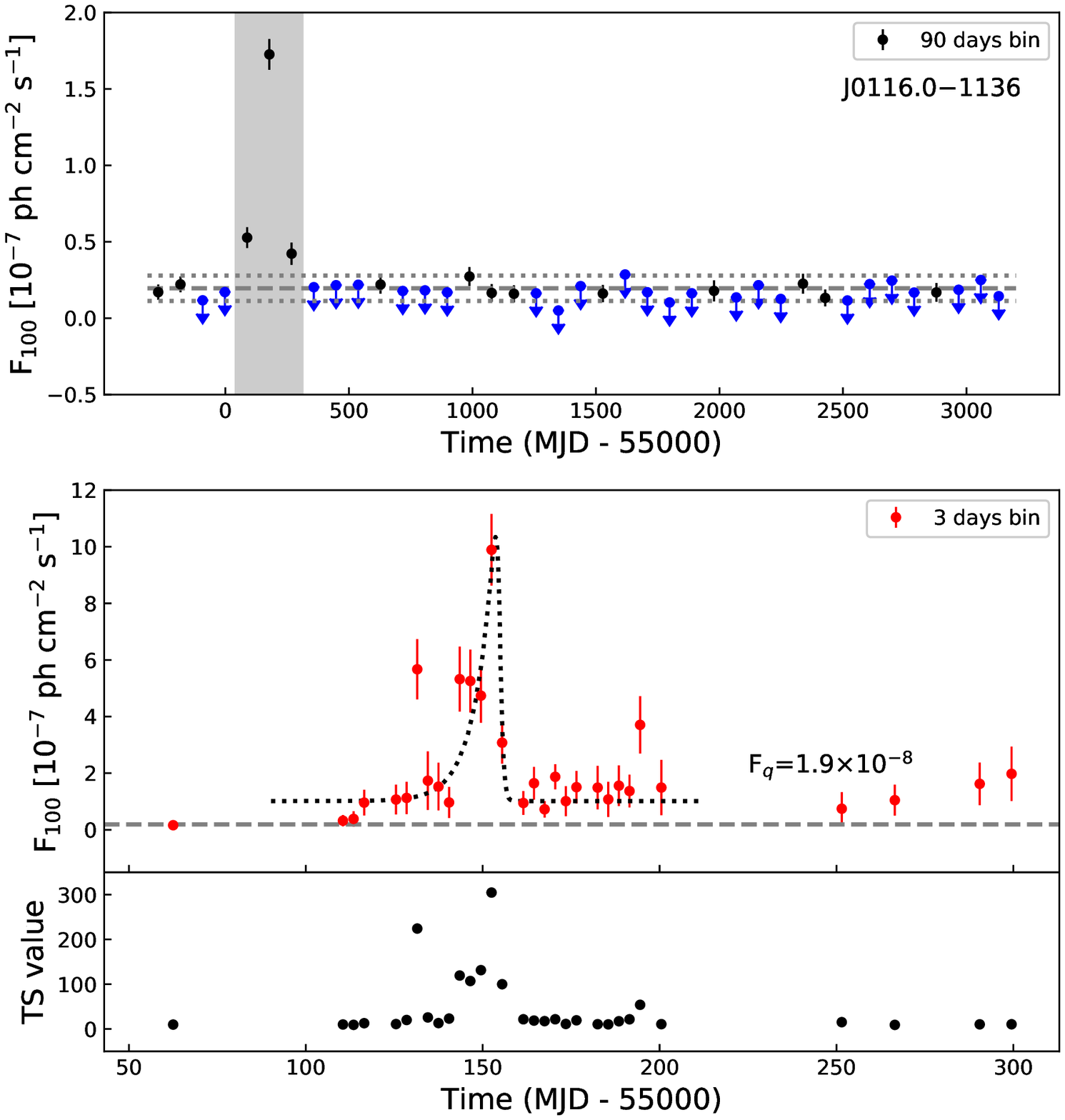}
\includegraphics[width=0.3\textwidth]{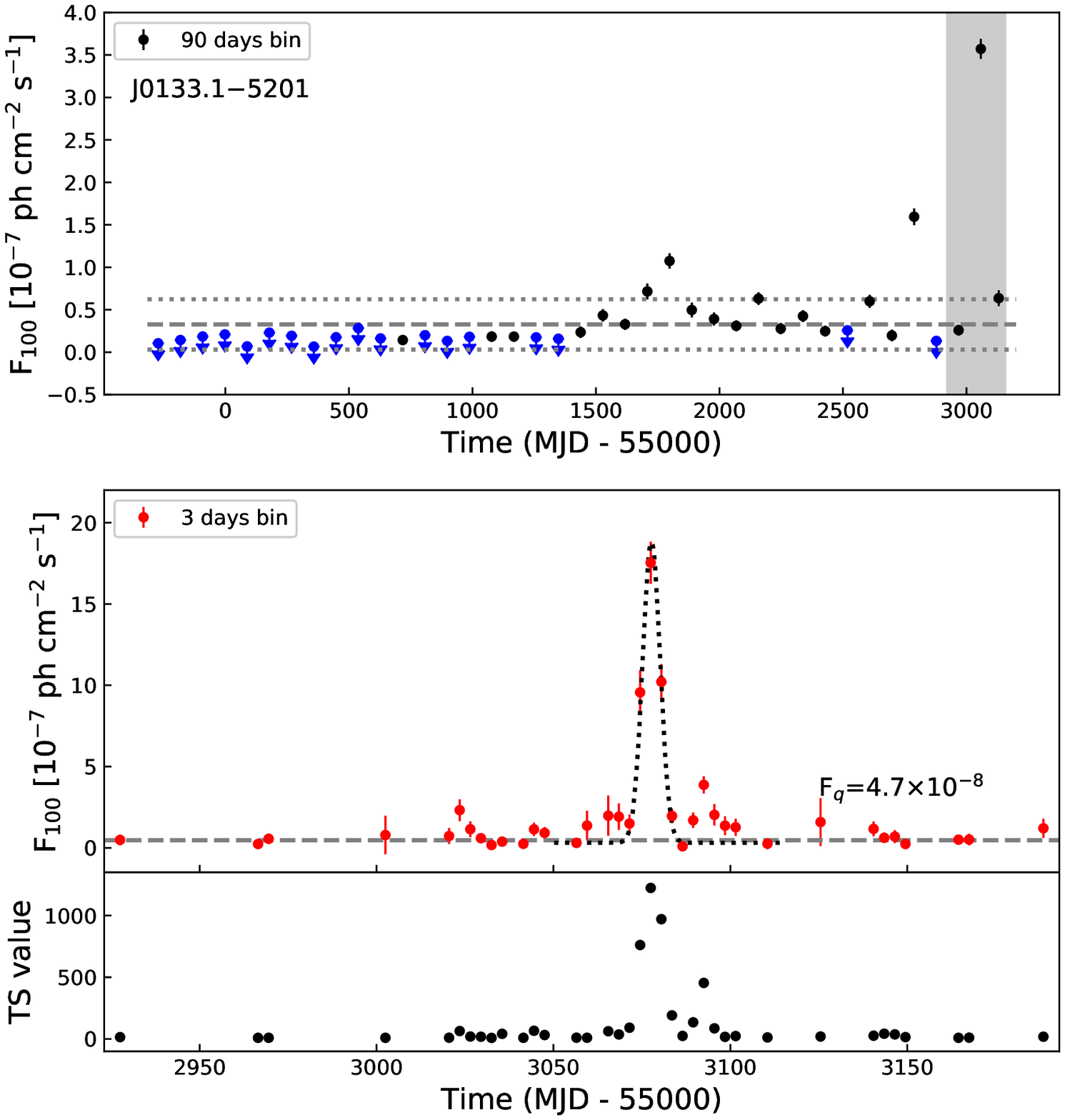}

\includegraphics[width=0.3\textwidth]{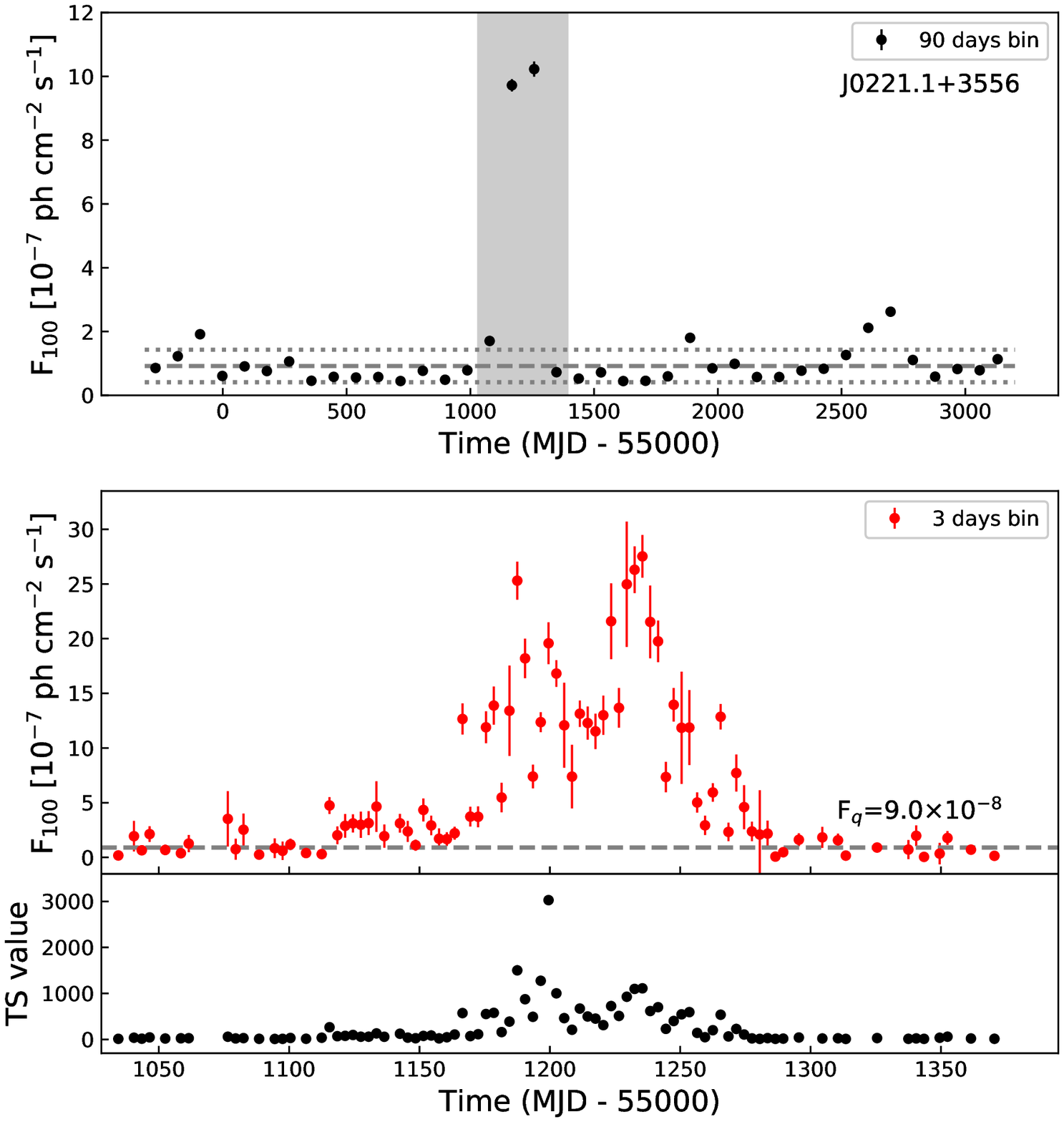}
\includegraphics[width=0.3\textwidth]{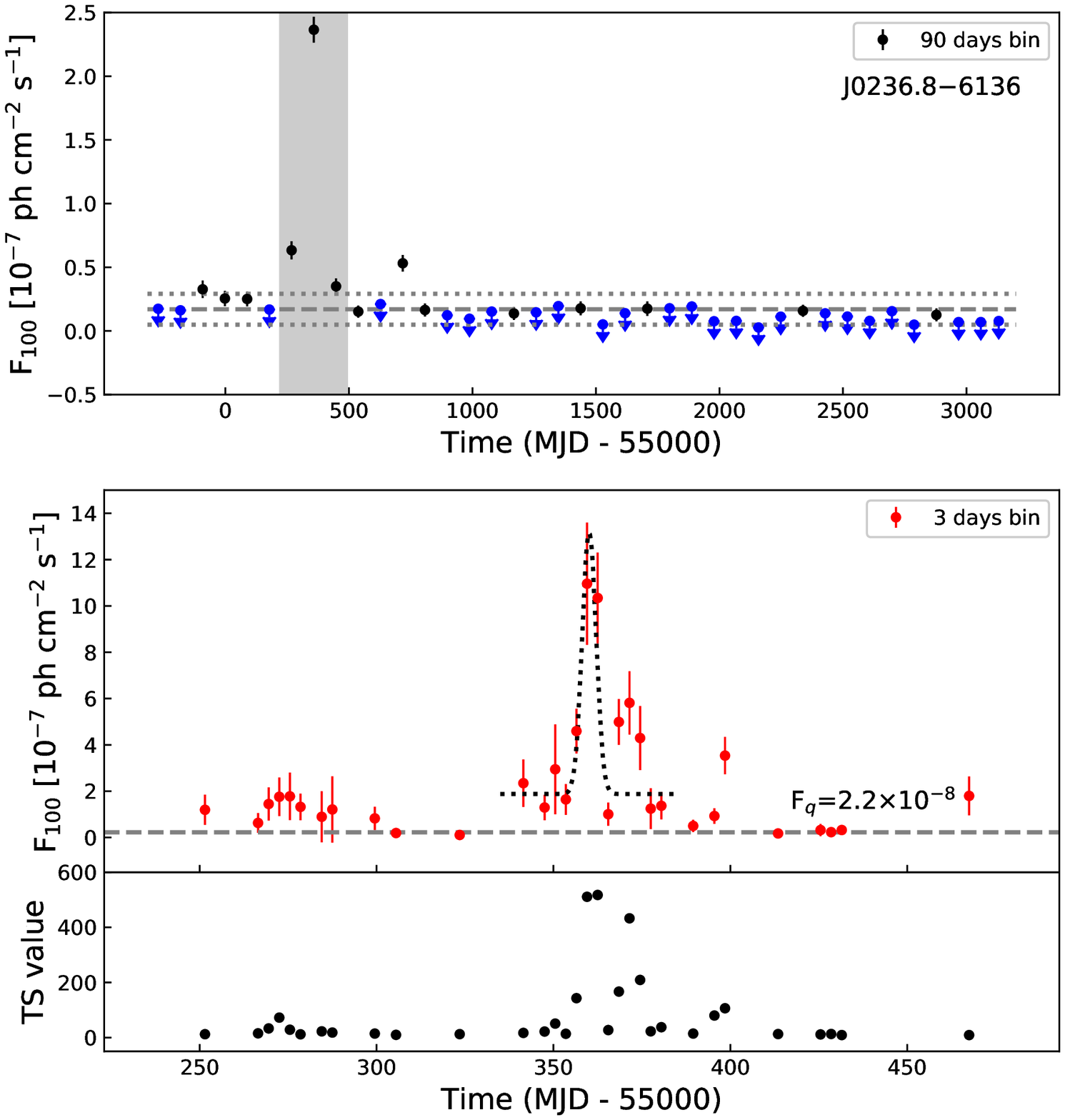}
\includegraphics[width=0.3\textwidth]{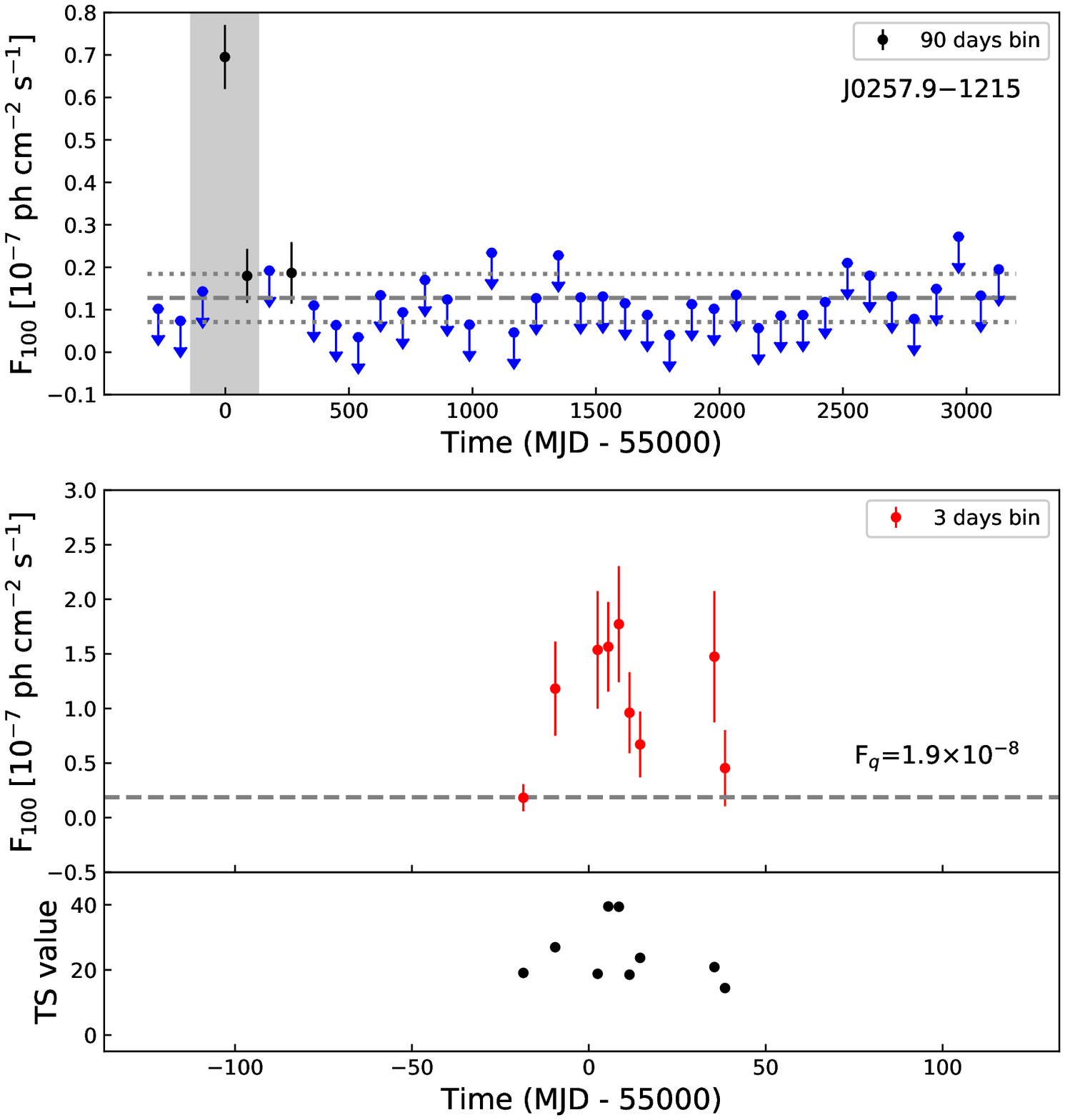}

\includegraphics[width=0.3\textwidth]{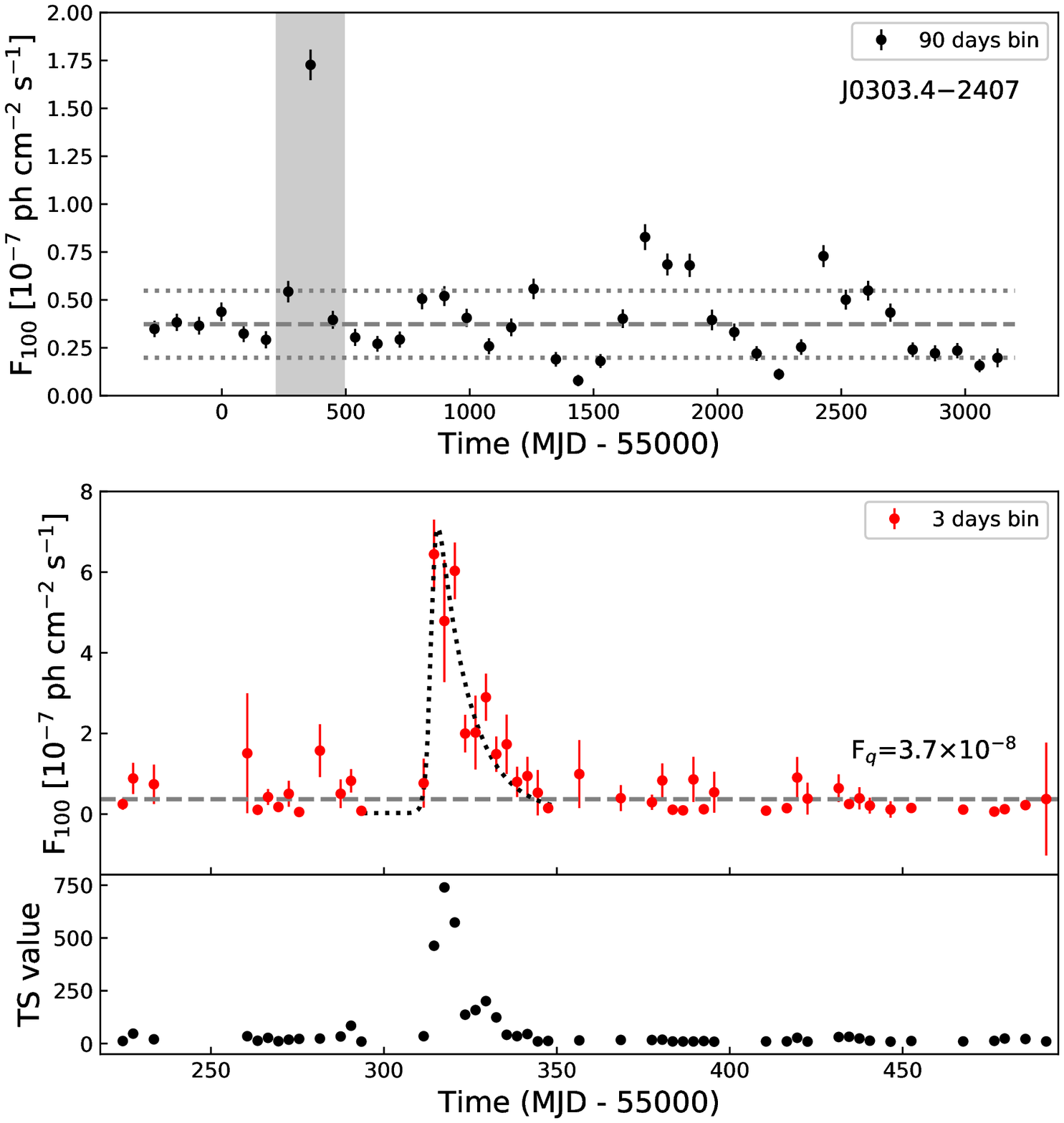}
\includegraphics[width=0.3\textwidth]{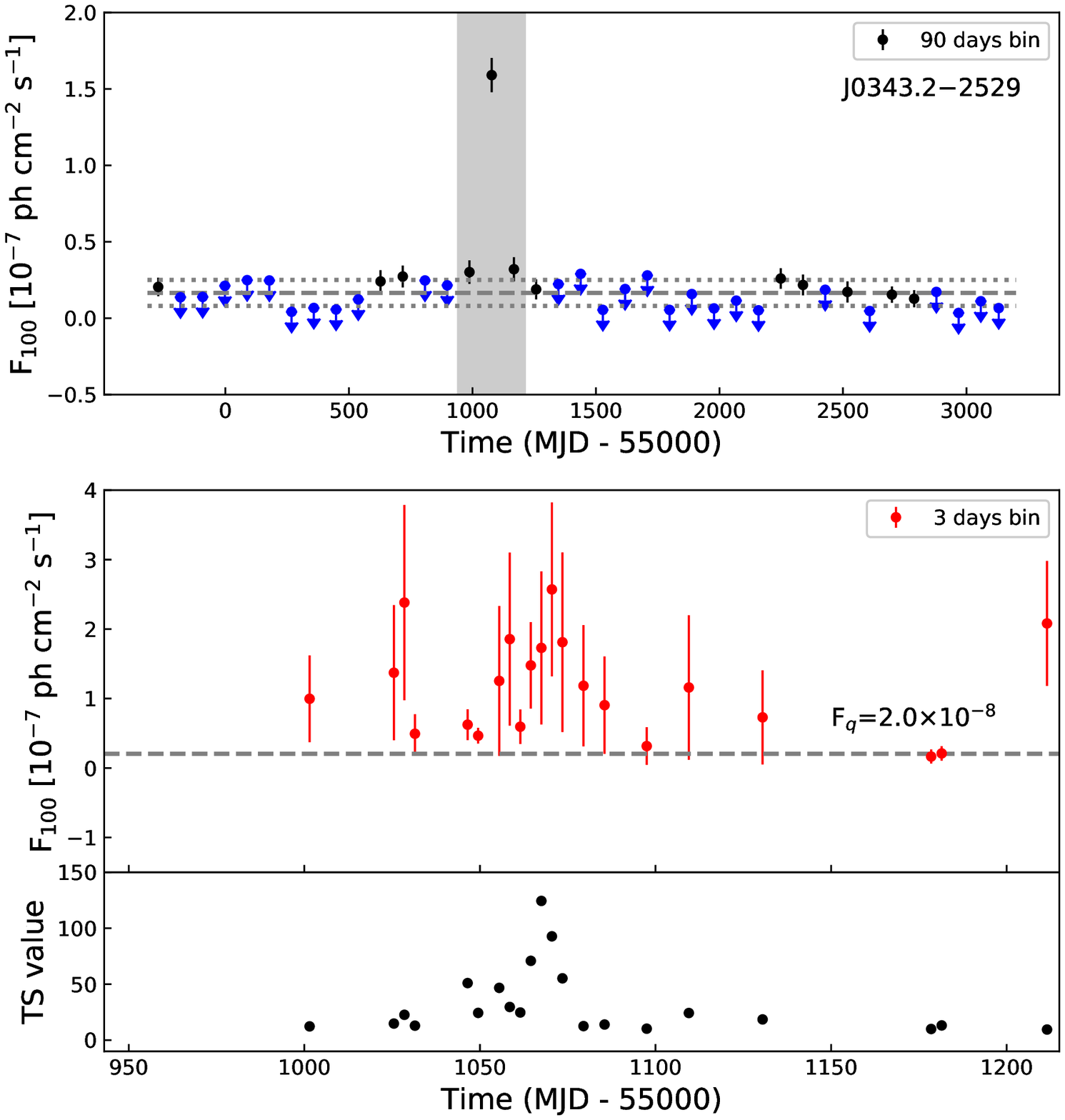}
\includegraphics[width=0.3\textwidth]{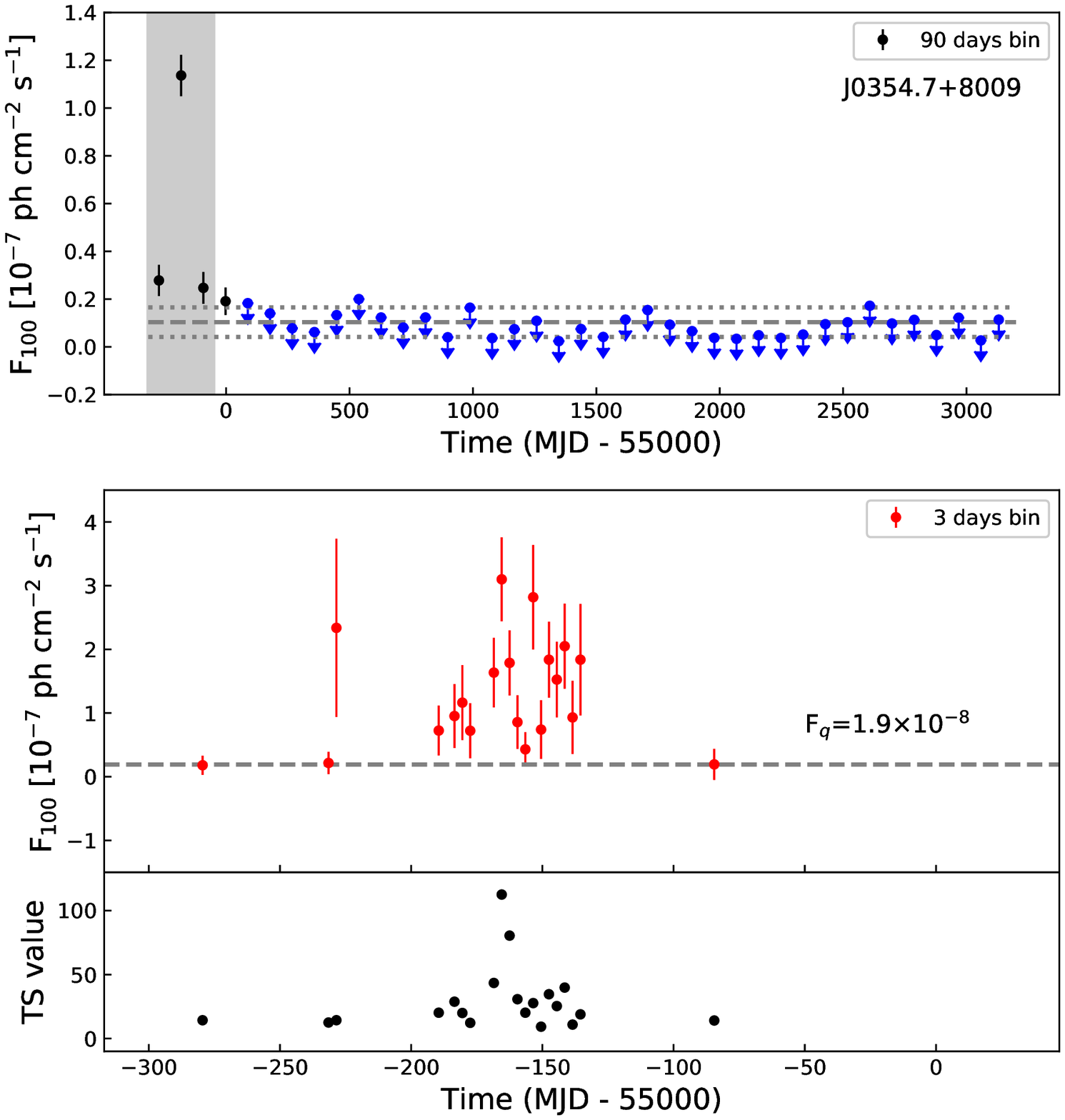}

\includegraphics[width=0.3\textwidth]{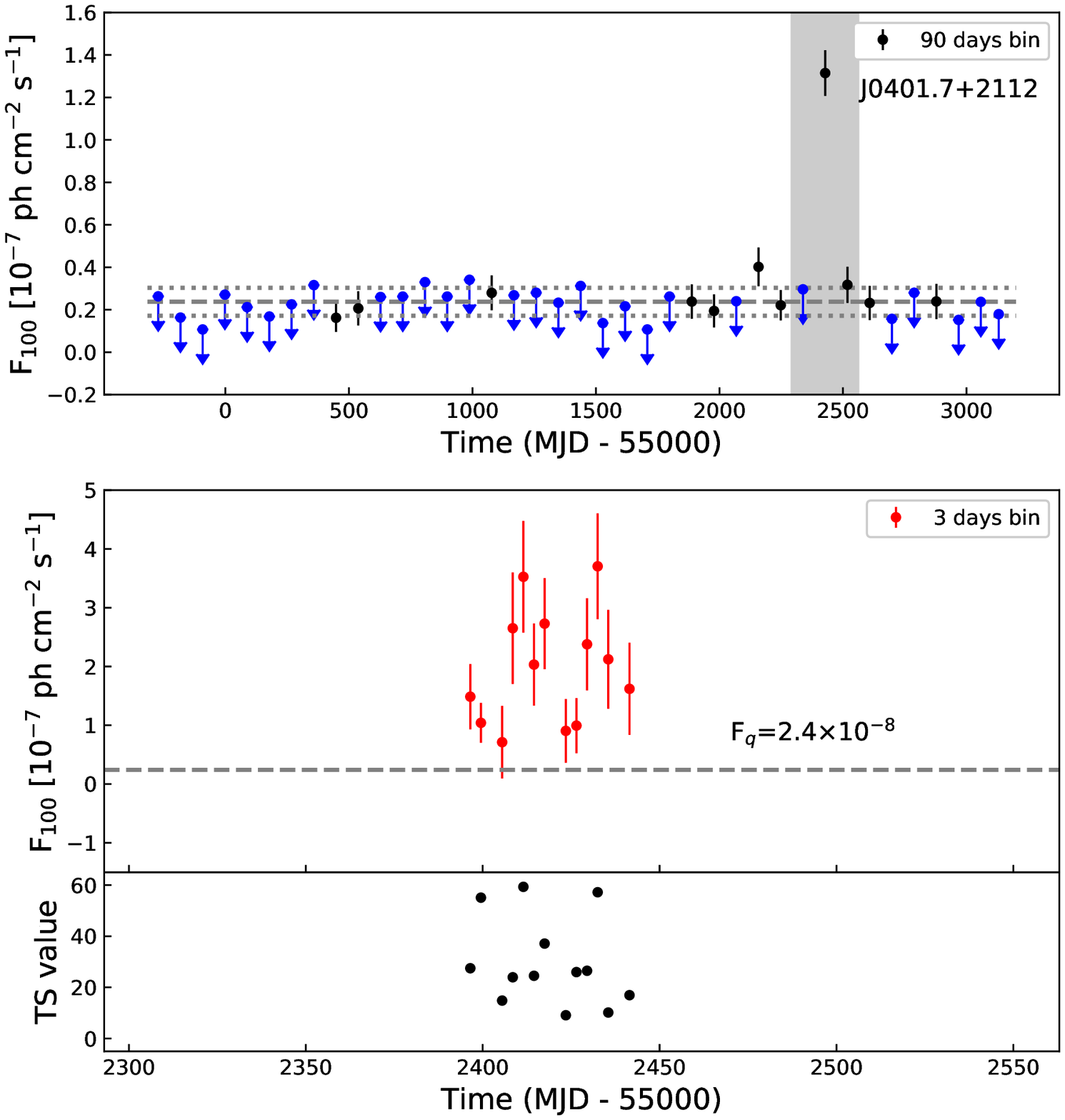}
\includegraphics[width=0.3\textwidth]{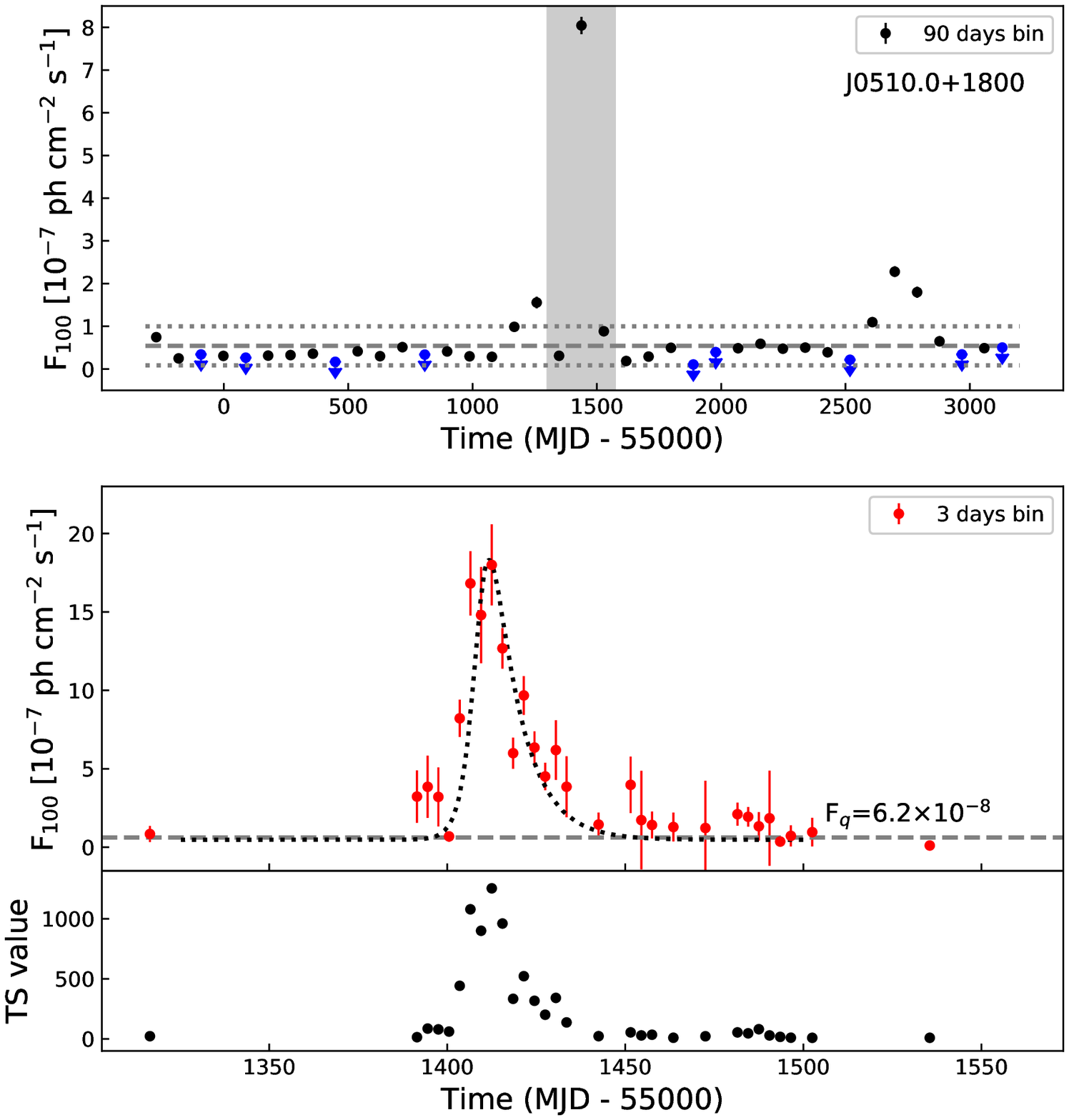}
\includegraphics[width=0.3\textwidth]{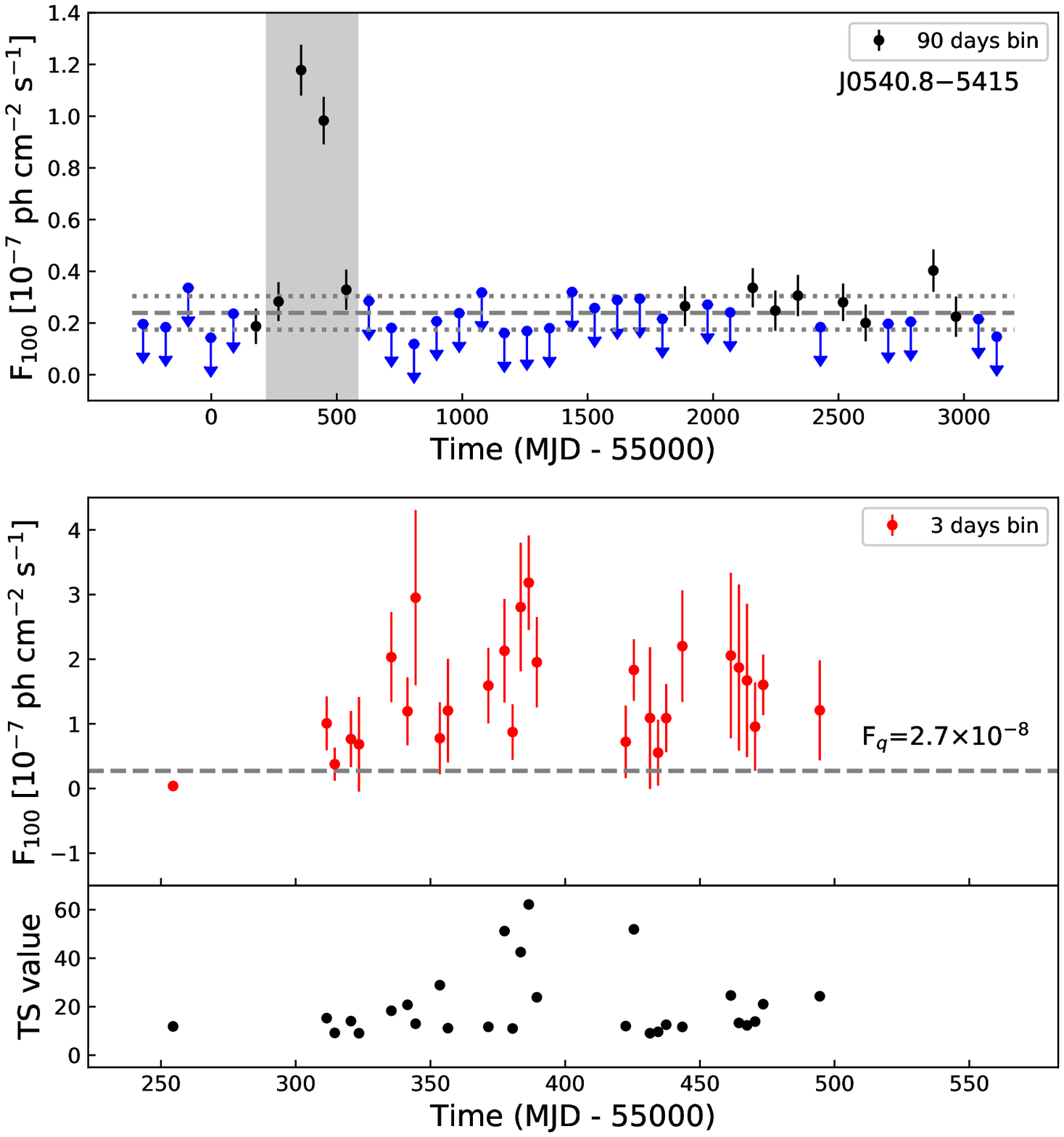}
\caption{{\it Top panels}: light curves binned in 90 day intervals.
 The average and 1$\sigma$ range of the quiescent
data points are indicated by dashed and dotted lines, respectively.
{\it Bottom panels:} light curves binned in 3 day intervals, 
the time ranges of which are given by the gray areas in the top panels. 
When there is a sharp-peak profile contained in a flare,  an
analytic function
(dotted curve) was used to fit the profile. The flux levels of the quiescent 
states are indicated by dashed lines, with the flux values ($F_q$) provided.}
\label{fig:lc}
\end{figure*}

\setcounter{figure}{1}
\begin{figure*}
\includegraphics[width=0.3\textwidth]{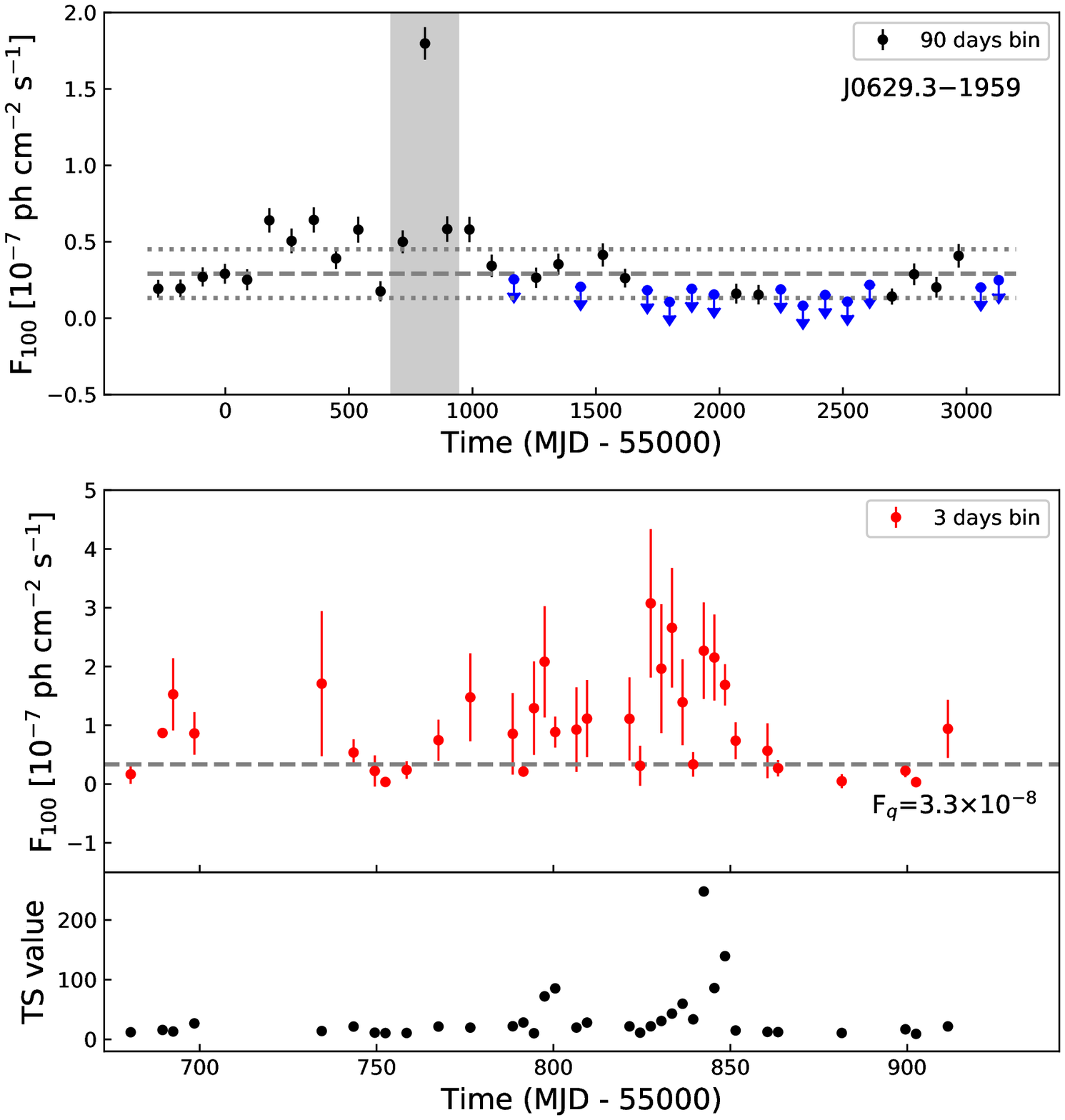}
\includegraphics[width=0.3\textwidth]{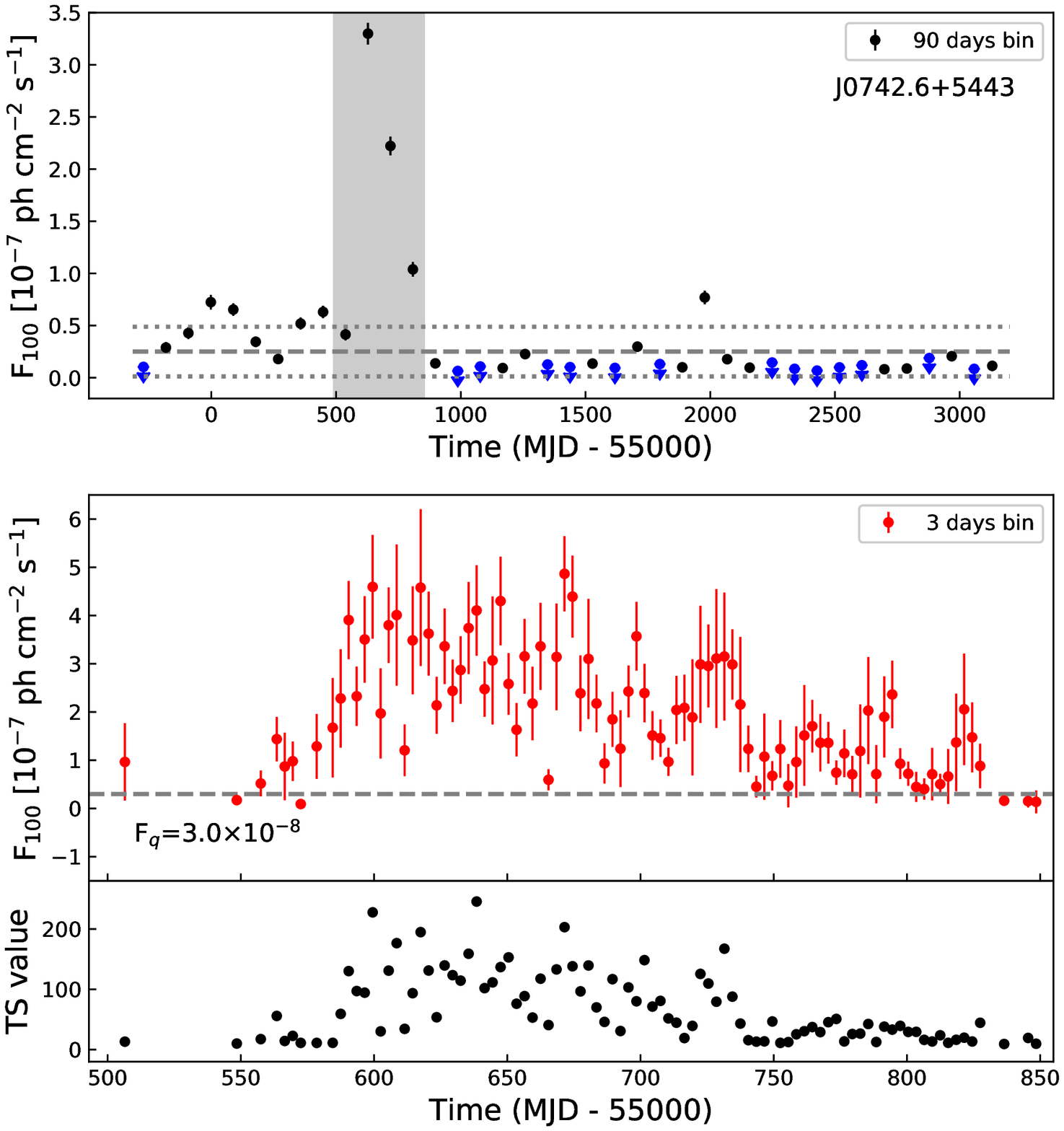}
\includegraphics[width=0.3\textwidth]{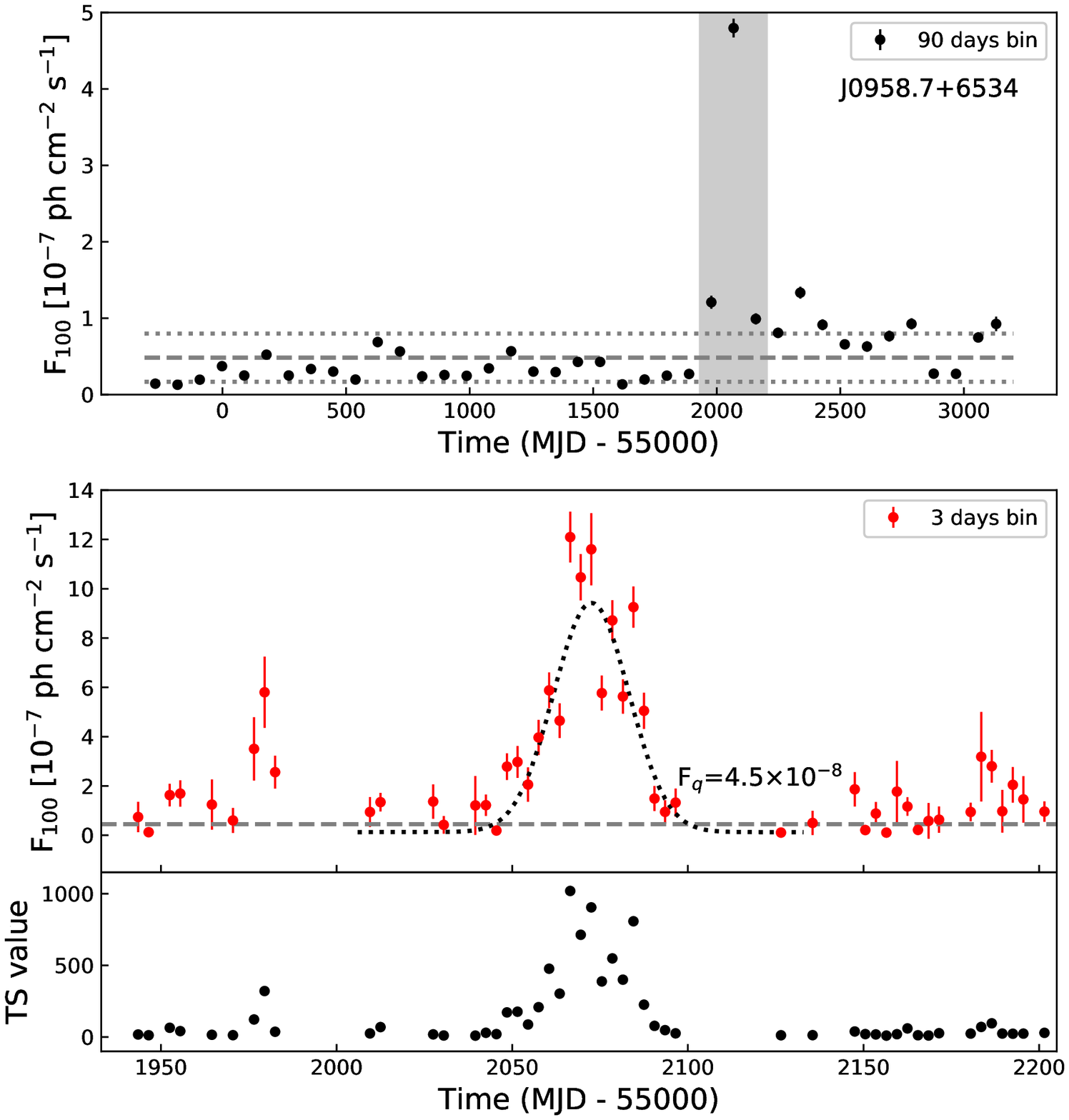}

\includegraphics[width=0.3\textwidth]{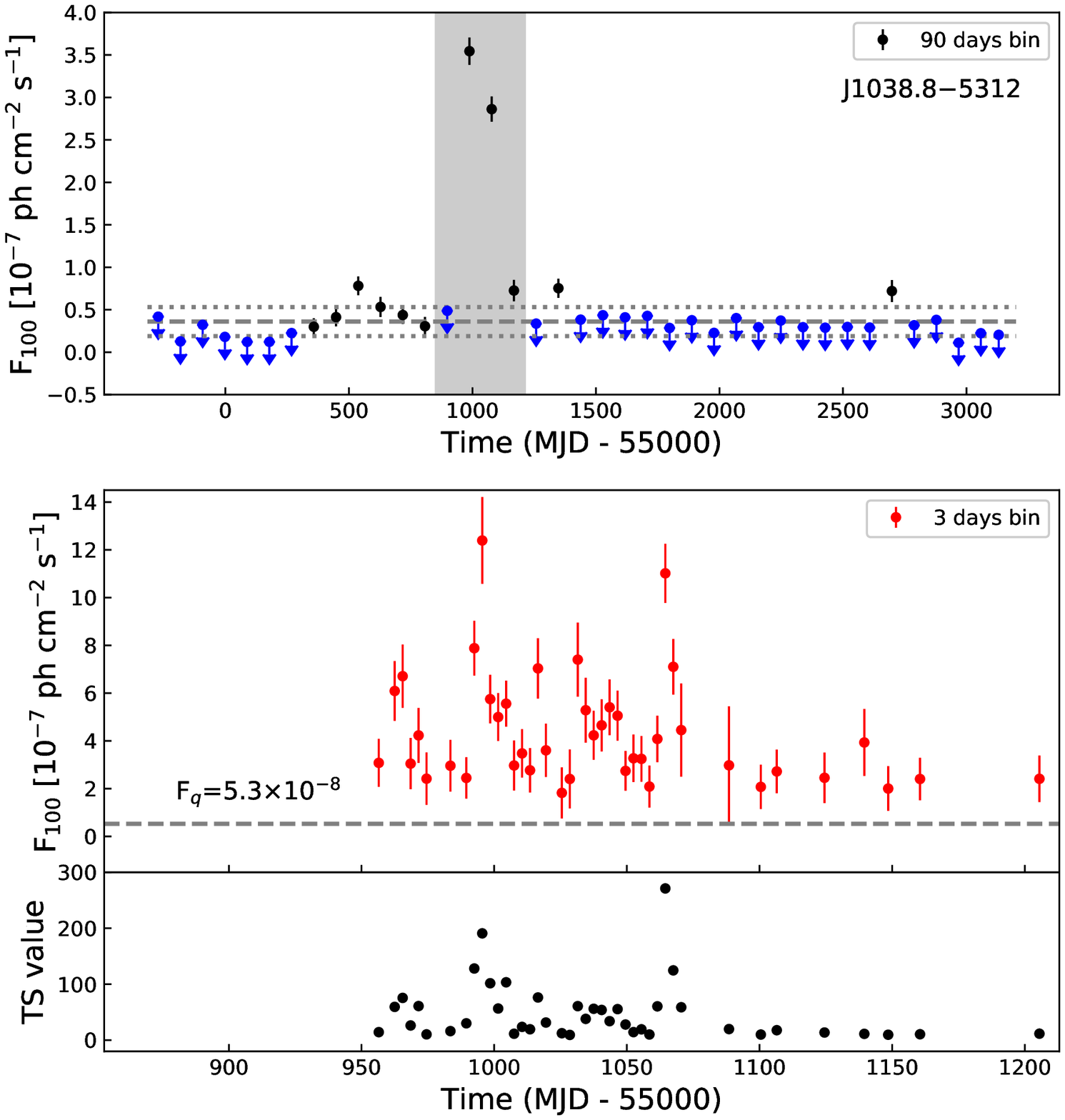}
\includegraphics[width=0.3\textwidth]{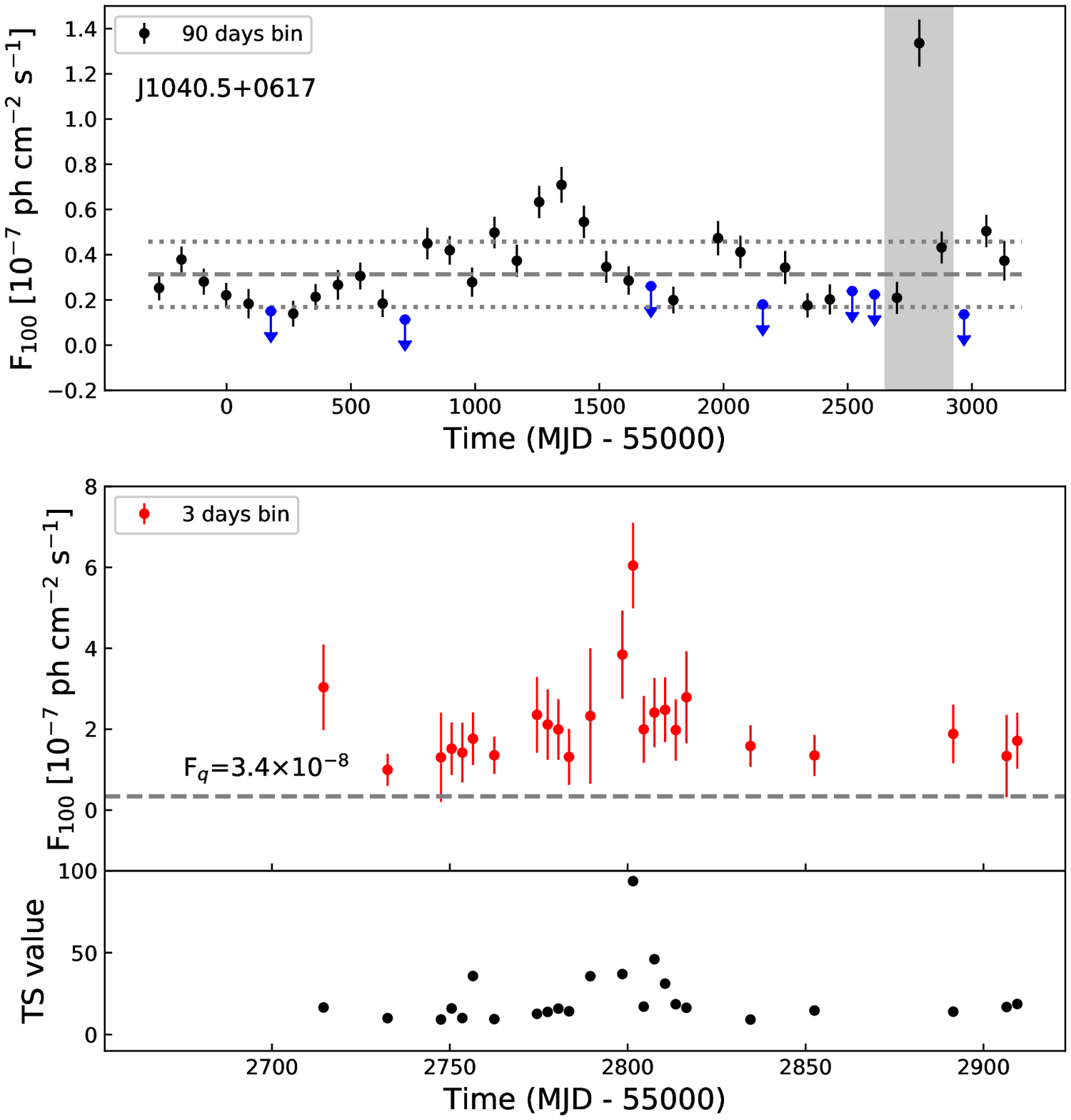}
\includegraphics[width=0.3\textwidth]{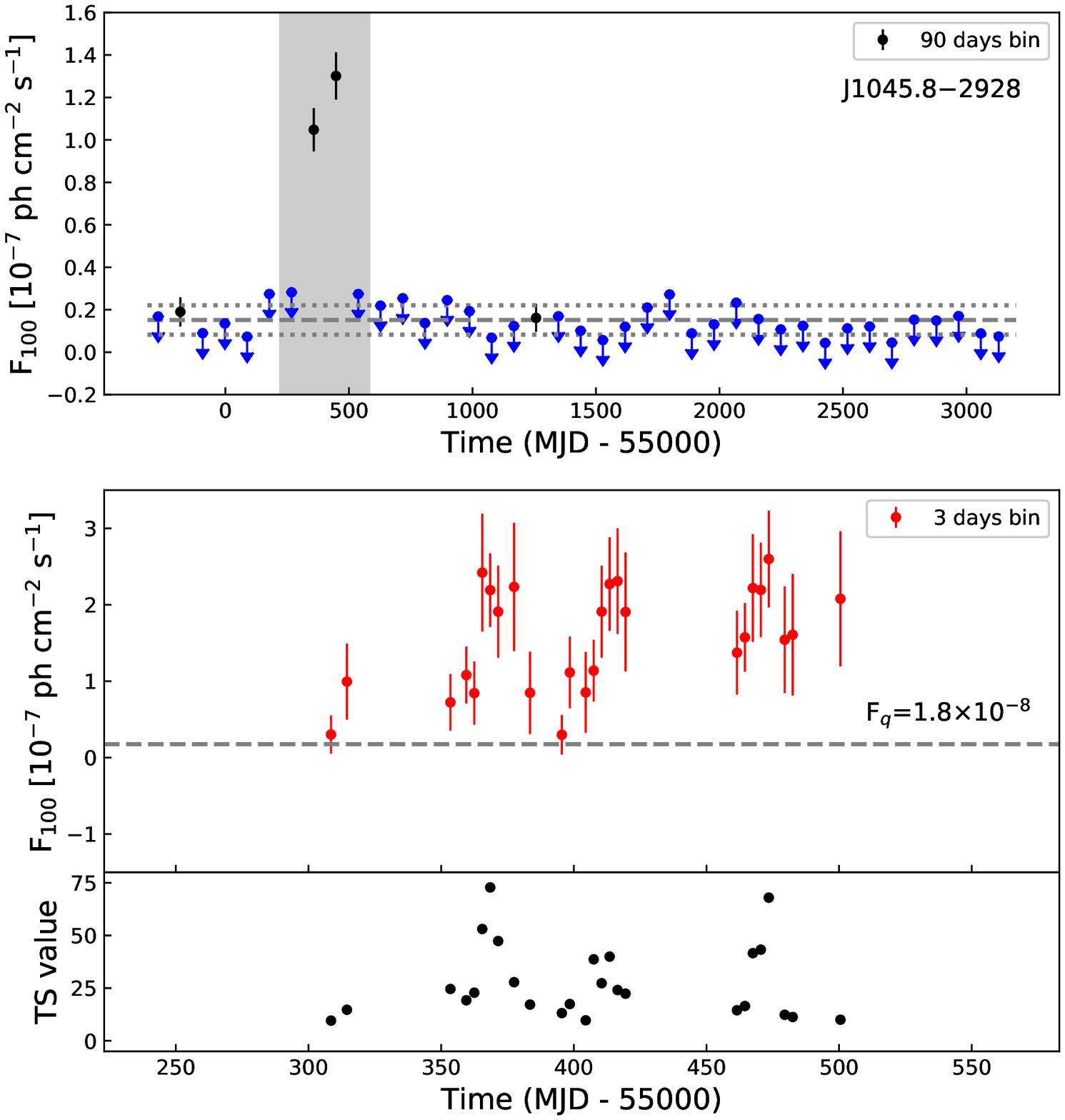}

\includegraphics[width=0.3\textwidth]{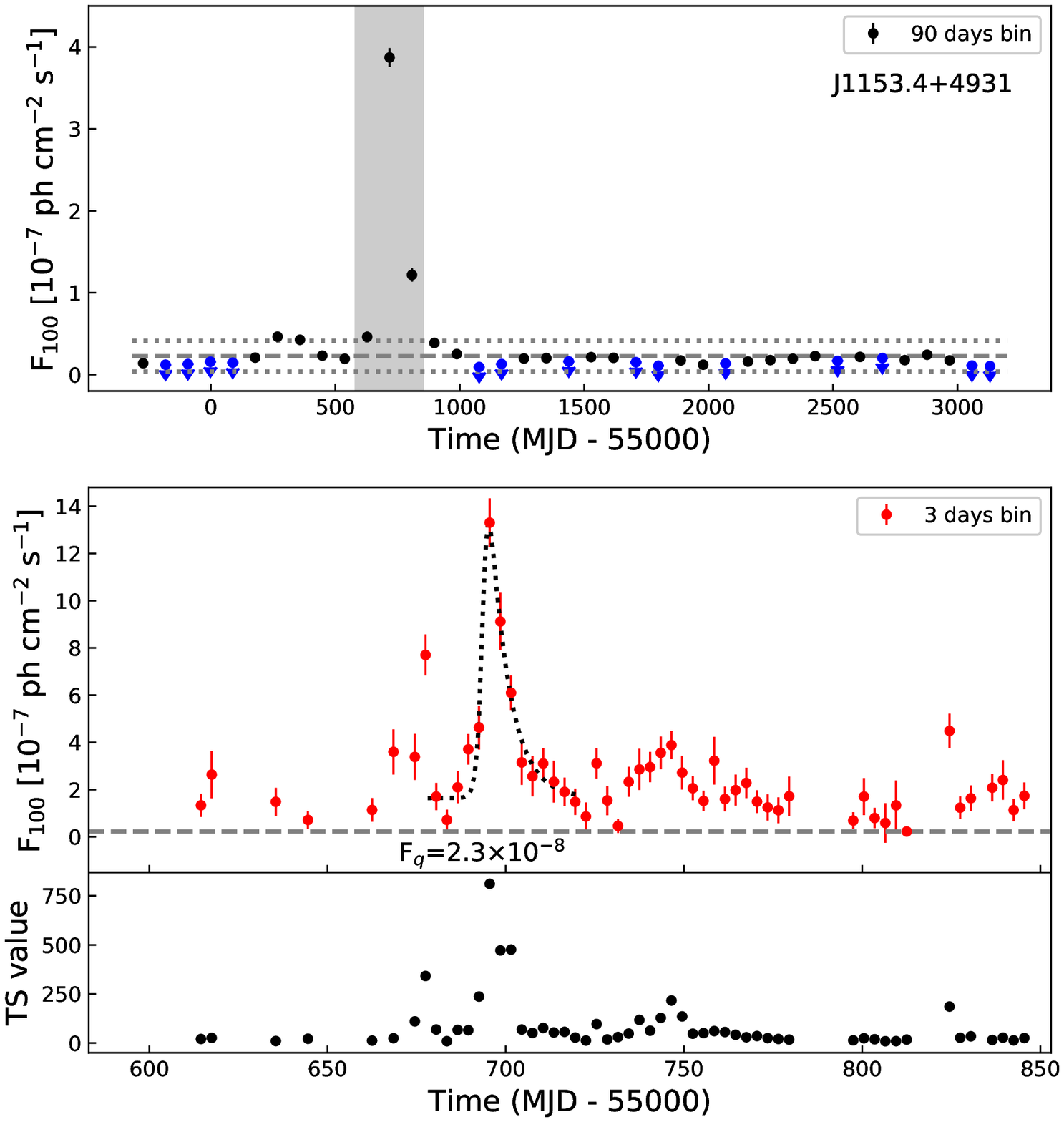}
\includegraphics[width=0.3\textwidth]{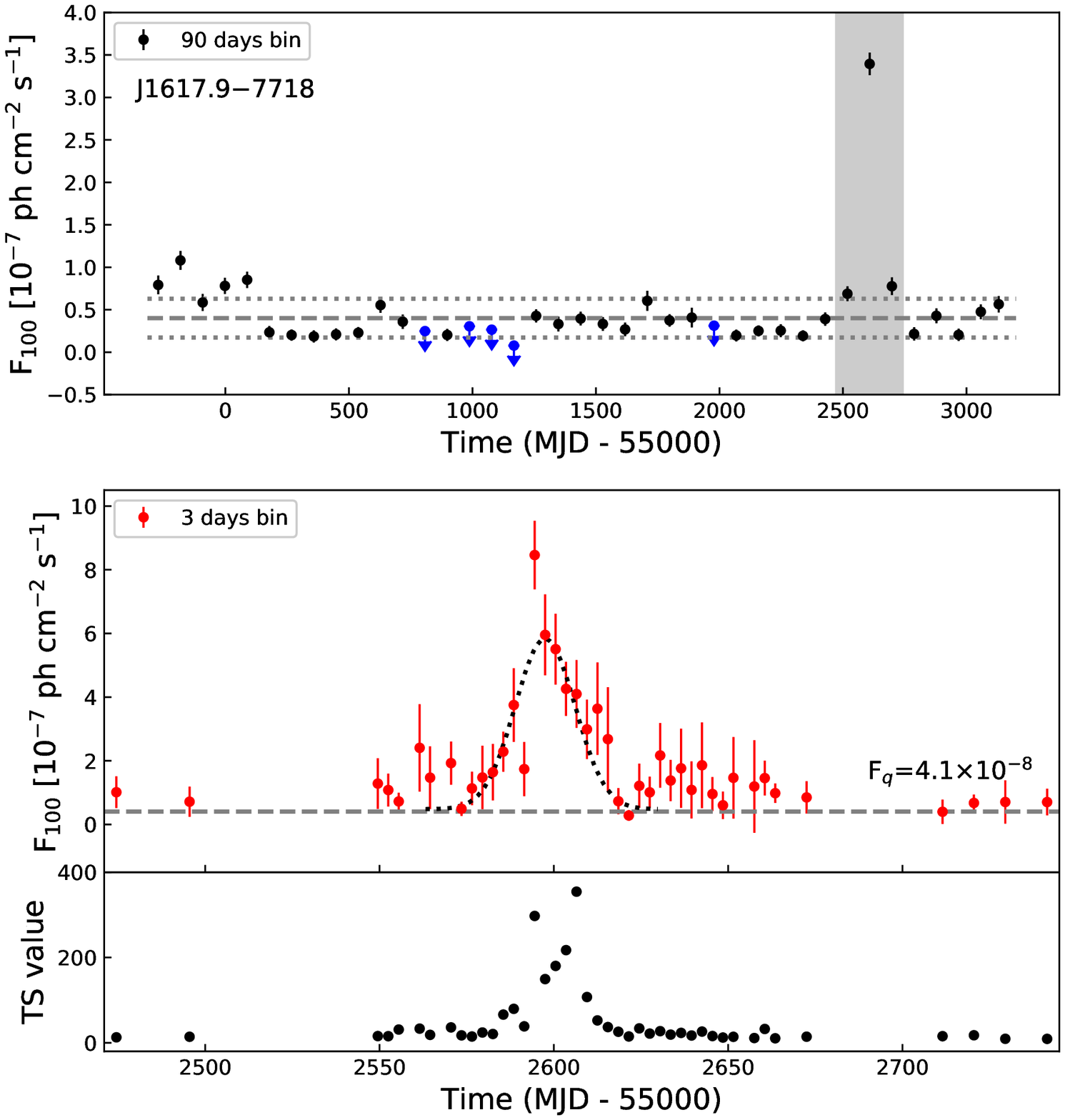}
\includegraphics[width=0.3\textwidth]{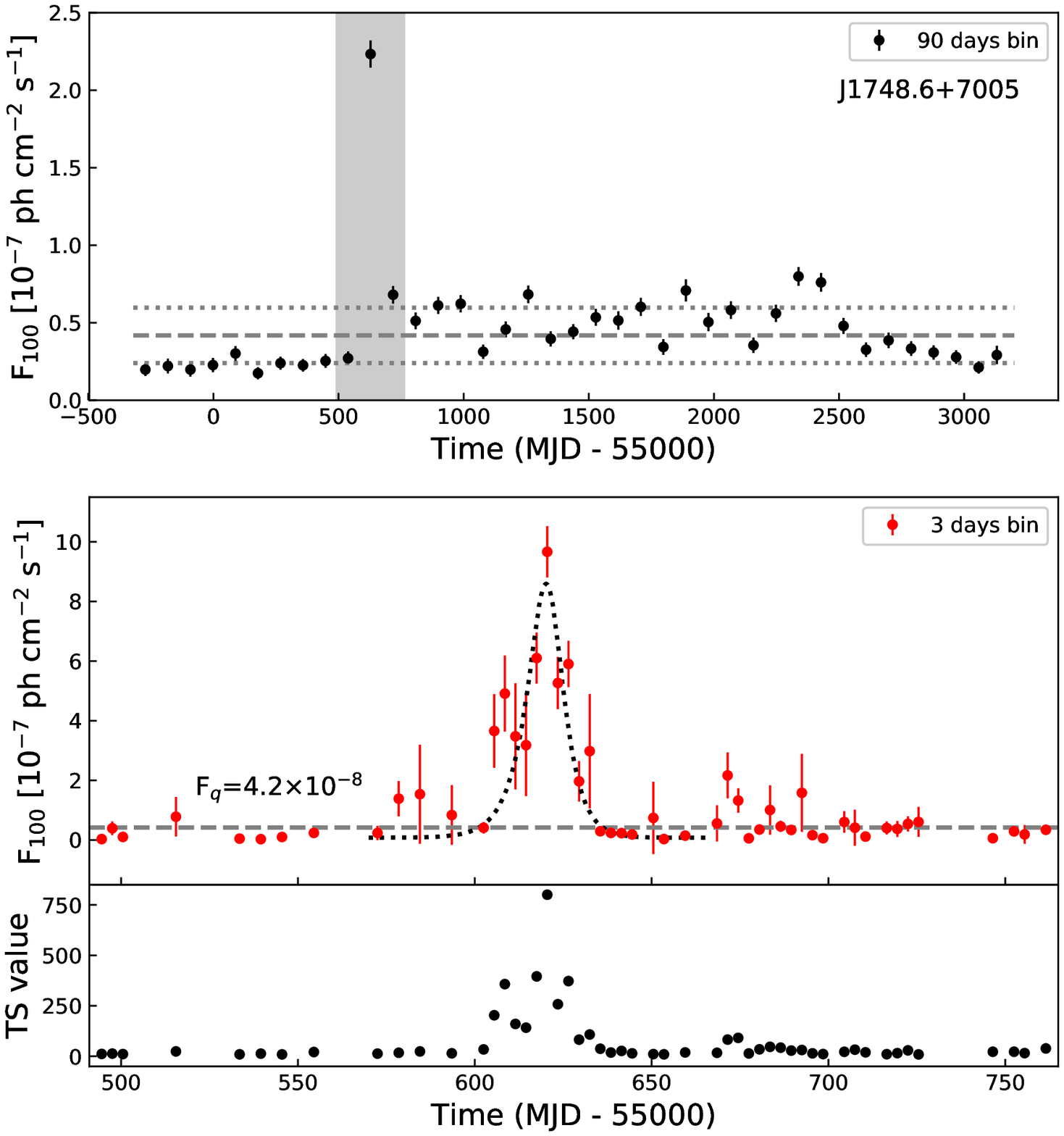}

\includegraphics[width=0.3\textwidth]{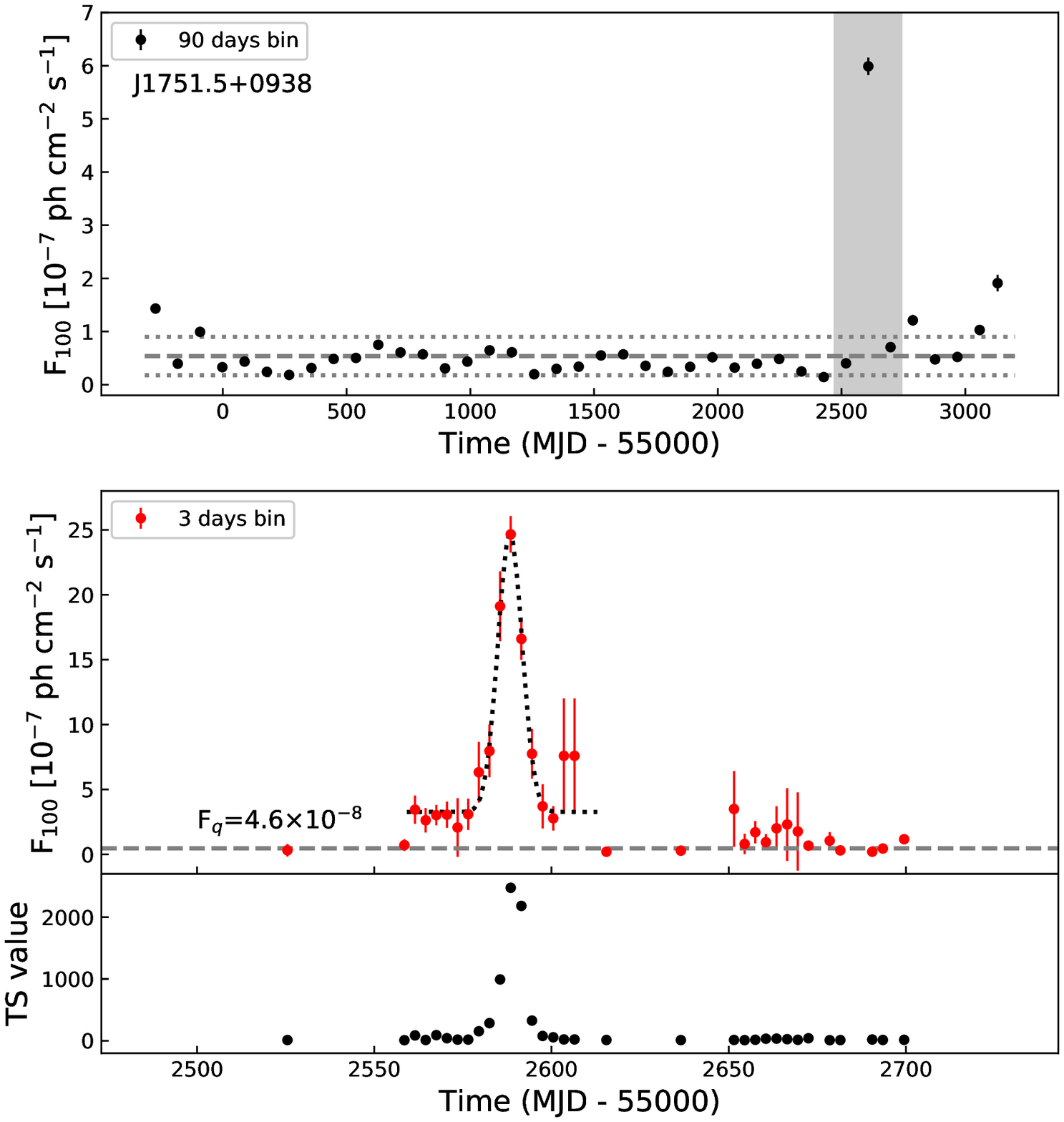}
\includegraphics[width=0.3\textwidth]{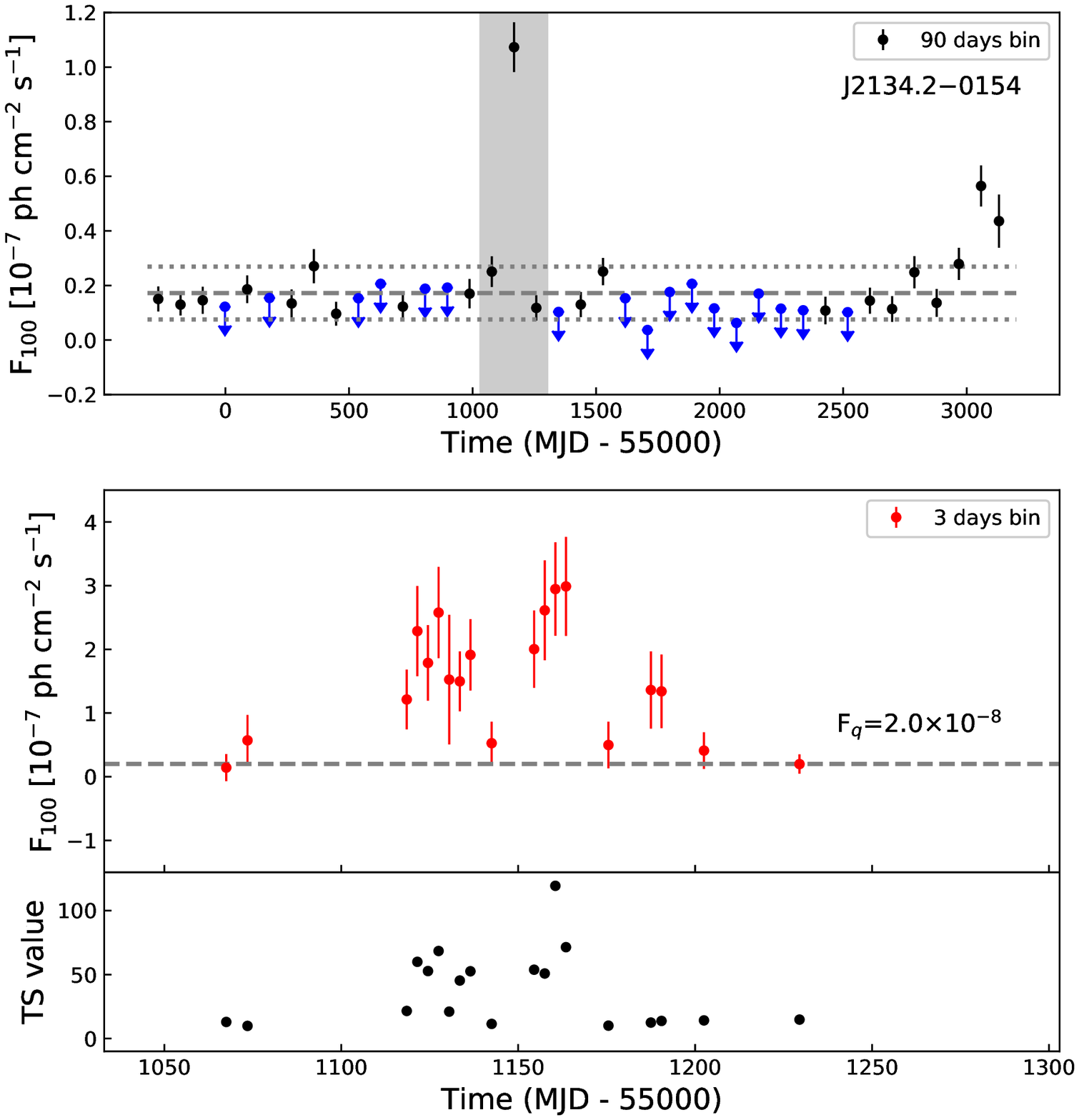}
\includegraphics[width=0.3\textwidth]{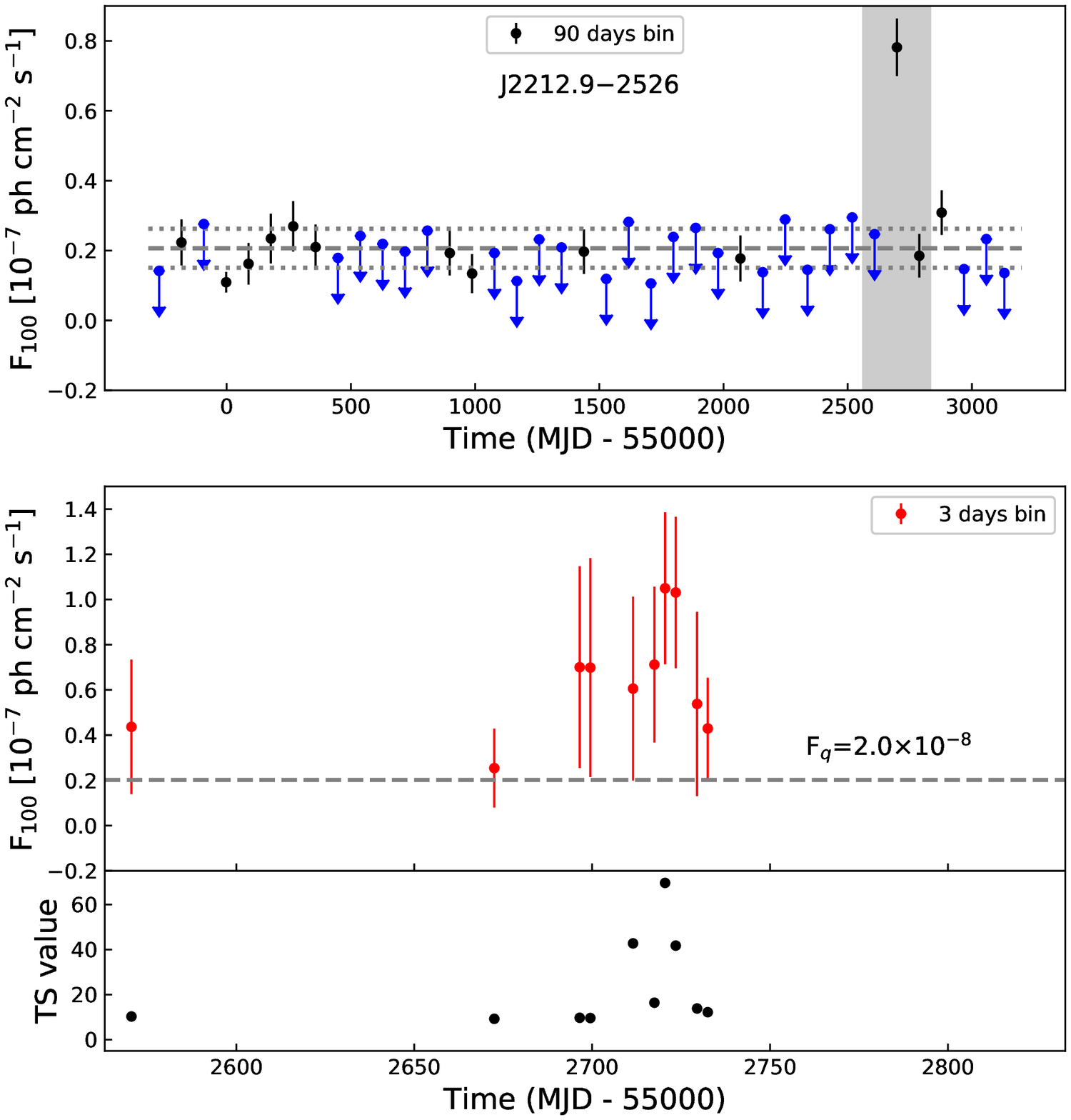}
\caption{Continued.}
\end{figure*}

\begin{figure*}
\centering
\includegraphics[width=0.3\textwidth]{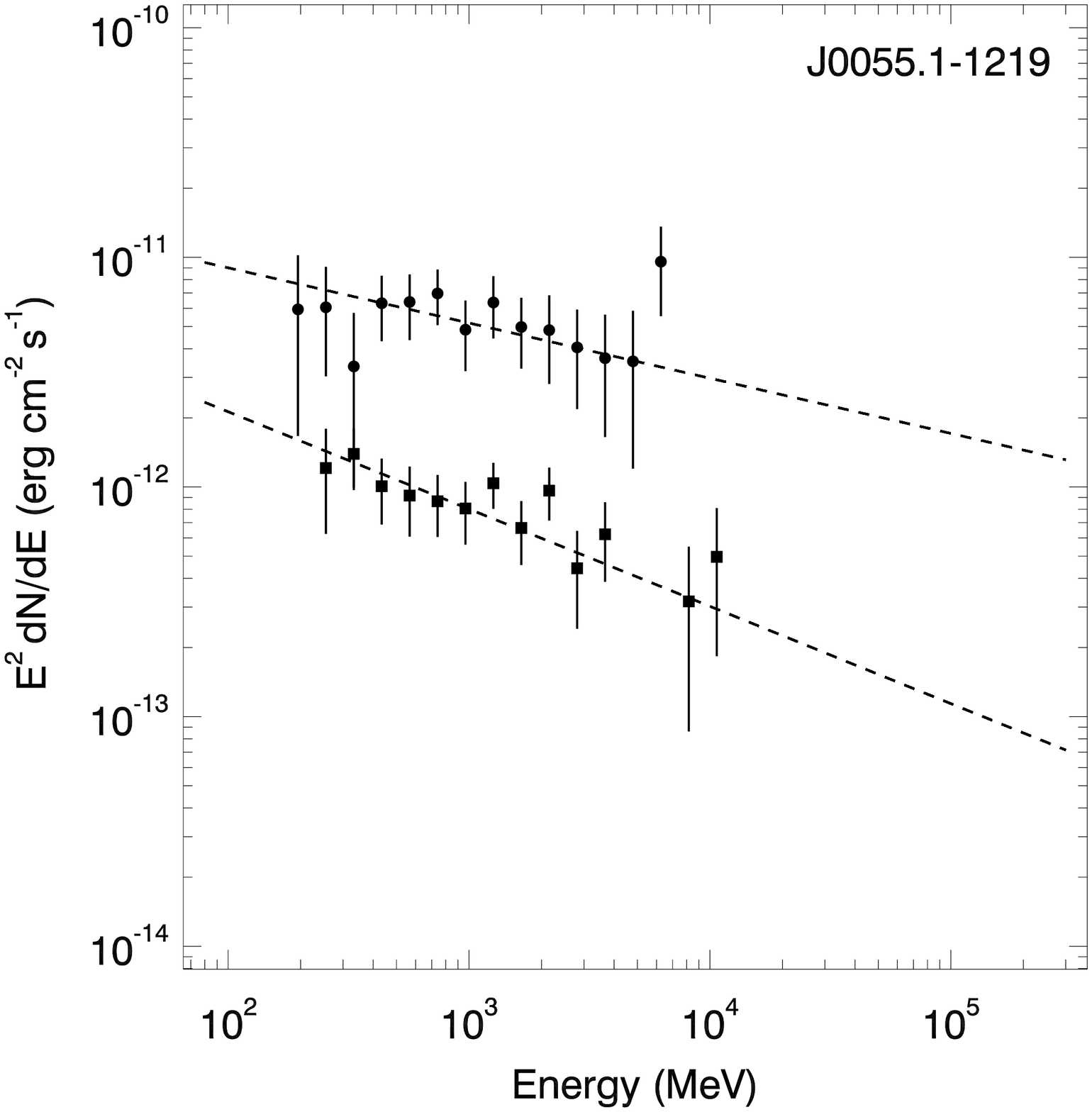}
\includegraphics[width=0.3\textwidth]{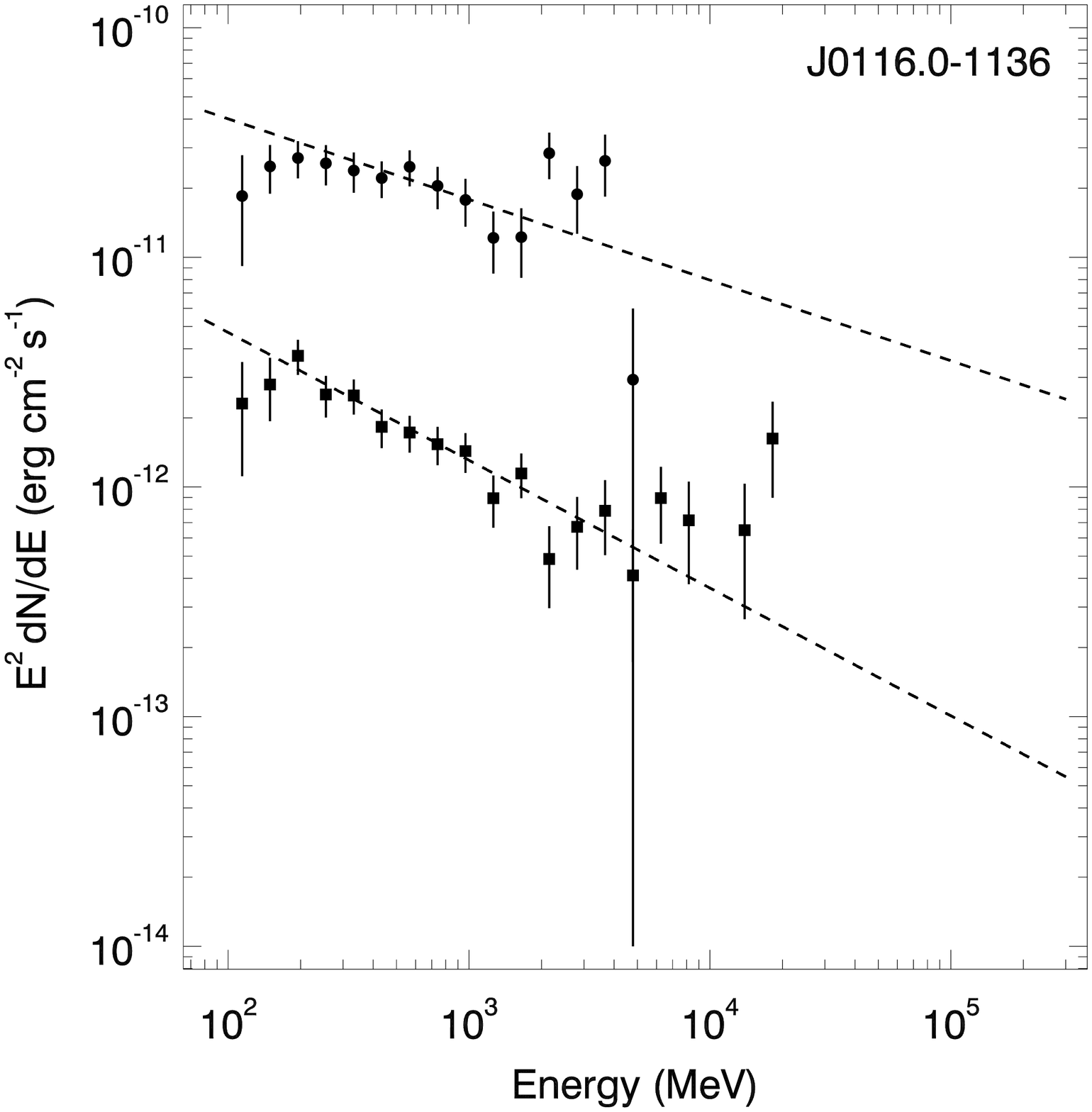}
\includegraphics[width=0.3\textwidth]{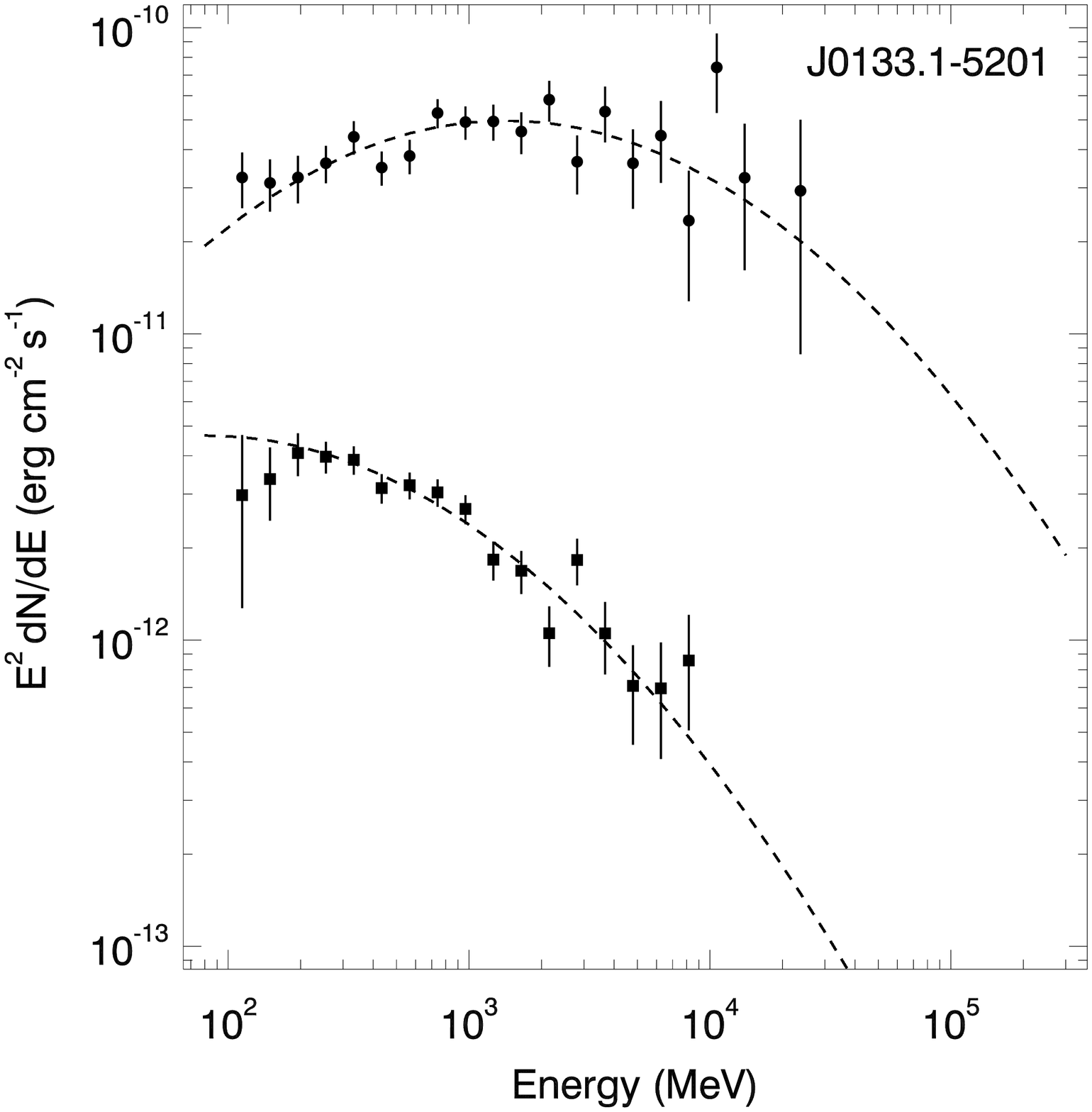}

\includegraphics[width=0.3\textwidth]{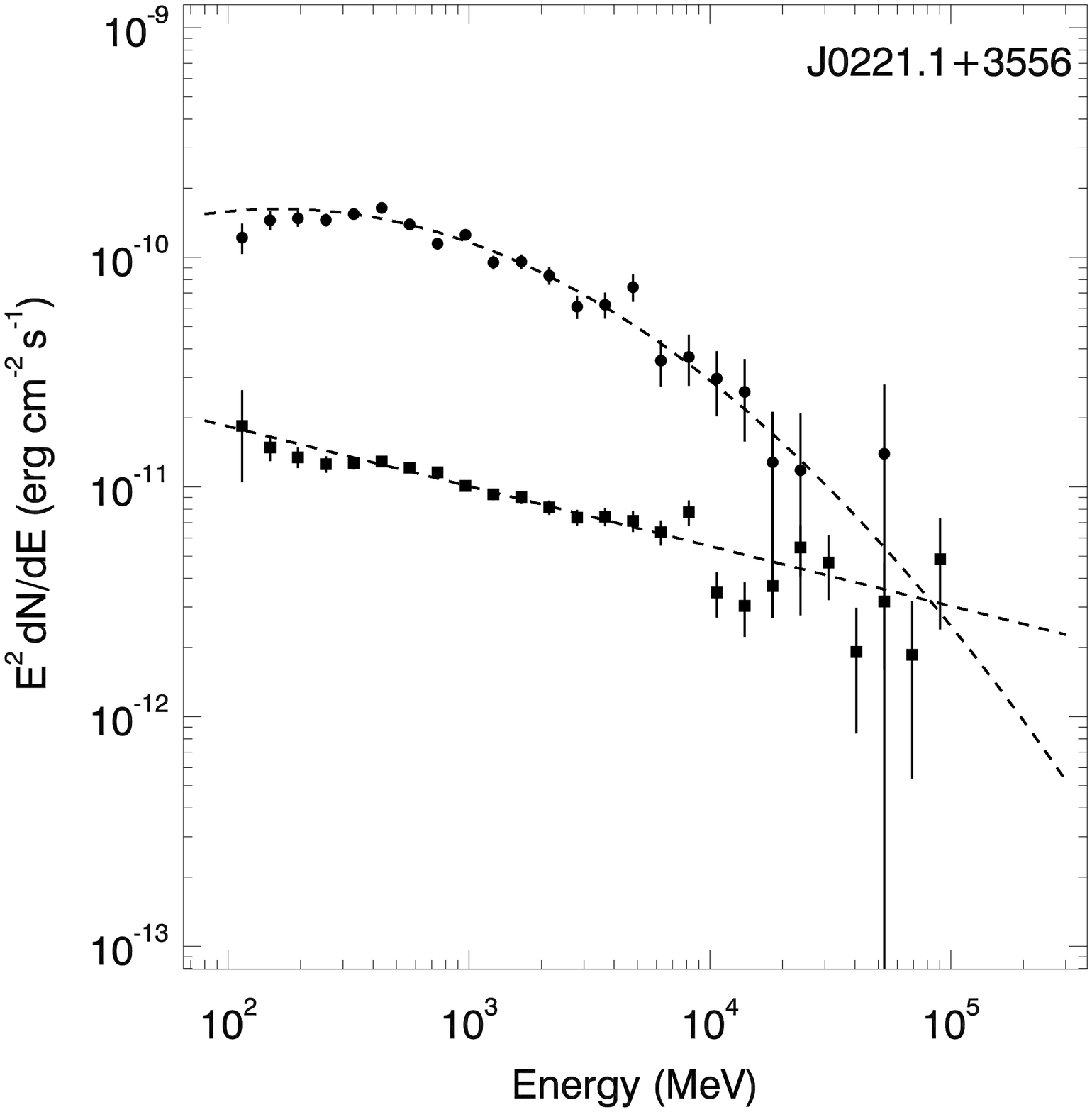}
\includegraphics[width=0.3\textwidth]{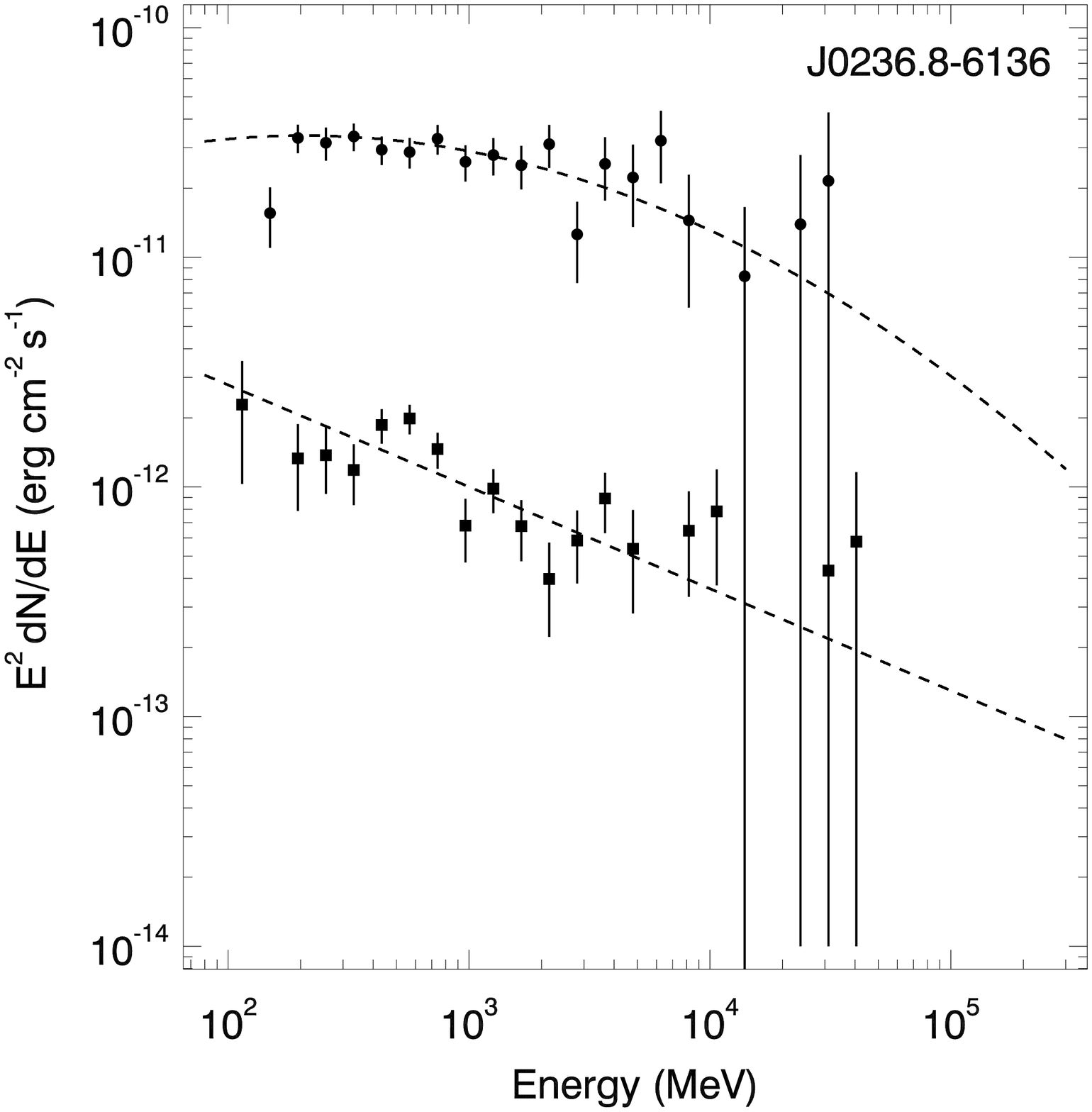}
\includegraphics[width=0.3\textwidth]{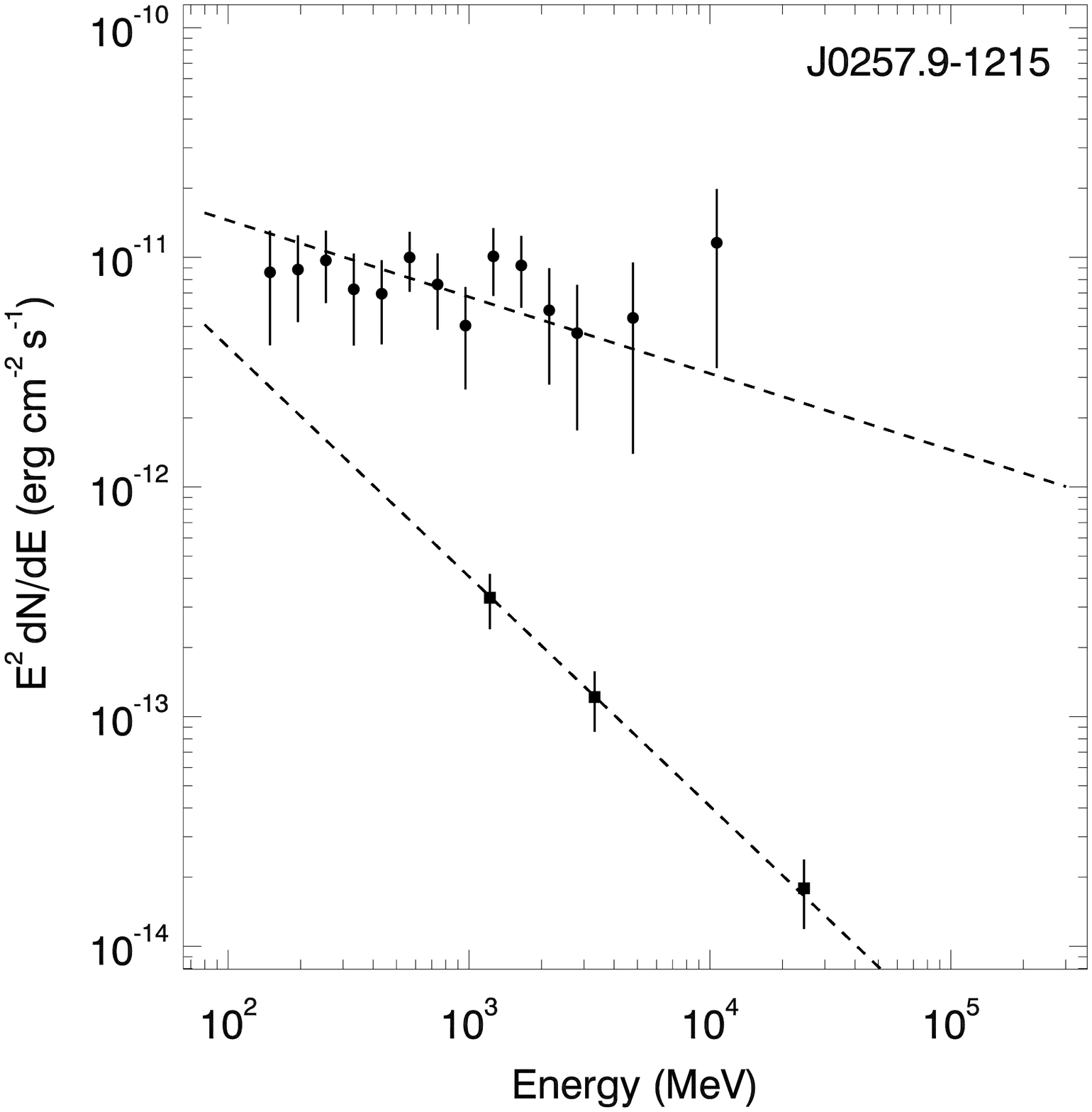}

\includegraphics[width=0.3\textwidth]{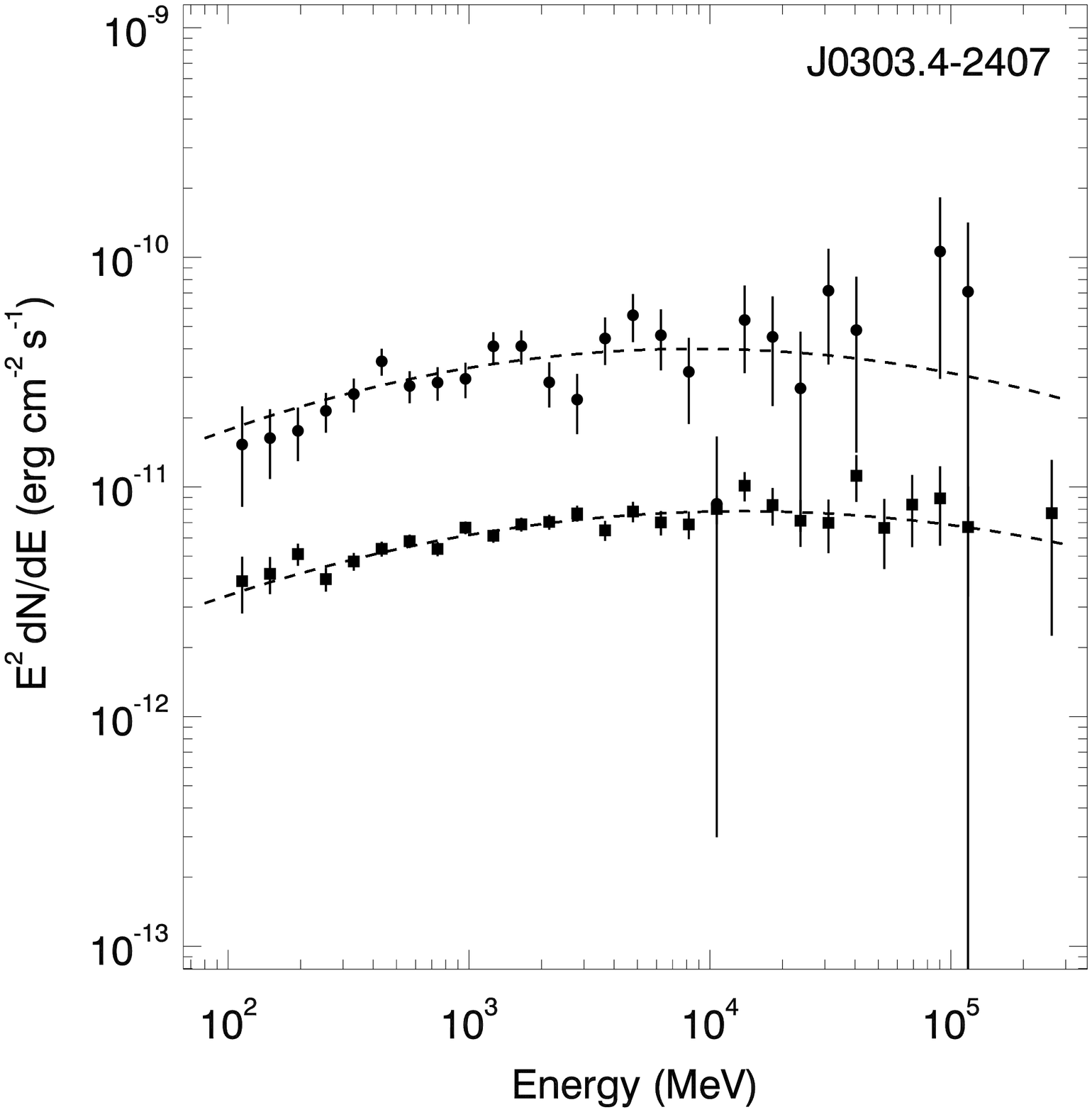}
\includegraphics[width=0.3\textwidth]{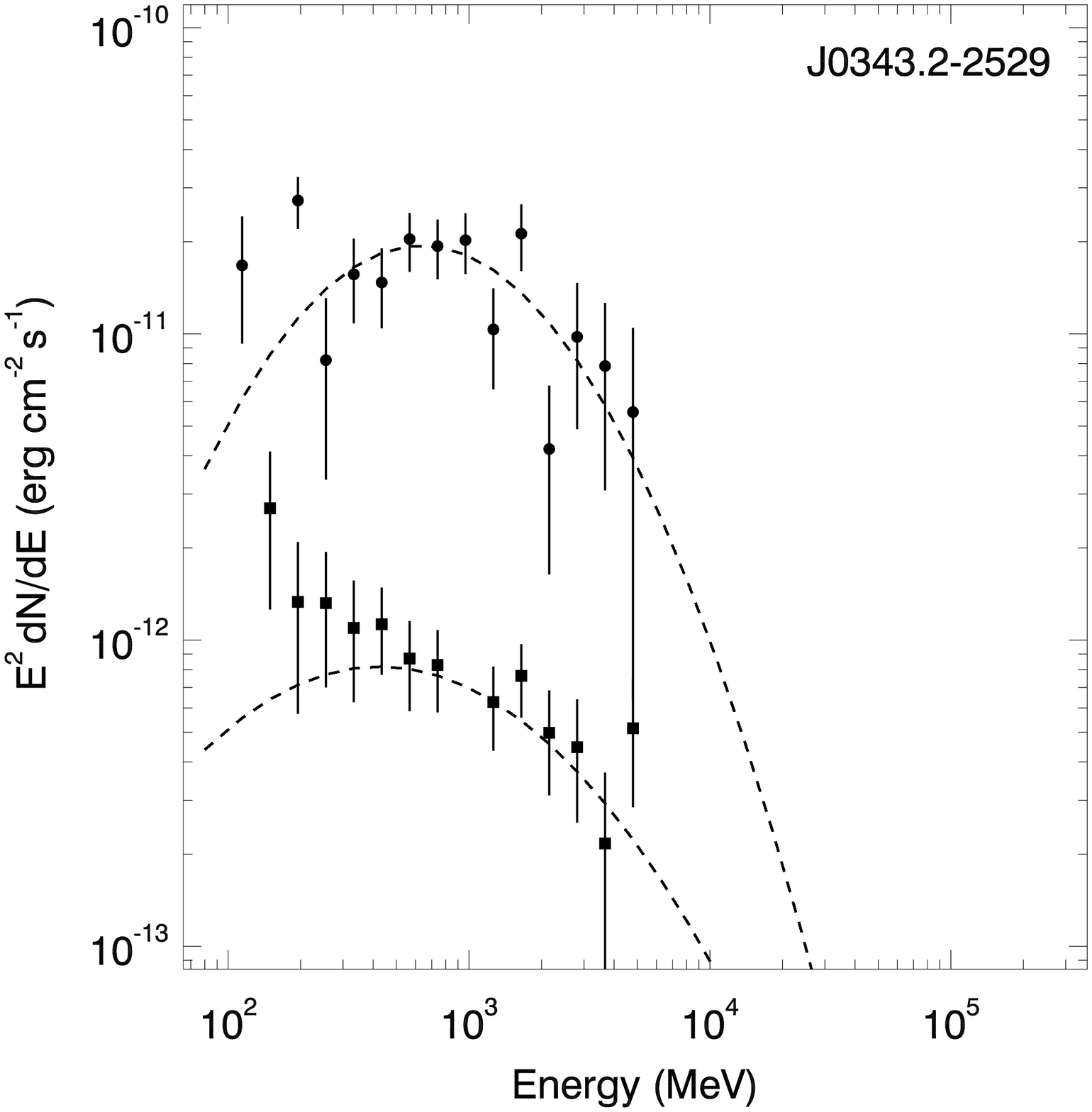}
\includegraphics[width=0.3\textwidth]{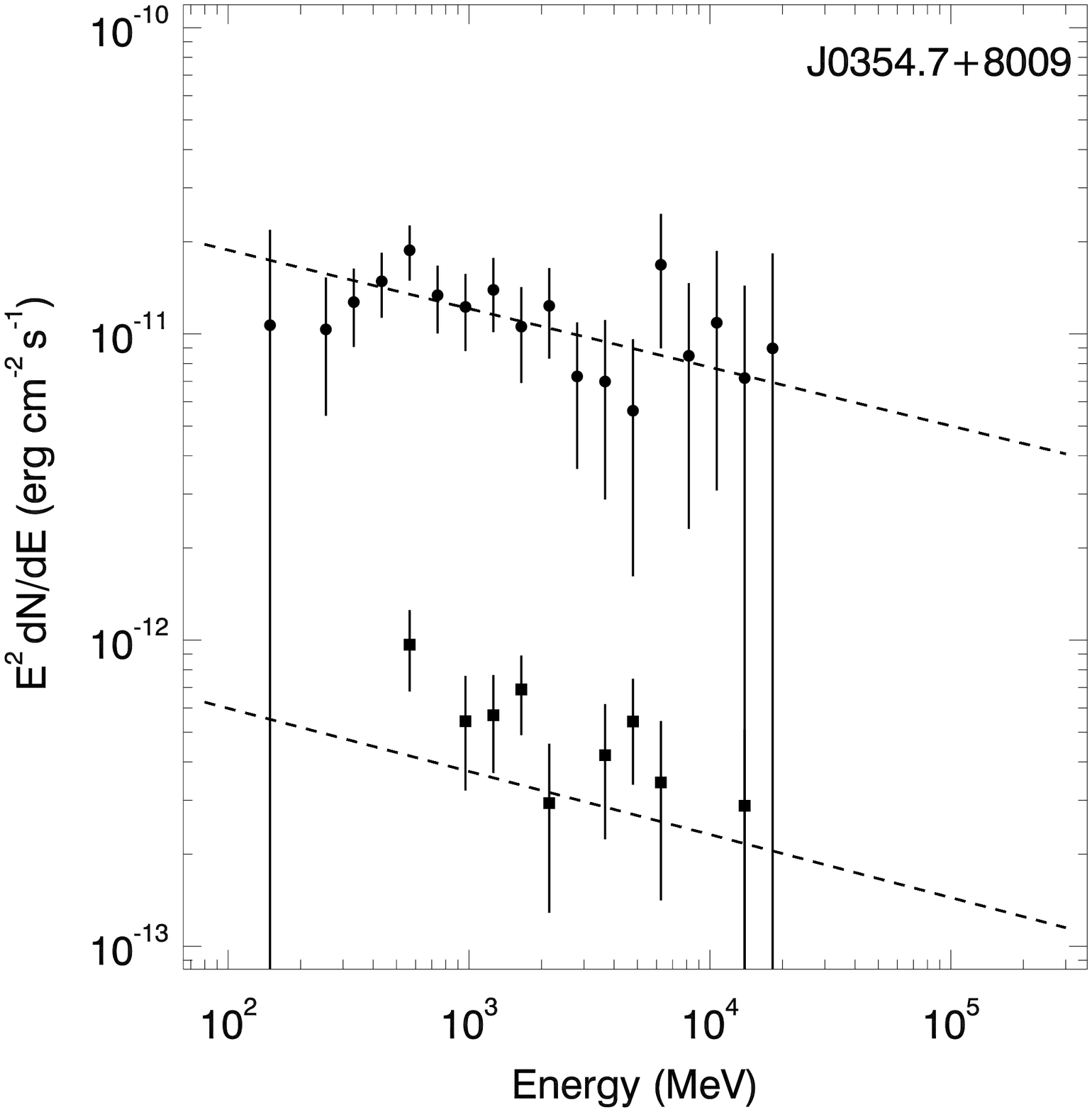}

\includegraphics[width=0.3\textwidth]{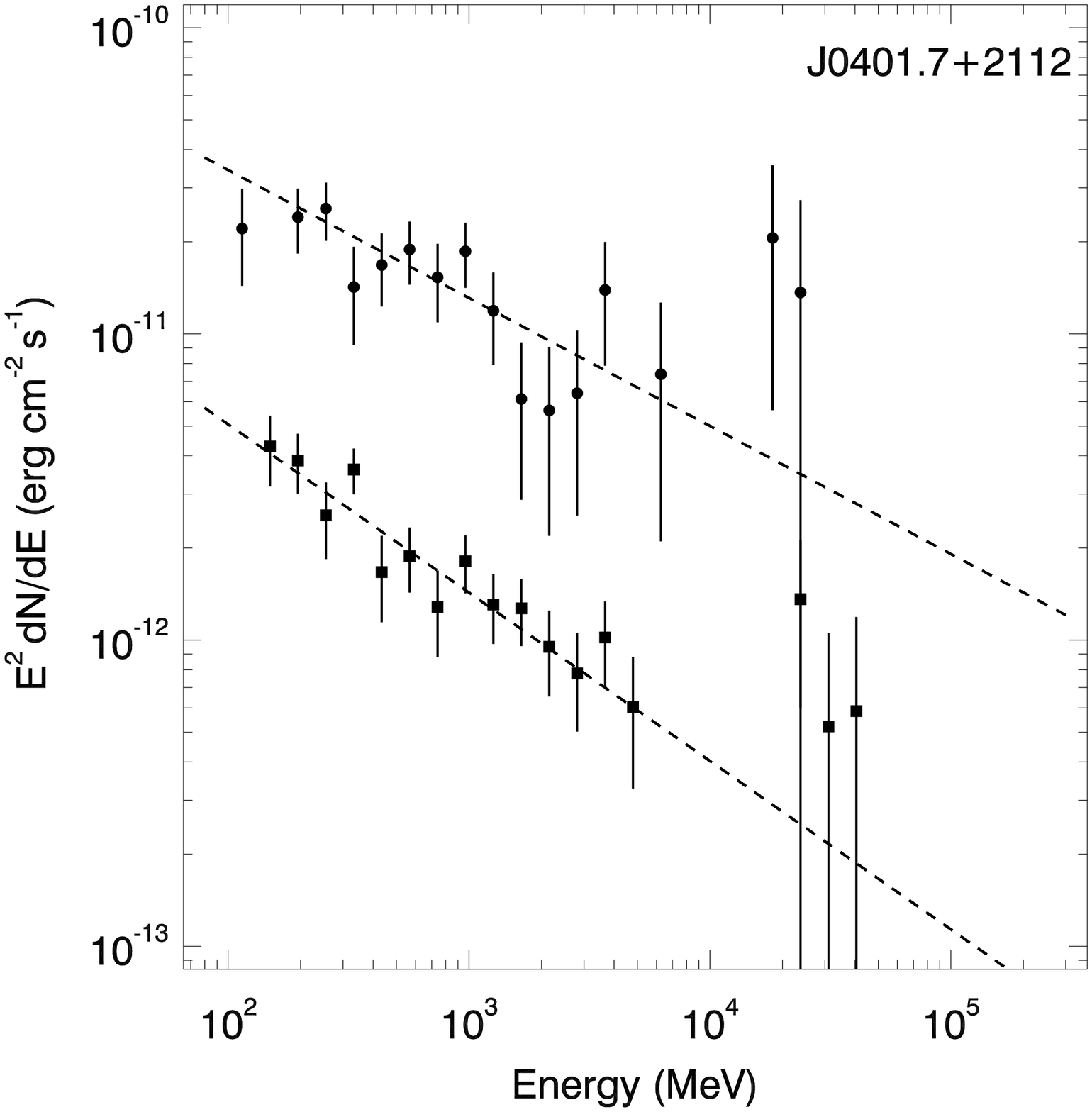}
\includegraphics[width=0.3\textwidth]{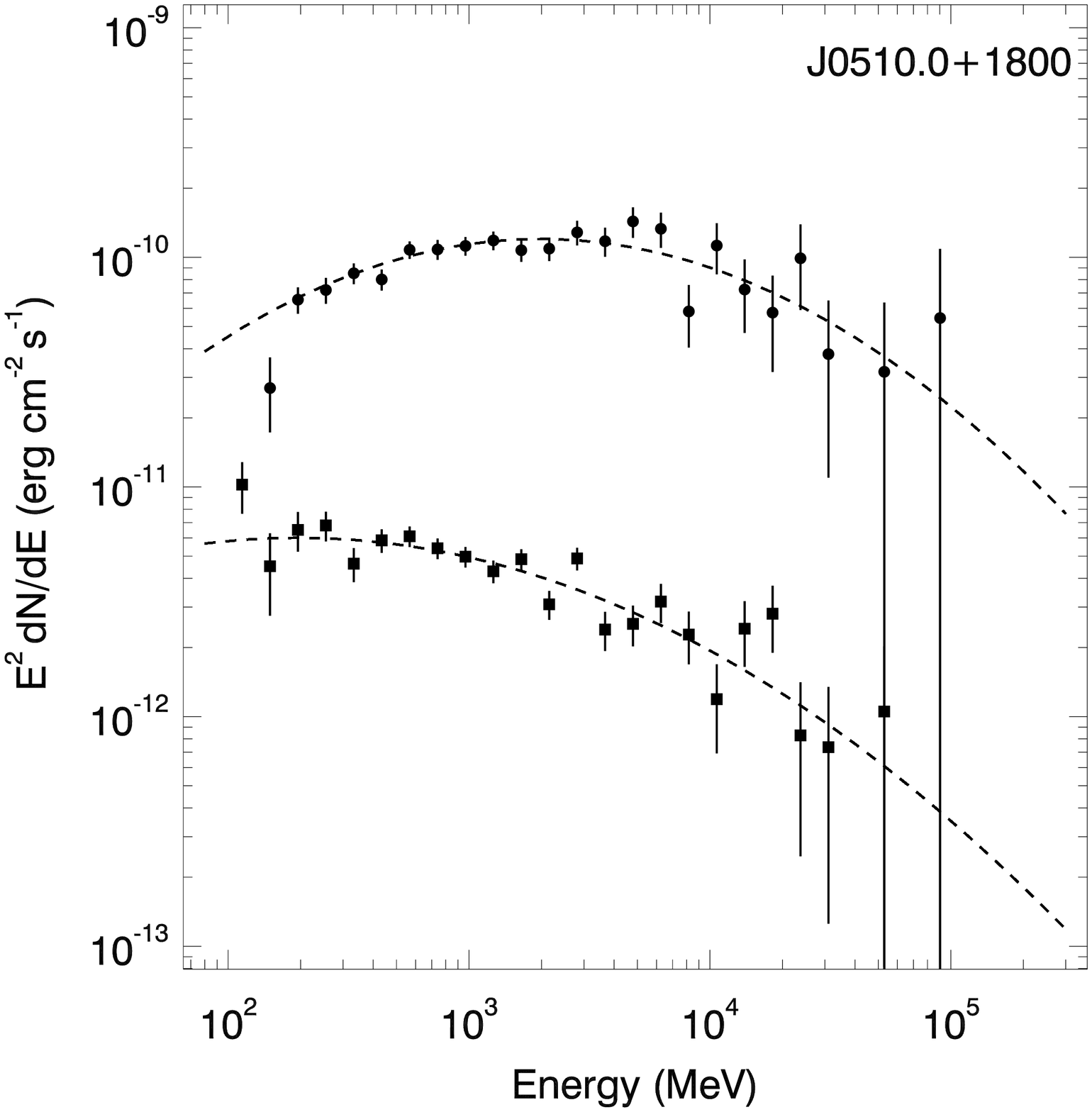}
\includegraphics[width=0.3\textwidth]{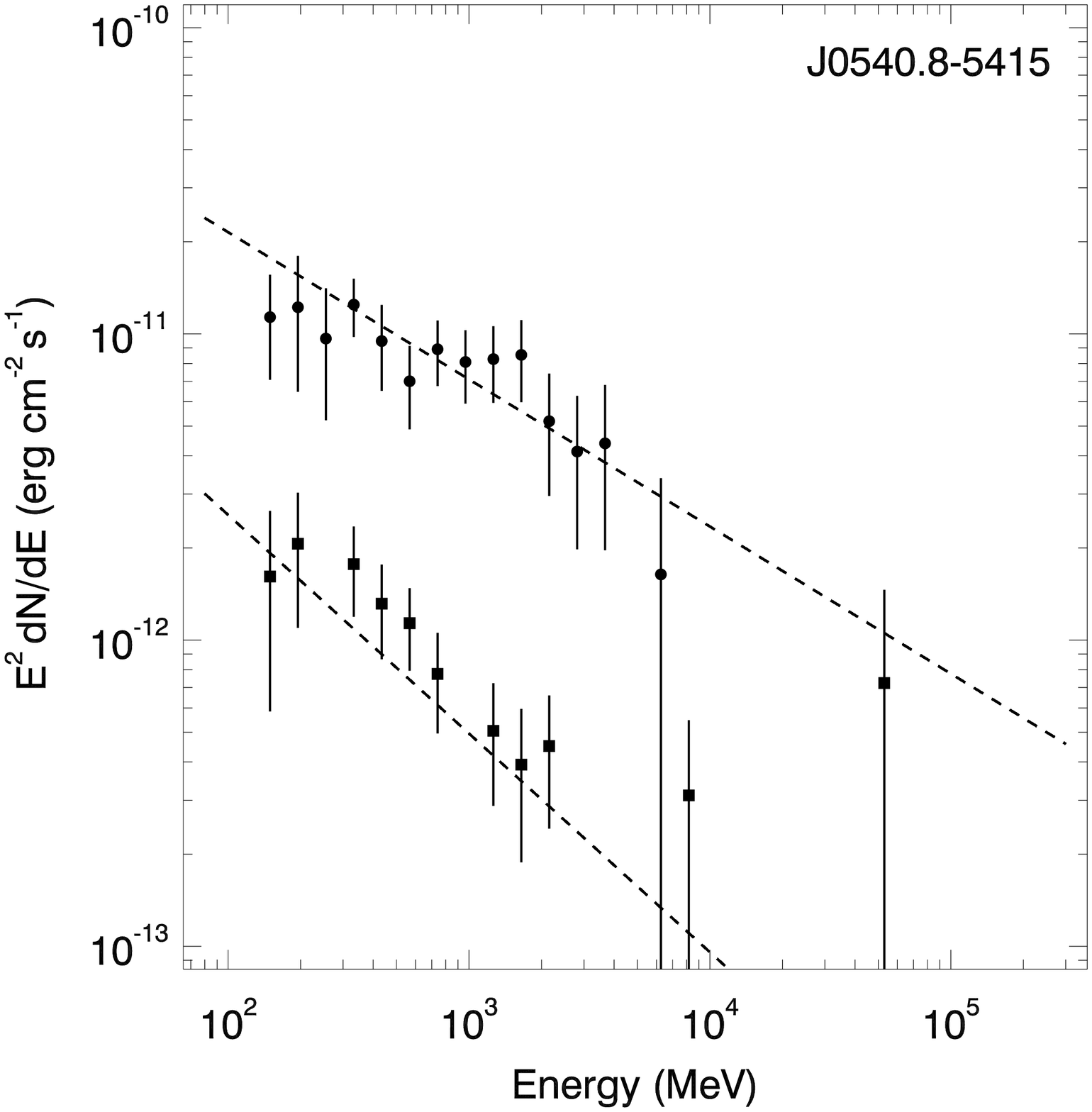}
\caption{Spectra of the 24 blazar targets in the flaring (dots) and quiescent
(squares) states. The models of either a power law or a log-parabola from
the likelihood analysis are shown as dashed lines. 
}
\label{fig:spec}
\end{figure*}
\setcounter{figure}{2}
\begin{figure*}
\includegraphics[width=0.3\textwidth]{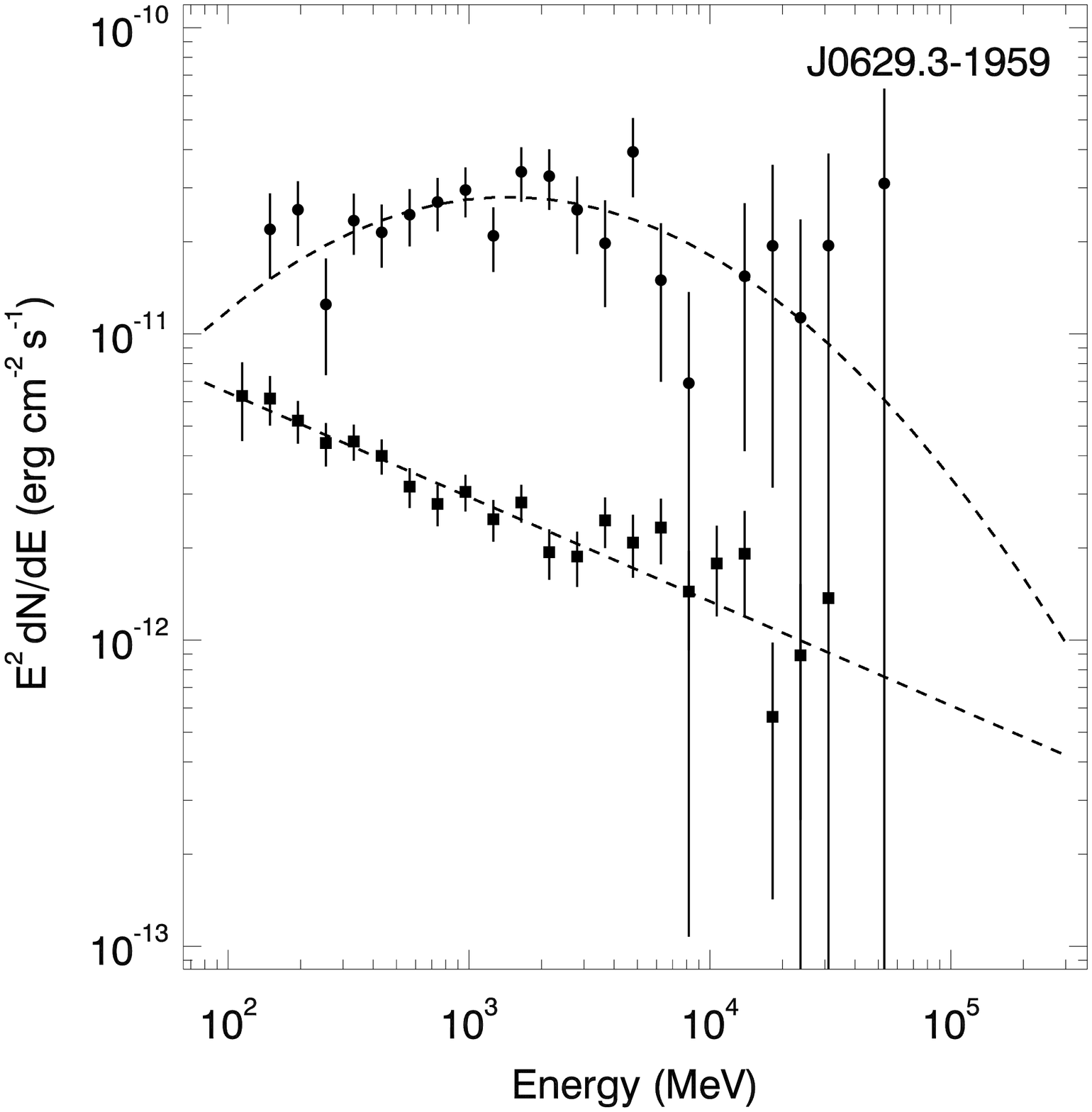}
\includegraphics[width=0.3\textwidth]{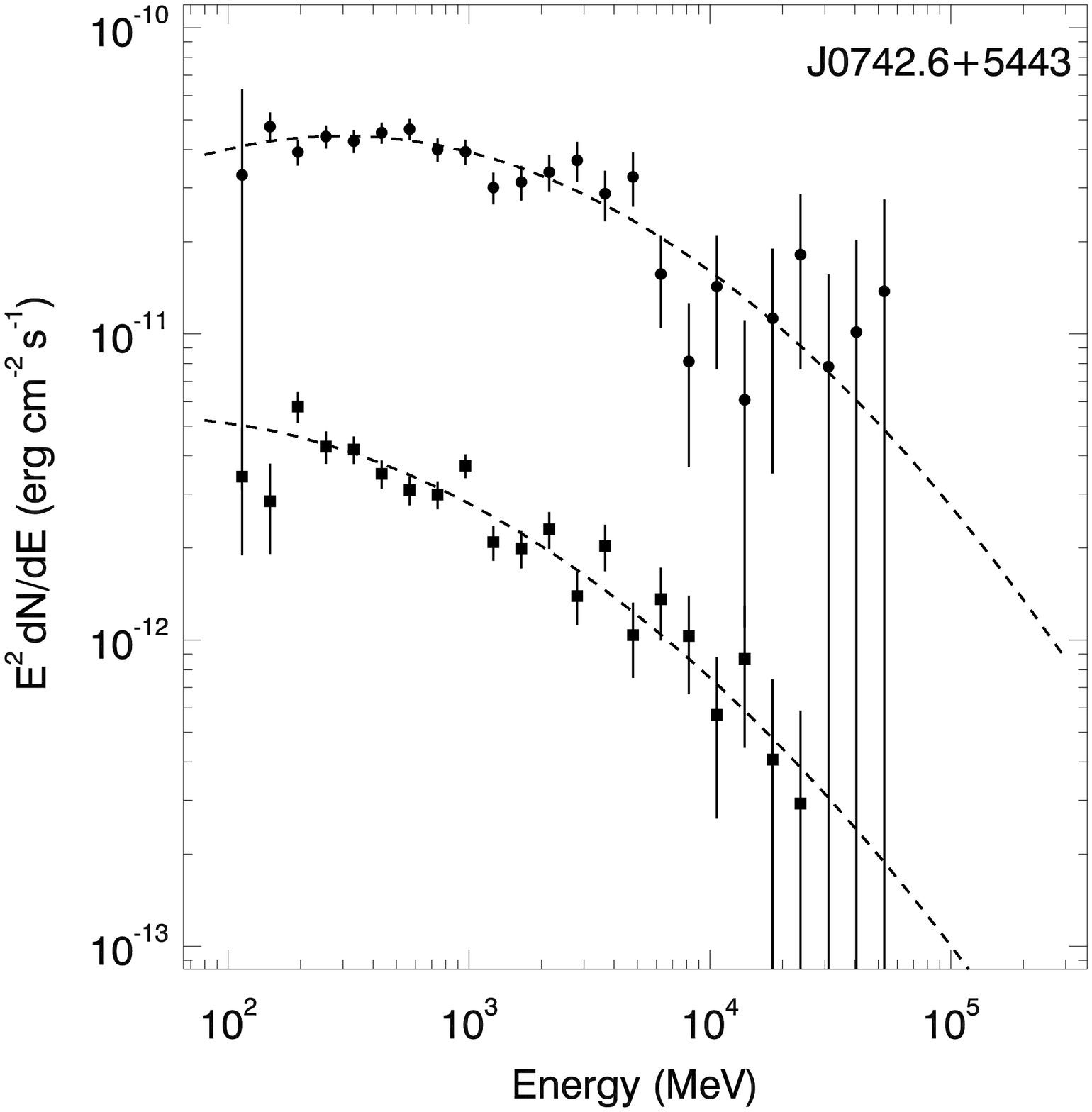}
\includegraphics[width=0.3\textwidth]{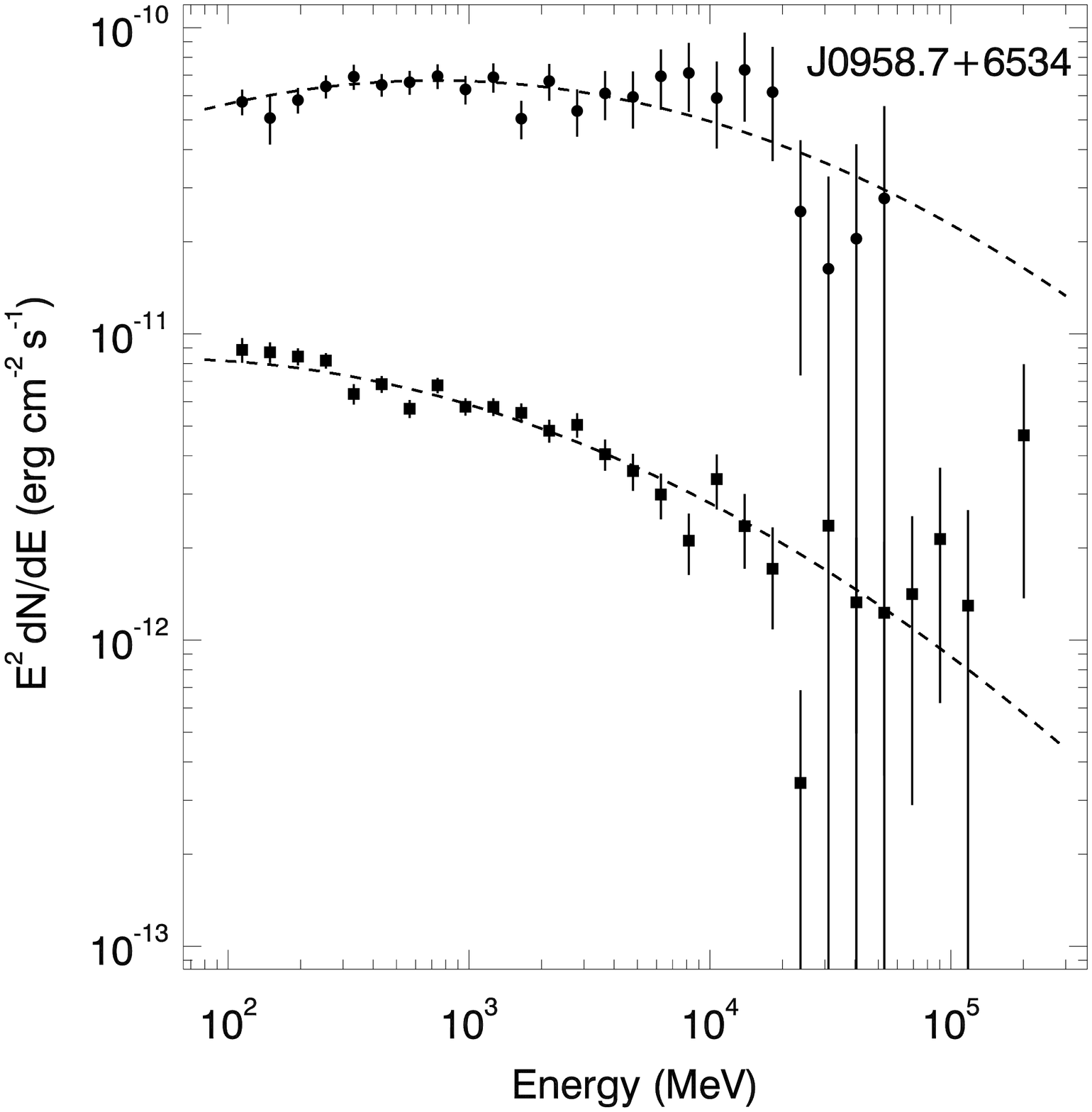}

\includegraphics[width=0.3\textwidth]{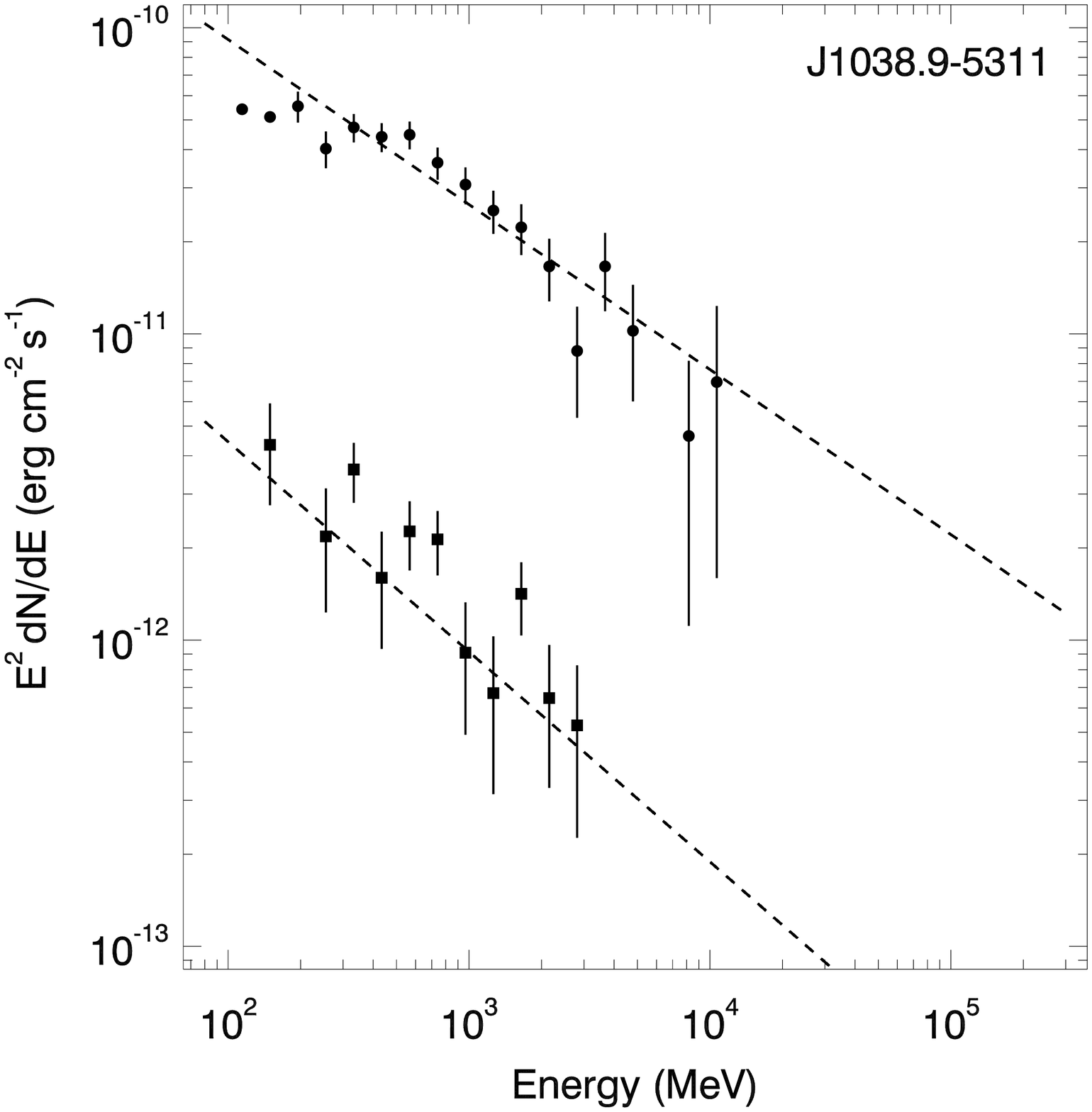}
\includegraphics[width=0.3\textwidth]{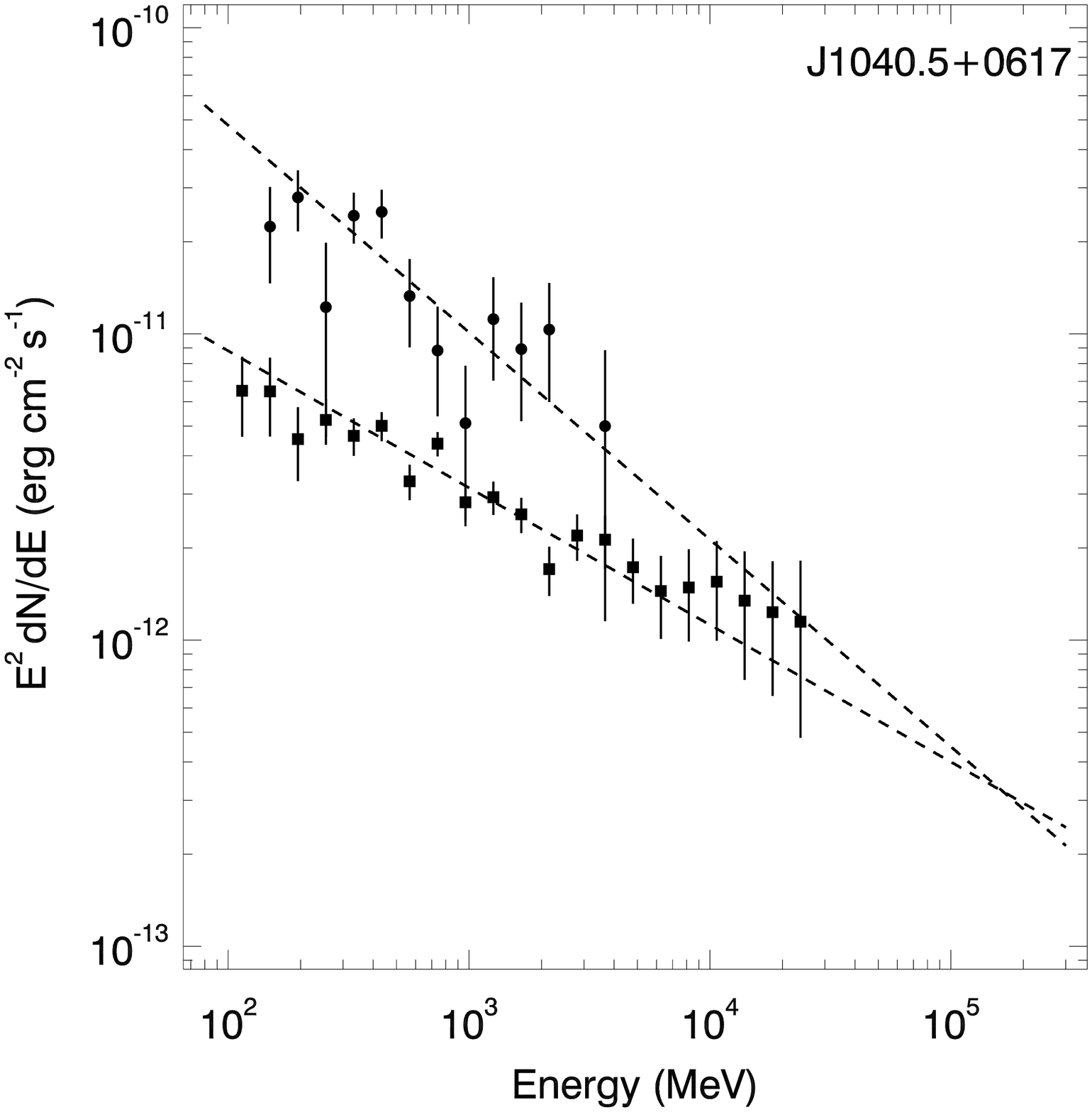}
\includegraphics[width=0.3\textwidth]{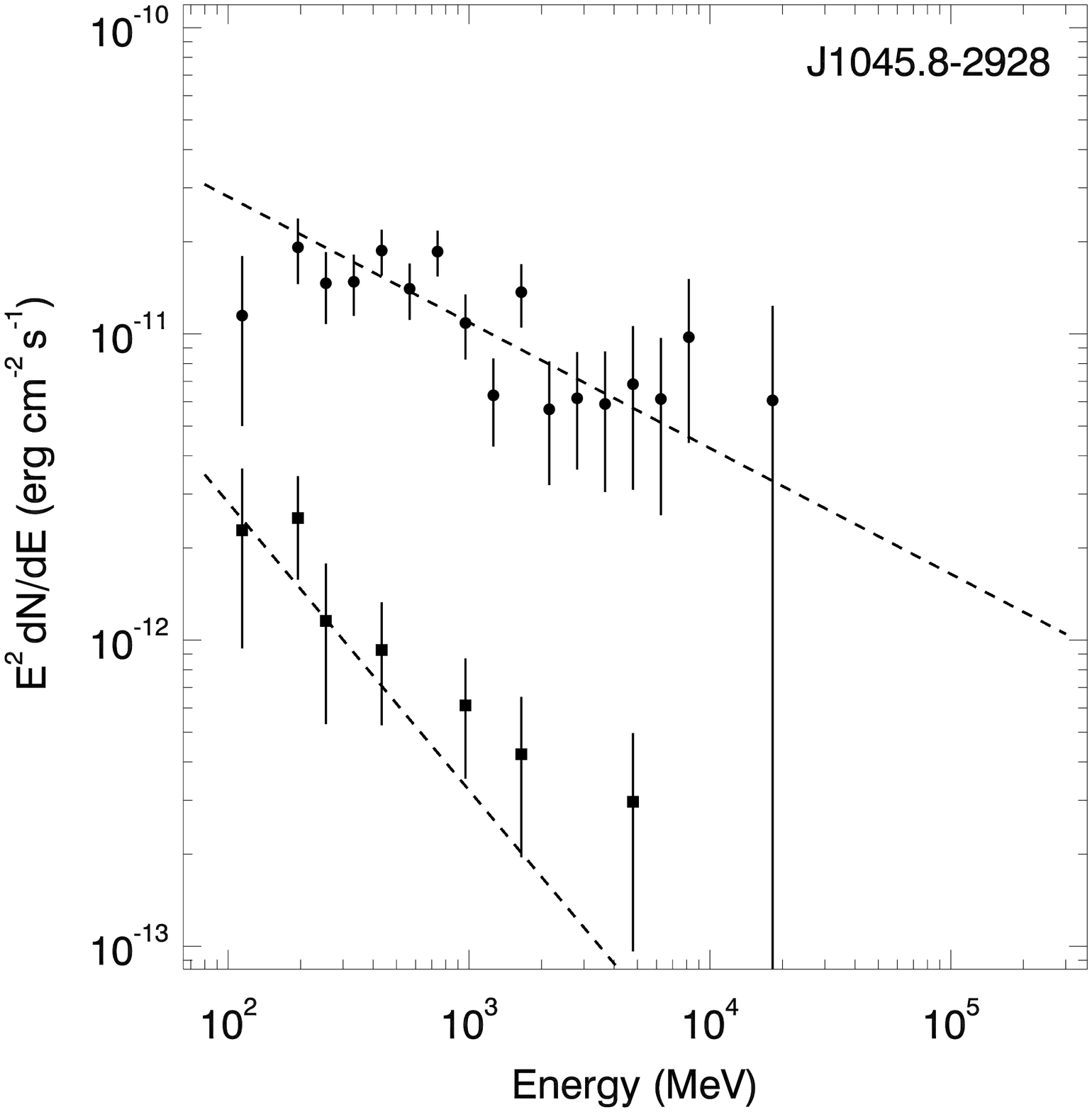}

\includegraphics[width=0.3\textwidth]{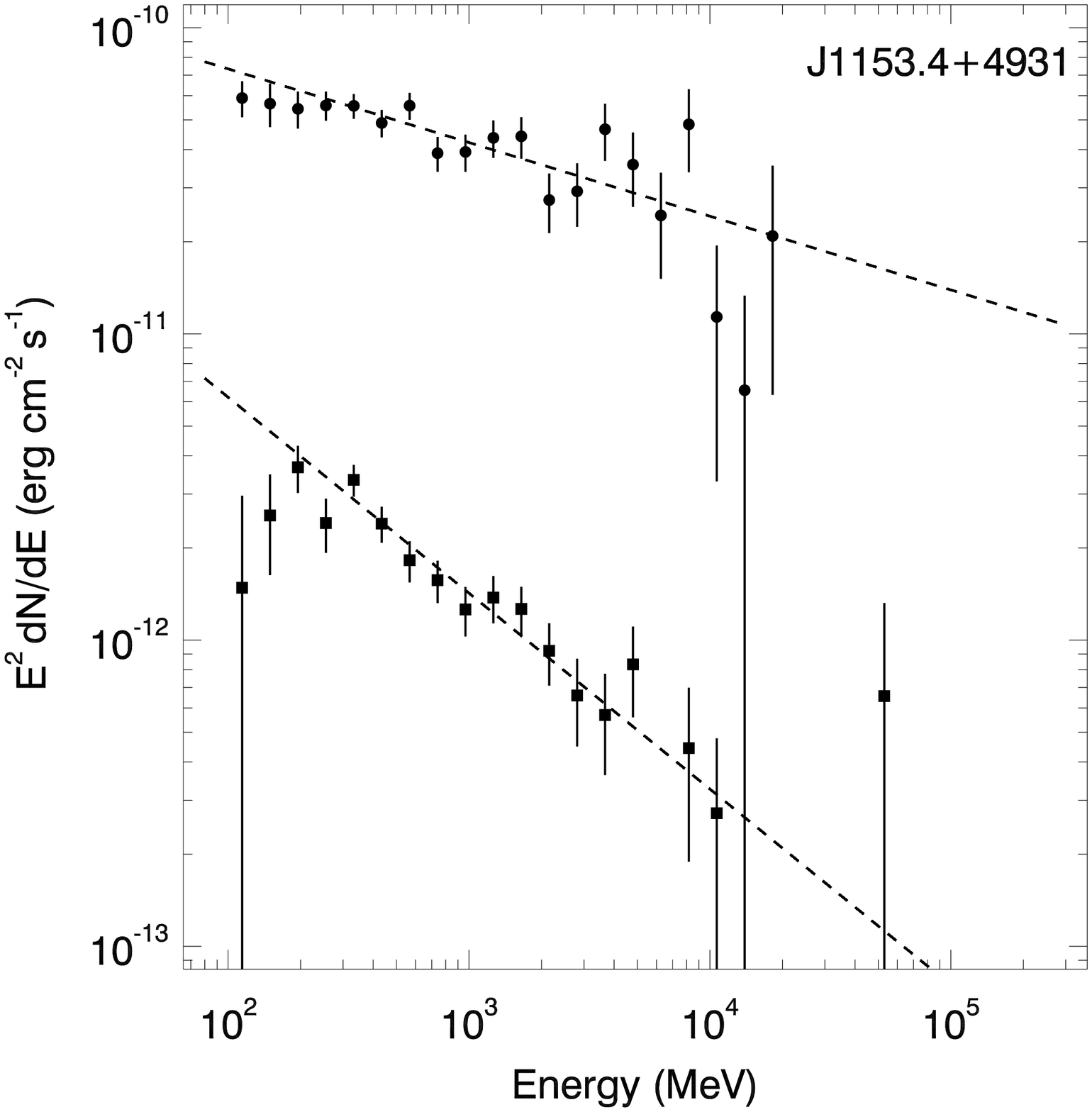}
\includegraphics[width=0.3\textwidth]{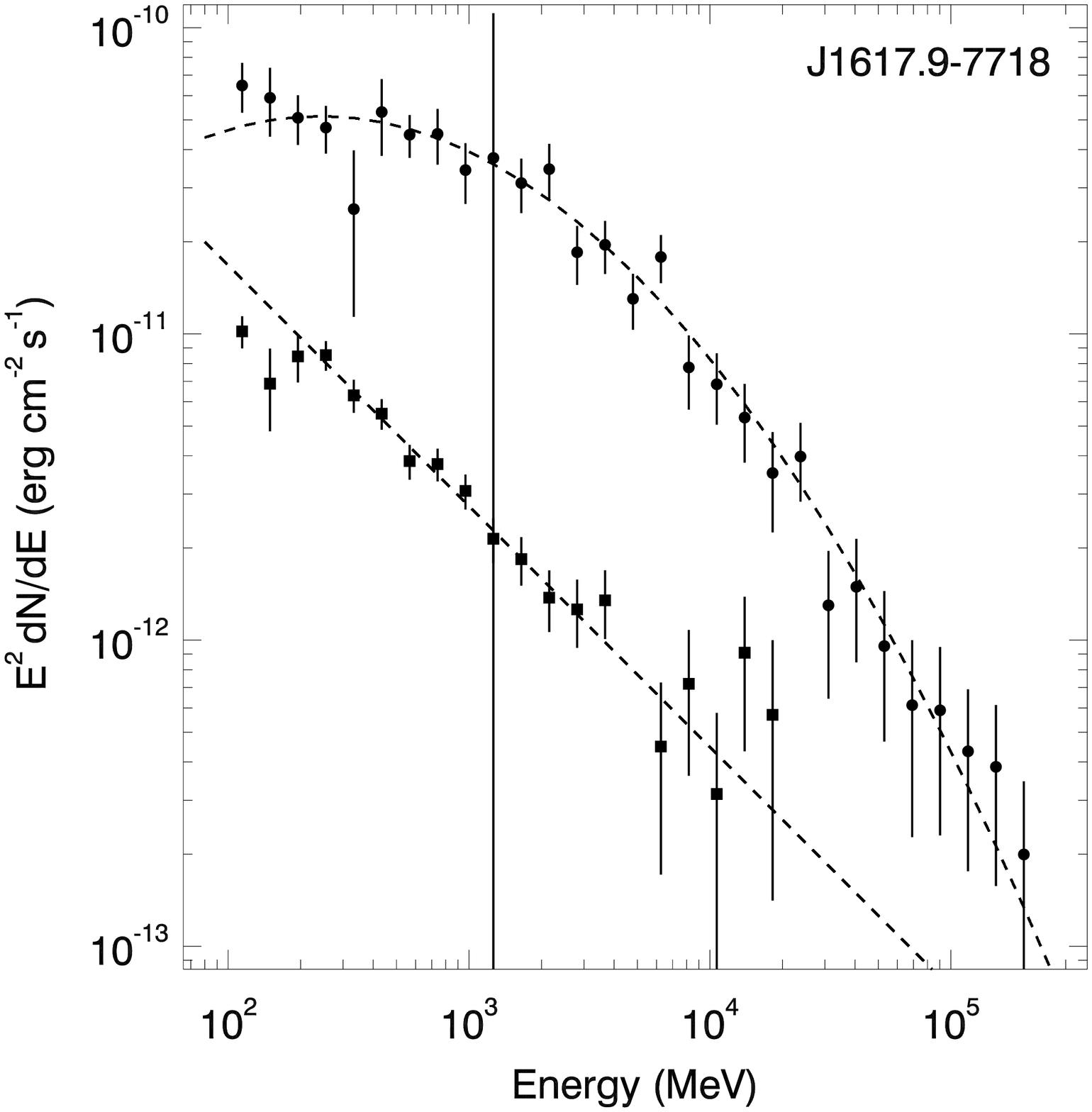}
\includegraphics[width=0.3\textwidth]{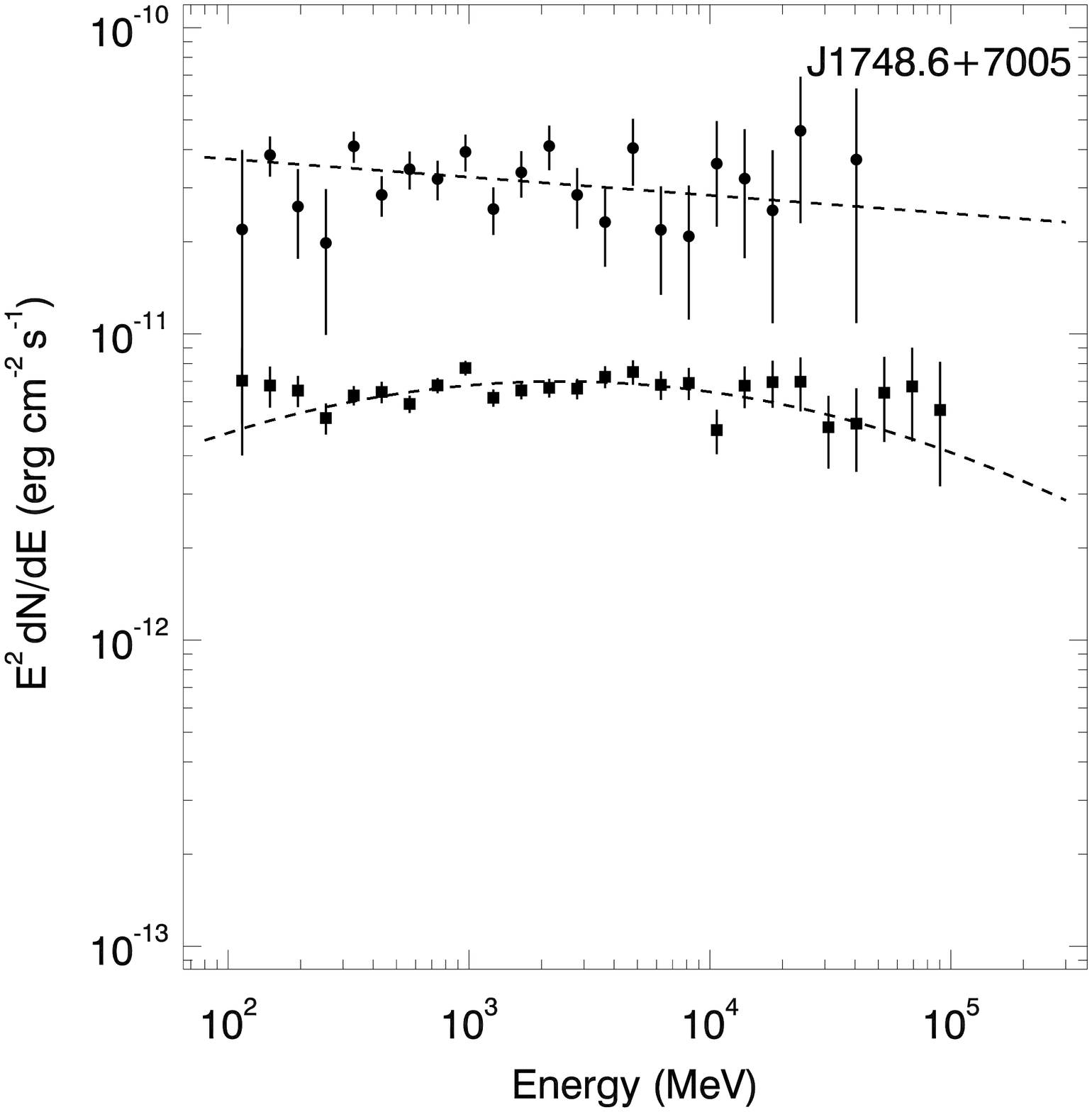}

\includegraphics[width=0.3\textwidth]{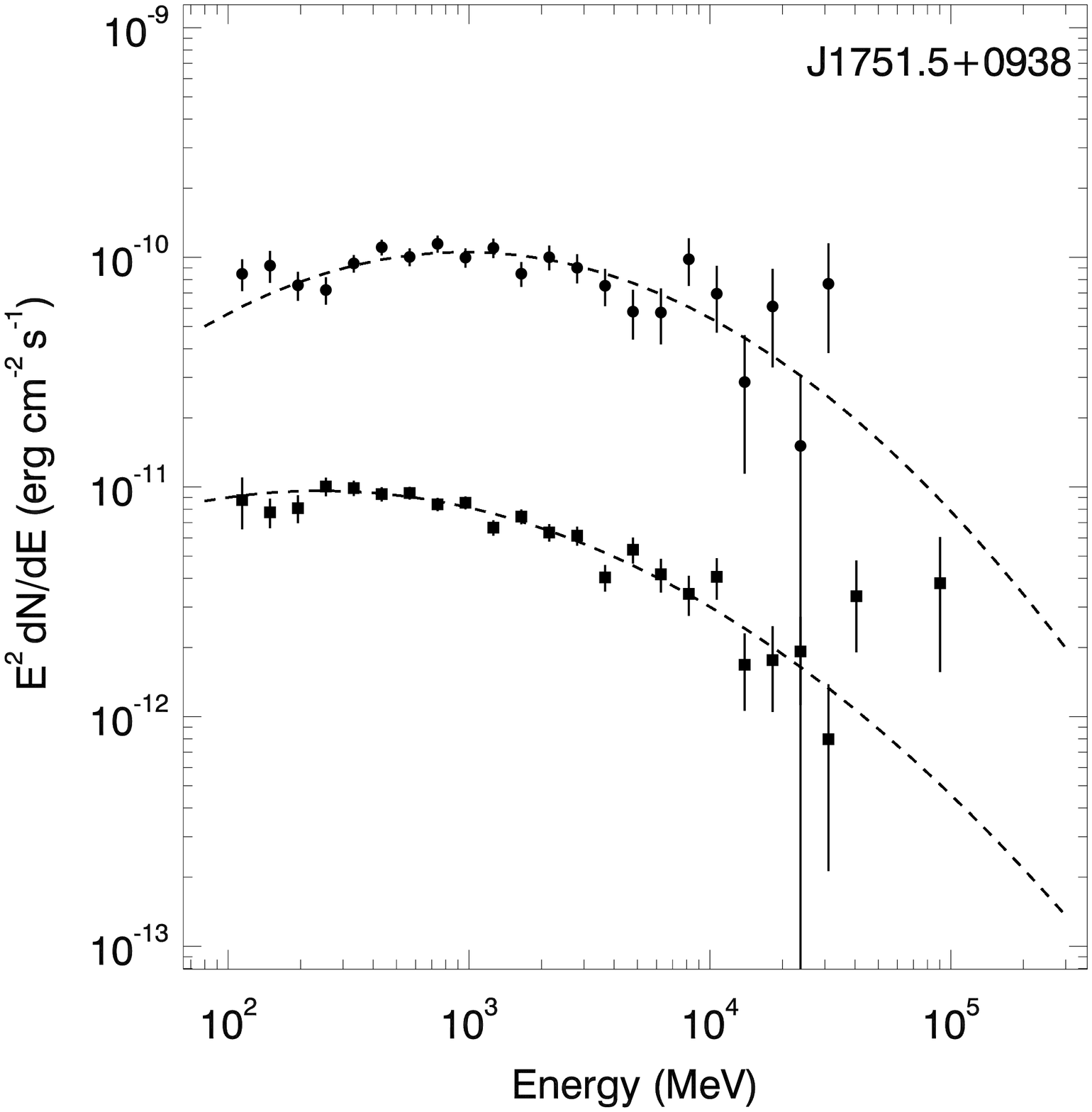}
\includegraphics[width=0.3\textwidth]{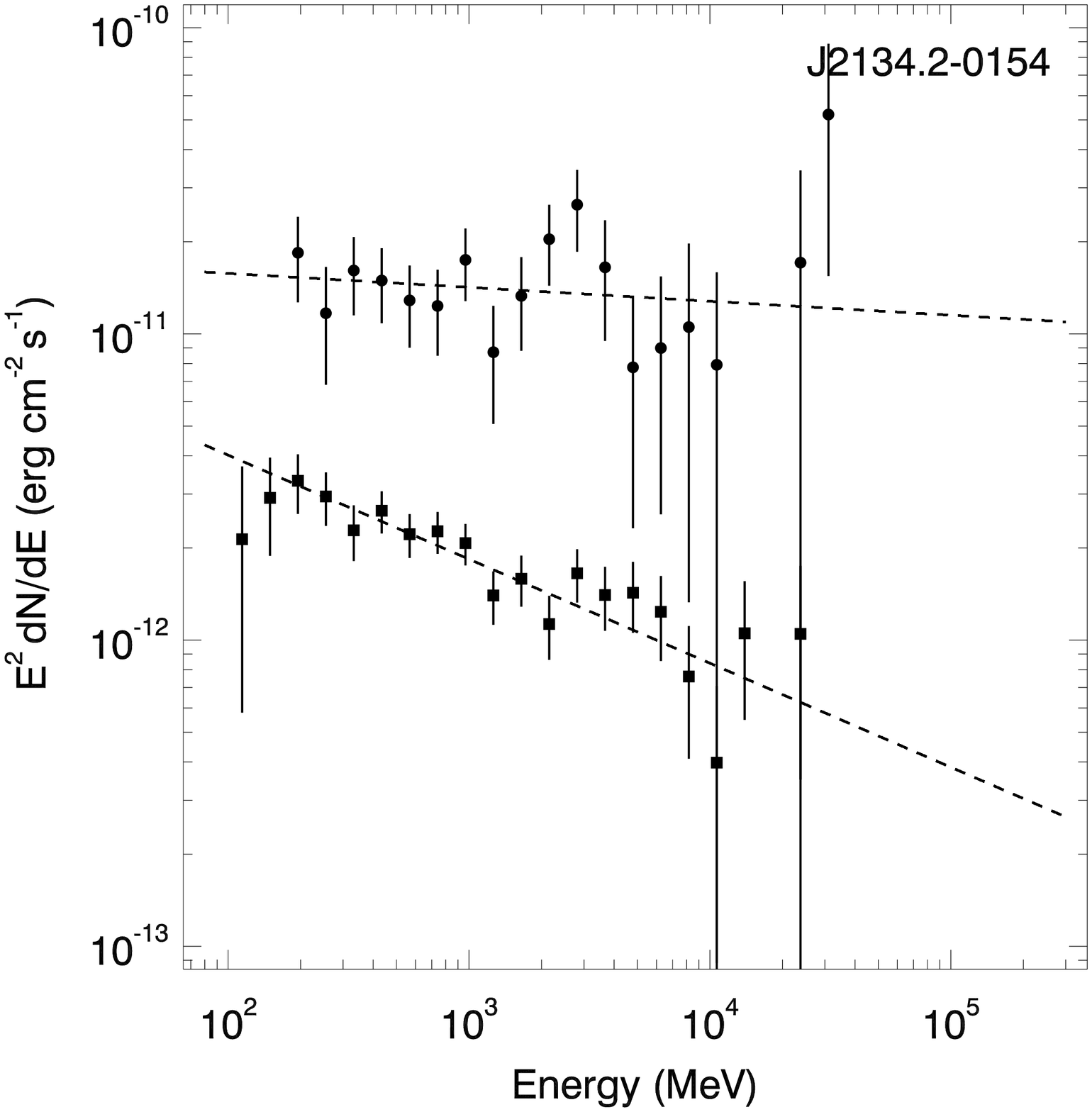}
\includegraphics[width=0.3\textwidth]{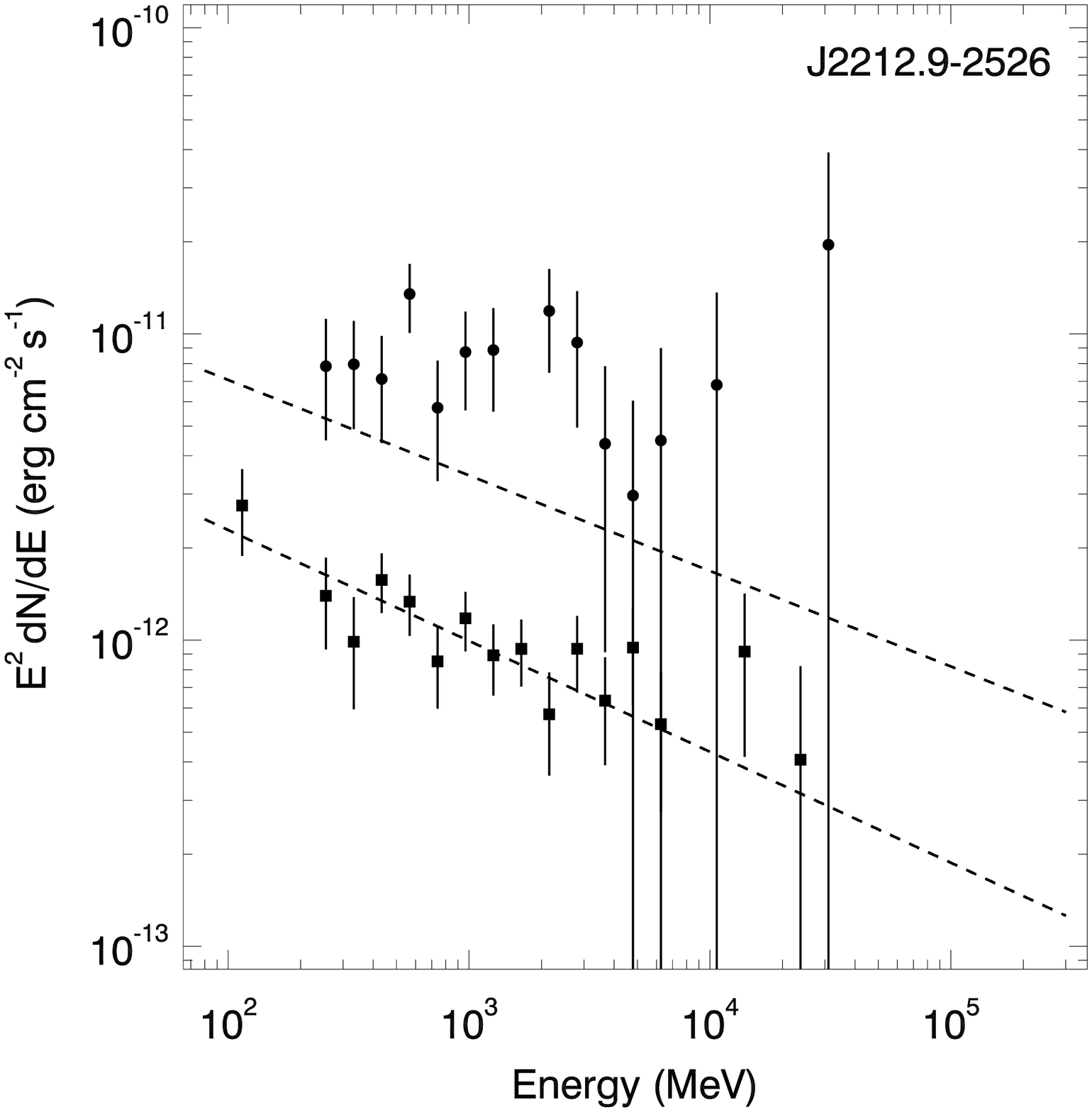}
\caption{Continued.}
\end{figure*}

The class type and redshift information for the blazars are from 
Chen (2018; see also \citealt{ace+15})
and three of them do not have redshifts.
In Figure~\ref{fig:gl}, we show their photon indices 
and luminosities in the quiescent states
(three sources without redshifts are not included).
Comparing to the whole \fermi\ blazars (cf. Figure 14 in \citealt{3fagn15}), 
our targets are in a slightly-brighter range, since
their luminosities are 
$\sim 10^{46}$--10$^{48}$\ erg\,s$^{-1}$, except one 
at $\sim 10^{43}$ erg\,s$^{-1}$.
The 90-day binned light
curves are shown in Figure~\ref{fig:lc}. 
As can be seen, 20 of them have only one single data point 
indicating a flare, and the other four have two or three data points in a flare.
We note that 20 of them were reported in the second catalog of flaring
\gr\ sources \citep{2fav}. Previously, the flare of 
J1153.4+4931 (4C +49.22) was studied with multi-wavelength observations
from radio to \gr\ \citep{cut+14}, and J0303.4$-$2407 was reported to
have a possible 2.1 yr quasi-periodic modulation in its \gr\ emission
\citep{zha+17}. 

\section{Data analysis} 
\label{sec:ab}

\subsection{Initial long-term light curve analysis of \fermi\ LAT data}
\label{sec:eve}

LAT scans the whole sky every three hours in the energy range from 
20 MeV to 300 GeV \citep{atw+09}. We selected 0.1--300 GeV LAT data from 
the \fermi\ Pass 8 database. For each of (candidate) blazars, 
a 20$^\circ\times 20^\circ$ square region of interest (ROI) was used,
centered at the position of the blazar given in 3FGL \citep{ace+15}. 
The time period of the LAT data is 9.5 years from 2008-08-04 15:43:36 (UTC) 
to 2018-02-08 23:52:17 (UTC). 
Following the recommendations of the LAT team\footnote{\footnotesize http://fermi.gsfc.nasa.gov/ssc/data/analysis/scitools/}, 
we selected events with zenith angles less than 90 deg to 
avoid possible contamination from the Earth's limb.

We constructed light curves binned in 90 day intervals by performing
standard binned maximum likelihood analysis.
The LAT science tools software package {\tt v10r0p5} was used.
The source model for each blazar was based on 3FGL, and
the normalization parameters and spectral indices of the sources 
within 5 deg from the target as well as sources within the ROI with variable 
index $\geq$ 72.44 \citep{ace+15} were set as free parameters.
All other parameters were fixed at their catalog values in 3FGL.
We used the original spectral models in 3FGL for the blazars. 
Galactic and extragalactic diffuse emission models
used were gll\_iem\_v06.fits and  iso\_P8R2\_SOURCE\_V6\_v06.txt, 
respectively.  The normalization parameters of the 
two diffuse emission components were set as free parameters. 

Long-term 90-day binned light curves of more than 1800 (candidate) blazars 
were obtained. We went over them and determined 24 blazars with a loner
flare.  Their light curves are shown in the top panels 
of Figure~\ref{fig:lc}, in which when
flux data points have the Test Statistic (TS)
values smaller than 9, we calculated their flux upper limits at 
a 95\% confidence level and plotted the upper limits in the light curves. 

In order to study detailed flux variations during each flare,
we chose a time period around each flare based on the 90-day binned light 
curves. The region of the time period of a source is 
indicated by the gray area in Figure~\ref{fig:lc}. 
 Excluding the flare data points in each gray region, those
far above the quiescent flux levels (mostly one or two data points),
we calculated the average
flux and standard deviation of each quiescent light curve. 
As shown in Figure~\ref{fig:lc}, the flare data points are outstanding,
most of which are at least 10$\sigma$ above the quiescent light curves.
We cautiously note that for several cases, 
there appear to be minor activities, such as in J0133.1$-$5201, J0303.4$-$2407,
and J1040.5+0617. However for these activities, only a few data points 
are slightly above the 1$\sigma$ ranges of the light curves. We thus still
considered them as in the quiescent states.

\subsection{Detailed Likelihood analysis}

Because of the recently released 4FGL, and updated database
(P8R3) and background files,
we performed detailed likelihood analysis based on them to the LAT data
for the 24 selected targets. The same source models were built but
based on 4FGL, with Galactic and extragalactic diffuse emission models
being gll\_iem\_v07.fits and iso\_P8R3\_SOURCE\_V2.txt, respectively.

For the 24 targets, 12 have emission that was described with a simple power
law ($dN/dE \propto E^{-\Gamma}$, where $\Gamma$ is the photon index),
and 12 have emission described with a log-parabola
($dN/dE \propto (E/E_0)^{-\alpha - \beta\log (E/E_0)}$, where
$\alpha$ and $\beta$ are spectral parameters. Note that for
the latter model when $\beta$ is small, it is close to being a power law.

We determined the spectral parameters in the quiescent state and for
the flare data points from the likelihood analysis. 
The values of the spectral parameters
$\Gamma_h$ (or $\alpha_h$ and $\beta_h$) in the flaring states 
and $\Gamma_l$ (or $\alpha_l$ and $\beta_l$) in the quiescent states
are given in Table~\ref{tab:src}. 

However from the \gr\ spectra we obtained 
(see below Section~\ref{sec:spe}),
we noted that five sources (J0221.1+3556, J0236.8$-$6136, J0629.3$-$1959,
J1617.9$-$7718, and J1748.6+7005), provided with spectral models of 
a log-parabola in 4FGL, might have power-law emission instead, particularly in 
the quiescent states. We re-ran the likelihood analysis assuming a power-law
model for them in the source models, and found that a  log-parabola
is not significantly preferred over a power law in the quiescent states of the first four 
sources, while for
J1748.6+7005, its emission in the flaring state can be described with
a power law. These comparisons were conducted by calculating 
$\sqrt{-2\log(L_{pl}/L_{logP})}$, 
where $L_{pl}$ and $L_{logP}$ are the maximum likelihood values obtained 
from a power law and a log-parabola respectively \citep{abd+13}.
Therefore in Table~\ref{tab:src}, we changed these sources' spectral parameters
accordingly.

\subsection{Fine light curves in flares}

For each time period around a flare we chose above, a fine light curve binned 
in 3-day intervals was constructed,  with the middle time of the light curve
corresponding to the middle time of the 90-day binned data points of
a flare. The choice of 3 days was based on
tests of different time intervals, which could well show the details of
a flare without having too many flux upper limits in a light curve.
For the data in each time bin, the same maximum likelihood analysis as
the above was performed.
Both light curves and TS curves are shown in the bottom
panels of Figure~\ref{fig:lc}.
Only the flux data points with TS$>$9 were kept in the 3-day
binned light curves. 
 Comparing to the average quiescent fluxes ($F_q$ given
in Figure~\ref{fig:lc}),
the light curves show that their high-flux data points in a flare are 
several tens of times brighter.
\begin{figure}
\centering
\includegraphics[scale=0.5]{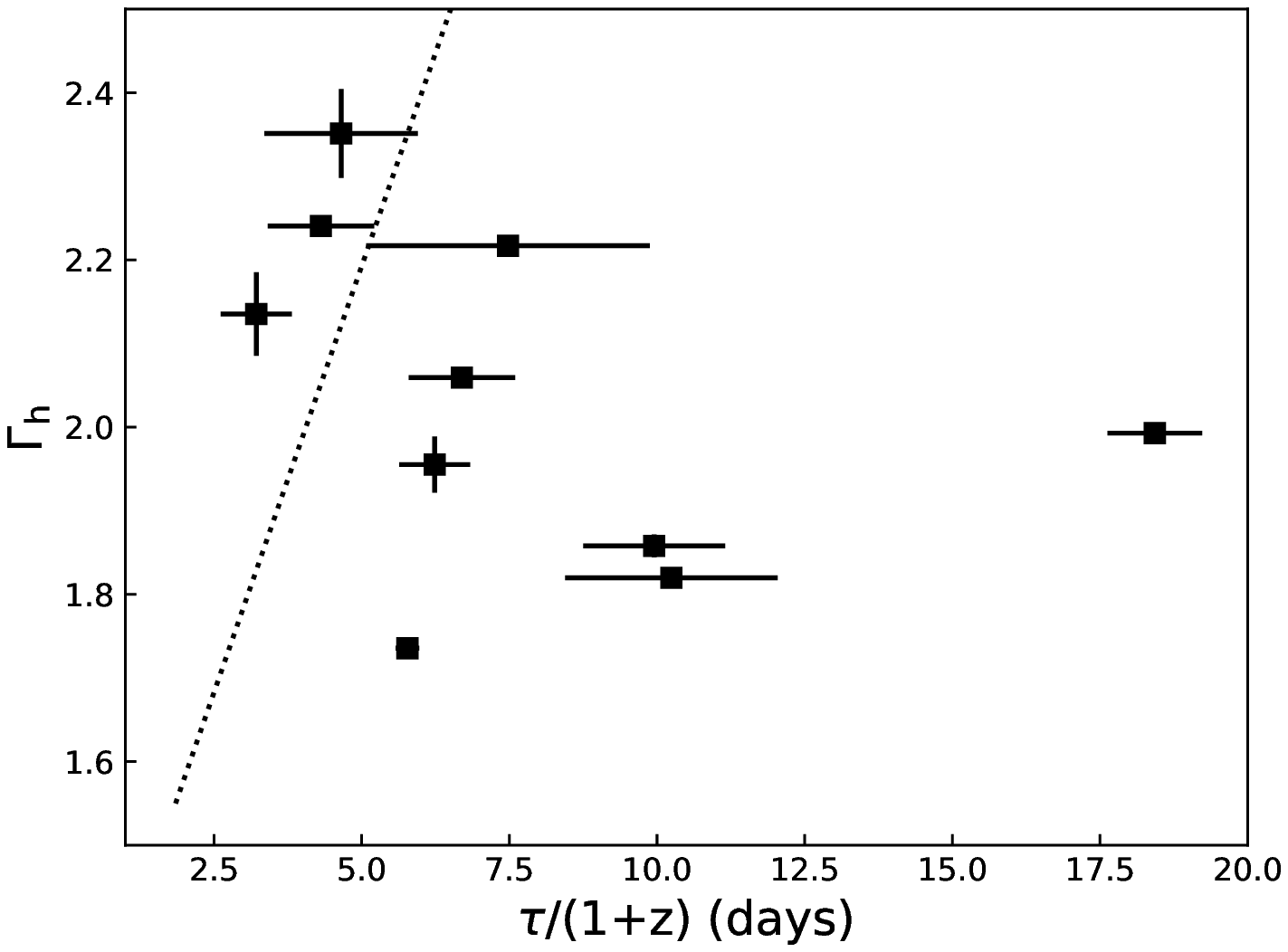}
\caption{Time durations of the sharp-peak flares at the local host galaxies.
The flares studied in \citet{nal13} are above the dotted line, showing
that most of our flares have longer time durations.}
\label{fig:td}
\end{figure}

\subsection{Spectral analysis}
\label{sec:spe}

We obtained \gr\ spectra for each target in its flaring and 
quiescent states. The energy range of 0.1 to 300 GeV 
was evenly divided logarithmically into 30 energy bands. 
The time ranges of the flaring high-flux data points in the 90-day binned light 
curves were taken as the flare time durations. In most cases, there is only
one data point, implying a 90-day time duration.
Excluding the flare data points in each light curves, the other data of 
a target were analyzed as in the quiescent state.
In this analysis, the normalization parameters of all the sources 
within an ROI
and the two diffuse emission components were set as free parameters. 
The obtained spectra are shown in Figure~\ref{fig:spec}, in which
only the spectral data points with TS$>$4 were kept. 
 As shown by the 3-day binned light curves (bottom panels of 
Figure~\ref{fig:lc}), the detailed flaring activities of a target often spread 
over part of the region that is not exactly around the center, which
is because of our coarse choice of 90-day time intervals.
The spectra in the flaring states thus
may not be as accurate as possible.
However, because the flares are dominantly bright,
the results are not significantly affected. We tested to perform likelihood
analysis to part of data containing the highest flux data points in
the 3-day binned light curves, the obtained
spectral results were nearly the same as those obtained from the flare data
points.

\section{Results}

While in 90-day binned light curves, the flares stand out above the low-flux
levels of the quiescent states, our detailed light curve analysis shows that
only 10 of them contain a sharp peak and
the other 14 generally do not have a clear profile.
We determined the properties of the 10 sharp-peak cases by
fitting their profiles with both
a Gaussian function  and an analytic function given in \citet{abd+10a}.
The former is given by
\begin{equation}
F(t)=F_{\rm c}+F_{0}e^{-(t-{t_{0})^{2}}/2\sigma_t^{2}}\ \ \ ,
\label{lcfiteq}
\end{equation}
where $F_{\rm c}$ and $F_{0}$ are the constant flux and height of a peak,
respectively, $t_{0}$ is the flux peak time,
and $\sigma_t$ is the standard deviation for measuring the peak time duration. 
 The latter can have an asymmetric profile, given by 
\begin{equation}
F(t)=F_{\rm c}+F_{0}(e^{(t_{0}-t)/T_r} + e^{(t-t0)/T_d})^{-1}\ \ \ ,
\end{equation}
where $T_r$ and $T_d$ are used to measure the rise and decay time separately.
The fitting results are given in Table~\ref{tab:fit}. We chose
the better ones based on the reduced $\chi^2$ values, which are overplotted
in the bottom panel of Figure~\ref{fig:lc}.  Although the fits seem 
poor (based on the $\chi^2$ values) and can not describe fine structures of 
the flares, they allowed us to estimate the time duration of a flare, which is 
the parameter commonly obtained for blazar flares. The time durations $\tau$
of these peaks, defined as when the flux is half of the peak value,
were calculated. They are in a range of approximately 
4--25 days. In Figure~\ref{fig:td}, the time durations at the host galaxies 
($\tau/(1+z)$) are shown.
In addition, comparing $F_{\rm c}$ and $F_0$ values with those of 
the quiescent states, the sharp peaks appear to have contributed primarily 
to the flaring events.

 In 4FGL, our 24 sources have emission half described with a power law and
half with a log-parabola. Our detailed analysis 
shows that in the quiescent states, four of 
the log-parabola sources are actually well described with a power law,
and J1748.6+7005 in the flaring state had emission consistent with being
a power law. We note that the spectrum of the last source in the quiescent
state also appears flat, although the likelihood analysis indicates that
the spectrum is better described with a log-parabola. In any case,
we have found four clear cases showing spectral form changes, from a power
law in the quiescent state to a log-parabola in the flaring state,
which are supported by the their spectra (Figure~\ref{fig:spec}) we obtained.
This result is slightly different from
previous studies such as in \citet{hcd14} and \citet{kn15}, since
they have found that nearly all the sources kept the same form of emission
during flares.

As indicated by both Figure~\ref{fig:spec} and Table~\ref{tab:src}, 
most sources (16 of them) had harder
emission when in flares, which is a pattern often seen in bright \gr\ emission
of blazars (e.g., \citealt{nal13,hay+15,bri+16}). For the other 8 sources,
five did not have significant spectral changes but three of them
(J0221.1+3556, J1040.5+0617, and J1748.6+7005) had softer emission
when in flares. Examining them, the 3-day binned light curve of
J0221.1+3556 shows a complex flaring pattern with two major peaks, while
the latter two 
had relatively ``active" quiescent states: J1040.5+0617 had as large as a 
factor of 3 flux variations and 
J1748.6+7005 showed minor activities 
following the flare, and returned to the lowest-flux level at the
end of its long-term light curve. These features are noted when comparing 
them to other sources.

It has been pointed out that since long integration times have to be used
in \gr\ observations, in order to collect statistically sufficient photons 
for analysis, spectra of blazars obtained in most cases are the results of 
variable emission added together \citep{kn15}. In our cases, the spectra of
the sources in quiescence might particularly have this problem because their
spectra were obtained from long-term integrations. We therefore chose 
three relatively bright sources: J0221.1+3556, J0510.0+1800, and J0958.7+6534,
as they were detected in nearly all the 90-day bins.
We checked their spectral properties in one year 
time durations before and after the flares respectively. 
The results during the two time periods are consistent within uncertainties;
the sources did not have any significant changes in spectra.
We thus conclude that the sources likely have relatively stable emission
in quiescence before and after the flare.

\section{Discussion}
\label{sec:dis}

Analyzing the 9.5 year \gr\ light curves obtained for the \fermi\ blazars
and candidates, we have selected a sample of 24 sources from more 
than
1800 of them on the basis of the detection of a clear high-flux flaring event 
in the long-term light curves. The sample differs from other blazars as
in most time they have been in a quiescent state with relatively stable 
emission. We obtained detailed light
curves for the flares, and found that 10 of them contain a sharp peak.
The time 
durations of the flares were determined to be in a range of 4--25 days. 
For most of the other sources, the flares
were revealed to be due to random high-flux activity or consist of multiple 
minor flares, which can not be described by a simple function of time.

\begin{figure}
\centering
\includegraphics[scale=0.5]{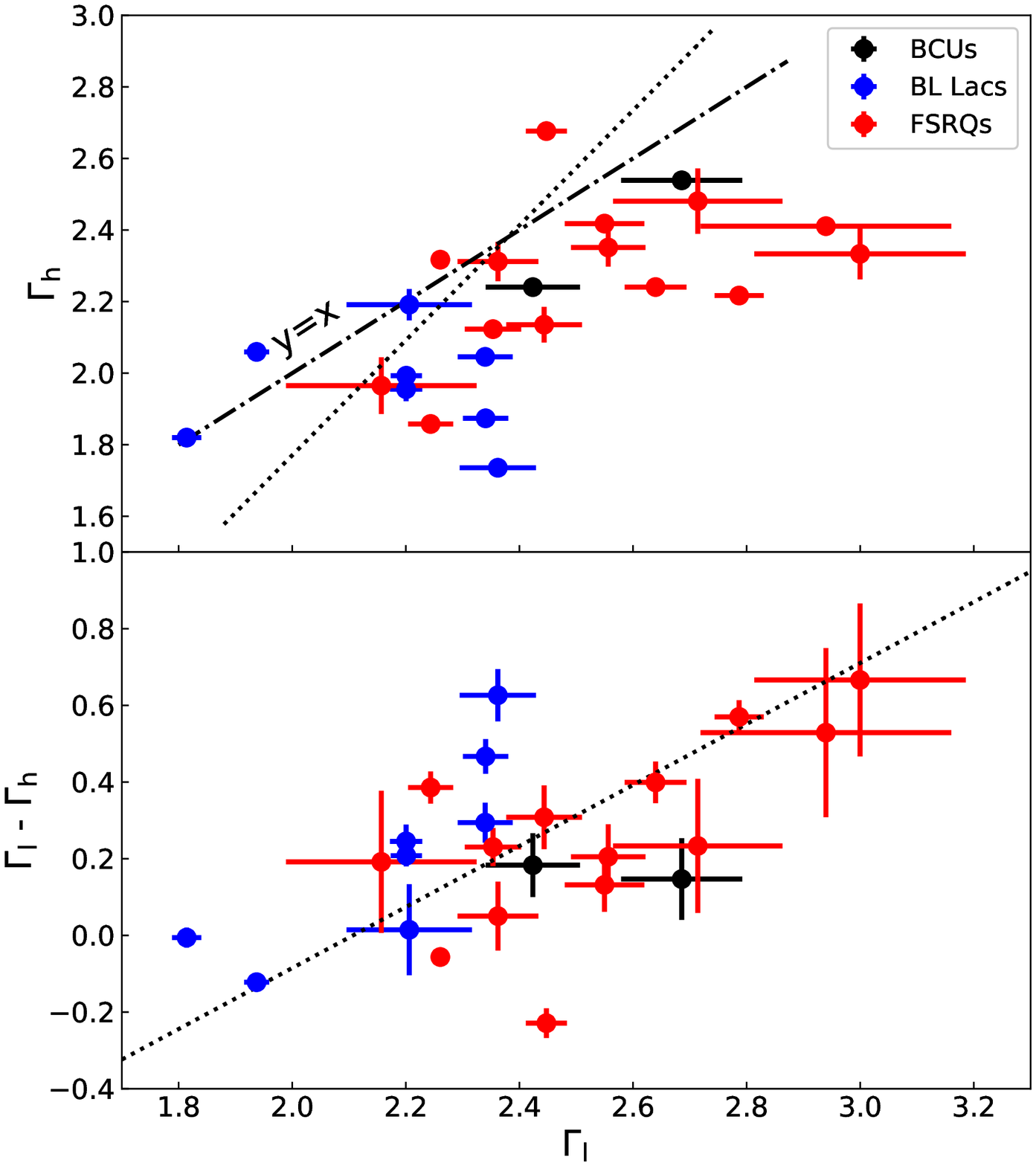}
\caption{{\it Top panel:} photon indices of 24 blazars in the quiescent 
($\Gamma_l$ or $\alpha_l$) and flaring ($\Gamma_h$ or $\alpha_h$) states. 
A possible correlation of
$\Gamma_h \sim 1.6\Gamma_l$ is indicated by a dotted line in the top panel.
For a comparison, a $\Gamma_h = \Gamma_l$ line (long-dash dotted line)
is also shown. {\it Bottom panel:} $\Gamma_l - \Gamma_h$ versus $\Gamma_l$,
with a possible relationship shown as a dotted line.}
\label{fig:gam}
\end{figure}

In the scenario of having many turbulent cells,
multiple subpulses/flares can be naturally produced. Different sizes of 
emission regions/cells or different magnetization parameters 
($\sigma=P_{B}/P_{k}$, where $P_{B}$ and $P_{k}$ are the magnetic and 
kinetic power, respectively) can result in various spike patterns on the 
overall light curve \citep{zz14}. This turbulent process may thus be the reason 
for the generally two types of profiles in the flares of our sample.
Multiple minor flares could be due to enhanced emission from multiple 
turbulent cells, and a sharp-peak flare would arise from
a single dominant turbulent cell.
We note that the sharp-peak events
have relatively long time durations.  For example, in Figure~\ref{fig:td},
a dotted line marks the approximate low boundary of the bright flares studied 
in \citet{nal13}. Note that the time durations of these flares were defined as
the flux doubling time plus the flux halving time, but in any case
only three of our sharp-peak flares are in the region (Figure~\ref{fig:td}). 
The other seven are not, with the time durations at the host galaxies 
in a range of 5--19 days. In addition, the emission in these seven cases
tended to be hard, five of them having $\Gamma_h$ (or $\alpha_h$) smaller than
2.0, among the lowest in our
sample (see Figure~\ref{fig:gam}). We thus suspect that these
hard-emission, sharp peak flares may provide additional insight into 
the physical properties of a single turbulent cell.

More than half of our sample sources showed harder emission during the flares, 
while five of them did not have any significant changes and three of them 
had softer emission in their flares. We searched through possible correlations 
between fluxes/luminosities and photon indices during the flaring and quiescent 
states,  where for log-parabola cases we used $\alpha_h$ or $\alpha_l$ as 
the indices since they define the slope of a spectrum.
Nothing significant was found except a possible correlation
between $\Gamma_h$ and $\Gamma_l$: $\Gamma_h \sim 1.6\Gamma_l$ 
(see the top panel of Figure~\ref{fig:gam}), 
The Pearson correlation coefficient is 0.62. However it is hard to derive
obvious physical connections between them, as why the enhanced emission
during flares would be in some way connected to quiescent emission. 
If we plot $\Gamma_l - \Gamma_h$ versus $\Gamma_l$ instead (the bottom panel
of Figure~\ref{fig:gam}), a trend of ($\Gamma_l-\Gamma_h) \sim 0.8\Gamma_l$ is 
seen, although with  a couple of data points significantly deviating away from
the  trend. The trend may be understood as that
the larger $\Gamma_l$, the larger the changes. In other words, as most
blazars turn to be harder when they are brighter in flares,  
and if $\Gamma_h$ are in a certain range, sources with
larger deviations from the range during quiescence would have
larger changes to jump back during flares. Further investigation on 
this possible property can
be conducted with studying a large sample of flares with clear quiescent levels.

In our cases, the spectra of FSRQs in either the flares or quiescence 
do not show any clear breaks (or cutoffs) at the GeV energy band,  and
half of them show spectra that extend beyond $\sim 20$\,GeV.
The presence of significant spectral breaks at GeV energies
in several blazars has been
suggested due to the absorption by photons 
from the broad line region (BLR) through the pair production process 
\citep{fd10,ps10},  while later large-sample studies have indicated
that the absorption is not universal and most blazars do not present 
the spectral feature (e.g., \citealt{cos+18,msb19}).
Therefore, the $\gamma$-ray emission regions of
our blazar sources are
expected to lie beyond the BLR \citep{lb06}, which are consistent 
with SED modeling of a large sample of blazars \citep{kcw14}.
The emission regions may still be within the dust torus,
and relativistic electrons in a jet will inverse Compton (IC) scatter 
the infrared (IR) photons from the dust torus, producing the observed
$\gamma$-rays. As a result, the electrons will be cooled with time scale 
$T_{cooling}\approx2.5\times10^{7}\left(\nu_{ext}/\nu_{IC}\right)^{1/2}\Gamma_j^{-3/2}\delta_j^{-1/2}U_{ext}^{-1}$~s, 
where $\nu_{ext}\sim 3\times 10^{13}$\,Hz and 
$U_{ext}\sim 3\times10^{-4}$ erg cm$^{-3}$ are the peak frequency and energy
density of the torus IR radiation. Applying the typical values for
the bulk Lorentz factor $\Gamma_j$ and Doppler beaming factor $\delta_j$
of a blazar jet, $\Gamma\approx\delta\approx10$, $T_{cooling}\sim 0.1$ days
for $\sim$1\,GeV $\gamma$-rays ($\nu_{IC}=1$ GeV=$2.4\times10^{23}$ Hz). 
The cooling time is much smaller than the observed time durations of 
the sharp-peak flares (4--25 days), indicating that the energetic electrons 
in these jet cases may be in-situ accelerated.

As a summary, our studies of these 24 loner flares have revealed the different 
variation profiles in their detailed light curves and the spectral property of 
having harder emission in flares in most cases. Different from
those obtained in bright flares, clear spectral form changes were seen
in four cases. 
We have also found a possible trend
between $\Gamma_l - \Gamma_h$ and $\Gamma_l$ and a tendency of having
long flaring time durations and hard emission in the sharp-peak flare cases. 
However, our sample is small, and no firm conclusions
can be drawn for these two possible features. 
For the purpose of enlarging our sample, similar studies can be carried out
by including those sources with multiple flares in their long-term light 
curves.  Given 1000 more blazars are preliminarily listed in the 
fourth \fermi\ LAT AGN catalog \citep{4fagn19}, similar work can also be 
extended to the new sources.
Thus statistically significant relations may be established, 
helping our understanding of blazar flares and related physical processes.  


\bigskip
This research made use of the High Performance Computing Resource in the Core
Facility for Advanced Research Computing at Shanghai Astronomical Observatory.
This research was supported by the National Program on Key Research 
and Development Project (Grant No. 2016YFA0400804) and the National Natural 
Science Foundation of China (11633007, U1738131).


\clearpage
\begin{sidewaystable}
  \caption{\mbox{24 blazar sources and their spectral properties}}
  \label{tab:src}
\centering
  \begin{tabular}{lccllccccc}
\hline
Source name    & Class  & $z$ &  Association &  Spectral model  & $\Gamma_h$ ($\alpha_h$  & $\beta_h$ & $\Gamma_l$ ($\alpha_l$ & $\beta_l$ & $\Delta \Gamma$  \\ 
\hline
J0055.1$-$1219 & BCU & - & TXS 0052$-$125 & PowerLaw         & 2.24 $\pm$ 0.002  & \qquad  & 2.42 $\pm$ 0.08  & \qquad     &   0.18 $\pm$ 0.08  \\
J0116.0$-$1136 & CF & 0.670 & PKS 0113$-$118 & PowerLaw         & 2.35 $\pm$ 0.05  & \qquad  & 2.56 $\pm$ 0.07 & \qquad     &   0.21 $\pm$ 0.08 \\
J0133.1$-$5201 & UB & 0.020 & PKS 0131$-$522 & LogParabola         & 1.74 $\pm$ 0.01 &  0.11 $\pm$ 0.004  & 2.36 $\pm$ 0.07 & 0.11  $\pm$ 0.03   &   0.63 $\pm$ 0.07  \\
J0221.1+3556$^{\ast}$  & CF & 0.685 & S3 0218+35 & LogParabola         & 2.32 $\pm$ 0.01  & 0.10  $\pm$ 0.004 & 2.26 $\pm$ 0.02  & \qquad    &  -0.06 $\pm$ 0.02  \\
J0236.8$-$6136$^{\ast}$ & CF & 0.465 & PKS 0235$-$618 & LogParabola         & 2.14 $\pm$ 0.05 &  0.06 $\pm$ 0.03  & 2.44 $\pm$ 0.07 & \qquad    & 0.31 $\pm$ 0.08   \\
J0257.9$-$1215 & CF & 1.391 & PMN J0257$-$1211 & PowerLaw         & 2.33 $\pm$ 0.07  & \qquad  & 3.00 $\pm$ 0.19  & \qquad   &  0.67 $\pm$ 0.20 \\
J0303.4$-$2407 & CB & 0.260 & PKS 0301$-$243 & LogParabola         & 1.82 $\pm$ 0.002 & 0.04  $\pm$ 0.001  & 1.81 $\pm$ 0.03 & 0.03   $\pm$ 0.01   &  -0.006 $\pm$ 0.03   \\
J0343.2$-$2529 & CF & 1.419 & PKS 0341$-$256 & LogParabola         & 1.97 $\pm$ 0.08 & 0.39  $\pm$ 0.06  & 2.16 $\pm$ 0.17 & 0.22  $\pm$ 0.09   &  0.19 $\pm$ 0.19  \\
J0354.7+8009 & CB & - & S5 0346+80 & PowerLaw         & 2.19 $\pm$ 0.04 & \qquad    & 2.21 $\pm$ 0.11  & \qquad    &  0.01 $\pm$ 0.12  \\
J0401.7+2112 & CF & 0.834 & TXS 0358+210 & PowerLaw         & 2.42 $\pm$ 0.001 & \qquad   & 2.55 $\pm$ 0.07 & \qquad    &  0.13 $\pm$ 0.07   \\
J0510.0+1800 & CF & 0.416 & PKS 0507+17 & LogParabola         & 1.86 $\pm$ 0.01  &  0.11 $\pm$ 0.01 & 2.24 $\pm$ 0.04 & 0.07  $\pm$ 0.02   & 0.39 $\pm$ 0.04  \\
J0540.8$-$5415 & CF & 1.185 & PKS 0539$-$543 & PowerLaw         & 2.48 $\pm$ 0.09 & \qquad   & 2.71 $\pm$ 0.15 & \qquad    &  0.23 $\pm$ 0.17   \\
J0629.3$-$1959$^{\ast}$ & CB & 1.724 & PKS 0627$-$199 & LogParabola         & 1.87 $\pm$ 0.02 &  0.12   $\pm$ 0.01 & 2.34 $\pm$ 0.04 & \qquad   &  0.47 $\pm$ 0.05   \\
J0742.6+5443 & CF & 0.720 & GB6 J0742+5444 & LogParabola       & 2.12 $\pm$ 0.001 &  0.08 $\pm$ 0.001  & 2.35 $\pm$ 0.05 & 0.07  $\pm$ 0.03   &  0.23 $\pm$ 0.05  \\
J0958.7+6534 & CB & 0.368 & S4 0954+65 & LogParabola         & 1.99 $\pm$ 0.004 & 0.04  $\pm$ 0.002  & 2.20 $\pm$ 0.03 & 0.04$\pm$ 0.01     &  0.21 $\pm$ 0.03  \\
J1038.8$-$5312 & BCU & - & MRC 1036$-$529 & PowerLaw     & 2.54 $\pm$ 0.001 & \qquad   & 2.69 $\pm$ 0.11  & \qquad   &  0.15 $\pm$ 0.11  \\
J1040.5+0617  & UF & 2.715 & GB6 J1040+0617 & PowerLaw         & 2.68 $\pm$ 0.01  & \qquad    & 2.45 $\pm$ 0.04  & \qquad    &  -0.23 $\pm$ 0.04   \\
J1045.8$-$2928 & CF & 2.128 & PKS B1043$-$291 & PowerLaw         & 2.41 $\pm$ 0.001 & \qquad   & 2.94 $\pm$ 0.22  & \qquad    &  0.53 $\pm$ 0.22 \\
J1153.4+4931 & CF & 0.925 & 4C +49.22 & PowerLaw         & 2.24 $\pm$ 0.002 & \qquad   & 2.64 $\pm$ 0.05  & \qquad    &  0.40 $\pm$ 0.05  \\
J1617.9$-$7718$^{\ast}$ & CF & 1.710 & PKS 1610$-$77 & LogParabola    & 2.22 $\pm$ 0.01 & 0.13  $\pm$ 0.004  & 2.79 $\pm$ 0.04  & \qquad  &  0.57 $\pm$ 0.04   \\
J1748.6+7005$^{\ast}$ & CB & 0.770 & S4 1749+70 & LogParabola         & 2.06 $\pm$ 0.002 & \qquad  & 1.94 $\pm$ 0.02 & 0.04 $\pm$ 0.01    &   -0.12 $\pm$ 0.02   \\
J1751.5+0938 & CB & 0.322 & OT 081 & LogParabola         & 1.96 $\pm$ 0.03 & 0.12  $\pm$ 0.02  & 2.20 $\pm$ 0.03 & 0.08 $\pm$ 0.02  &  0.25 $\pm$ 0.04   \\
J2134.2$-$0154 & CB & 1.283 & PKS 2131$-$021 & PowerLaw         & 2.05 $\pm$ 0.02 & \qquad   & 2.34 $\pm$ 0.05  & \qquad    &  0.29 $\pm$ 0.05  \\
J2212.9$-$2526 & CF & 1.833 & PKS 2210$-$25 & PowerLaw         & 2.31 $\pm$ 0.06  & \qquad  & 2.36 $\pm$ 0.07  & \qquad   &  0.05 $\pm$ 0.09  \\
\hline 
\hline
\end{tabular} \\
\begin{minipage}[]{160mm}
{\small Notes: Class and redshift information for the sources are 
from \citet{che18}, where ``UF" 
are blazar candidates of uncertain type (BCUs) classified 
as FSRQs, ``UB" the BCUs classified as BL Lacs, and ``CF" and ``CB" are 
confirmed FSRQs and BL Lacs, respectively. $^{\ast}$: sources without
$\beta_h$ or $\beta_l$ are described with a power law instead.}
\end{minipage}
\end{sidewaystable}

\clearpage
\begin{table}
  \caption{\mbox{Fitting results for the flaring peaks in 10 sources}}
  \label{tab:fit}
\centering
  \begin{tabular}{lccccccc}
\hline
Source name    & $F_{c}$                 & $F_{0}$               & $t_{0}$        & $\sigma_t$    & $T_{r}$ & $T_{d}$ & $\chi^2/dof$ \\
\qquad    &   ($\times10^{-7}$)   &   ($\times10^{-7}$)   & (MJD)         &  (day)      
& (day) & (day) & \qquad   \\
\hline
J0116.0$-$1136 & 0.8 $\pm$ 0.1          & 6.1 $\pm$ 0.8    &  55151.2 $\pm$ 0.9   & 4.8   $\pm$ 0.6  & & &  76.1/22   \\
    & 1.0 $\pm$ 0.1   & 12.7 $\pm$ 2.5   & 55154.8 $\pm$ 0.9  & & 5.4 $\pm$ 1.1 & 0.5 $\pm$ 0.2 & 64.0/21  \\
J0133.2$-$5201 & 0.3 $\pm$ 0.1          & 18.5 $\pm$ 1.3  &  58077.7  $\pm$  0.5   & 2.5   $\pm$ 0.1  &  & &  96.0/12   \\
    & 0.1 $\pm$ 0.1   & 24.0 $\pm$ 2.3   & 58080.0 $\pm$ 0.7  &
    & 4.1 $\pm$ 0.4   & 1.3 $\pm$ 0.2    & 121.8/11  \\
J0236.8$-$6136 & 1.9 $\pm$ 0.3          & 11.3 $\pm$ 2.7  &  55360.1   $\pm$ 1.3       & 2.0   $\pm$ 0.3  & & &  32.8/9   \\
    & 2.0 $\pm$ 0.4   & 13.1 $\pm$ 6.7   & 55361.6 $\pm$ 1.9  &
    & 3.0 $\pm$ 1.3   & 0.8 $\pm$ 0.7    & 39.5/8  \\
J0303.4$-$2407 & 0.5 $\pm$ 0.1          & 4.8 $\pm$ 0.8    &  55319.1  $\pm$ 1.5     & 4.7   $\pm$ 0.6  & & &  60.9/9   \\
    & 0.0 $\pm$ 0.2   & 9.5 $\pm$ 1.8    & 55313.8 $\pm$ 1.6  &
    & 0.8 $\pm$ 0.3   & 9.0 $\pm$ 1.5    & 21.5/8  \\
J0510.0+1800 & 1.4 $\pm$ 0.1          & 16.7 $\pm$ 1.7     &  56412.7   $\pm$  1.2   & 4.7   $\pm$ 0.3  & & &  170.0/23   \\
    & 0.5 $\pm$ 0.1   & 31.0 $\pm$ 4.7   & 56409.5 $\pm$ 1.3  &
    & 2.6 $\pm$ 0.2   & 8.1 $\pm$ 1.0    & 95.6/22 \\
J0958.7+6534 & 0.1 $\pm$ 0.1          & 9.3 $\pm$ 0.5    &  57072.5  $\pm$ 0.7      & 10.7  $\pm$ 0.4  & & &  135.9/21   \\
    & 0.0 $\pm$ 0.1   & 18.6 $\pm$ 2.1   & 57066.6 $\pm$ 1.6  &
    & 5.7 $\pm$ 0.4   & 12.2 $\pm$ 1.8   & 149.6/20  \\
J1153.4+4931 & 1.8 $\pm$ 0.3          & 10.6 $\pm$ 1.2   &  55696.8  $\pm$ 1.1       & 3.4   $\pm$ 0.3  & & &  23.6/10   \\
    & 1.6 $\pm$ 0.3   & 17.5 $\pm$ 2.2   & 55693.8 $\pm$ 1.1  &
    & 0.9 $\pm$ 0.2   & 5.4 $\pm$ 0.7    & 16.9/9  \\
J1617.9$-$7718 & 0.5 $\pm$ 0.3          & 5.3 $\pm$ 0.7   & 57597.5  $\pm$ 1.0       & 8.6   $\pm$ 1.2  & & &  34.8/17   \\
    & 0.3 $\pm$ 0.3   & 11.8 $\pm$ 1.8   & 57597.9 $\pm$ 1.5  &
    & 6.9 $\pm$ 1.4   & 6.6 $\pm$ 1.6    & 34.3/16  \\
J1748.6+7005 & 0.1 $\pm$ 0.0         & 7.8 $\pm$ 0.6    & 55620.4  $\pm$ 0.2     & 8.8   $\pm$ 0.3  & & &  43.4/18   \\
    & 0.1 $\pm$ 0.0   & 16.9 $\pm$ 1.3   & 55620.7 $\pm$ 0.3  &
    & 5.0 $\pm$ 0.5   & 4.0 $\pm$ 0.4    & 40.2/17  \\
J1751.5+0938 & 3.3 $\pm$ 0.4          & 21.3 $\pm$ 4.6   &  57588.3  $\pm$ 4.1      & 3.5   $\pm$ 0.3  & & &  4.3/12   \\
    & 3.4 $\pm$ 0.4   & 54.4 $\pm$ 16.4  & 57589.4 $\pm$ 5.8 &
    & 2.2 $\pm$ 0.4   & 2.1 $\pm$ 0.2    & 28.0/11  \\
\hline 
\hline
\end{tabular} \\
\begin{minipage}[]{160mm}
{\small Notes:
$F_{c}$ and $F_{0}$ are in units of ${\rm ph}~{\rm cm}^{-2}~{\rm s}^{-1}$. 
$\chi^2/dof$ are $\chi^2$ value and degrees of freedom for the fitting
to the sharp-peak flares.}
\end{minipage}
\end{table}

\end{document}